\documentclass[11pt]{article}
\pdfoutput=1
\usepackage[usenames,dvipsnames,svgnames,table]{xcolor} 
\usepackage[obeyspaces,hyphens,spaces]{url}
\usepackage{jcapmod}

\usepackage[english]{babel}
\usepackage{amsmath, amssymb, amsbsy, amstext, amsthm,mathtools}
\usepackage{hyperref}
\usepackage{graphicx}
\usepackage{amsfonts}
\usepackage{fouridx}
\usepackage{cases}
\usepackage{bm}
\usepackage{bbm}
\usepackage{cleveref}
\usepackage{slashed}
\usepackage{mdframed}
\usepackage{tensor}

\usepackage{setspace}

\usepackage{braket}

\usepackage{subcaption}

\usepackage{tikz}
\usetikzlibrary{calc,shadings,patterns,tikzmark,fadings, decorations.markings, decorations.pathmorphing}

\definecolor{Blue}{rgb}{0.25, 0.41, 0.88}
\definecolor{Red}{rgb}{0.92,0.,0.}
\definecolor{darkorange}{rgb}{1.0,0.549,0.}
\definecolor{cobalt}{RGB}{44, 98, 120}
\definecolor{Mathematica1}{rgb}{0.368417, 0.506779, 0.709798}
\definecolor{Mathematica2}{rgb}{0.880722, 0.611041, 0.142051}
\definecolor{Mathematica3}{rgb}{0.560181, 0.691569, 0.194885}
\definecolor{Mathematica4}{rgb}{0.922526, 0.385626, 0.209179}
\definecolor{Mathematica5}{rgb}{0.528488, 0.470624, 0.701351}
\definecolor{Mathematica6}{rgb}{0.772079, 0.431554, 0.102387}
\definecolor{Mathematica7}{rgb}{0.363898, 0.618501, 0.782349}
\definecolor{Mathematica8}{rgb}{1, 0.75, 0}
\definecolor{Mathematica9}{rgb}{0.647624, 0.37816, 0.614037}
\definecolor{plotBlue}{RGB}{94, 130, 181}
\definecolor{plotRed}{RGB}{233, 85, 54}
\definecolor{plotGreen}{RGB}{142, 176, 50}
\definecolor{plotPurple}{RGB}{135, 120, 178}
\definecolor{cornellRed}{HTML}{B31B1B}
\definecolor{cornellBlue}{HTML}{0068AC}
\definecolor{cornellGreen}{HTML}{6EB43F}

\newcolumntype{C}[1]{>{\centering\let\newline\\\arraybackslash\hspace{0pt}}m{#1}}

\def\e{{\lab{e}}}

\newcommand{\SHc}[1]{{Y^{\raisebox{1pt}{$\scriptstyle*$}}}_{\hspace{-7pt}#1}}

\newcommand{\VEV}[1]{\left\langle #1 \right\rangle}

\newcommand{\tFo}[4]{{}_2 F_1\!\left[\genfrac..{-1pt}{0}{\raisebox{-1pt}{$#1, \,\, #2$}}{\raisebox{1pt}{$#3$}} \, \bigg| \, #4\, \right]}
\newcommand{\pFq}[5]{{}_{#1} F_{#2}\!\left[\genfrac..{-1pt}{0}{\raisebox{-1pt}{$#3$}}{\raisebox{1pt}{$#4$}} \, \bigg| \, #5\, \right]}

\newcommand{\tFoS}[4]{{}_2 F_1\!\left[\genfrac..{0pt}{1}{#1, \,\, #2}{#3} \, \Big| \, #4\, \right]}

\newcommand{\G}[2]{\Gamma\!\left[\genfrac..{-1pt}{0}{\raisebox{-1pt}{$#1$}}{\raisebox{1pt}{$#2$}}\right]}

\newcommand{\GB}[2]{\Gamma\!\left[\genfrac..{0pt}{0}{#1}{#2}\right]}

\makeatletter
\newlength{\apb@width}
\newcommand{\autoparbox}[2][c]{\settowidth{\apb@width}{#2}\parbox[#1]{\apb@width}{#2}}

\makeatother

\makeatletter
\newsavebox\myboxA
\newsavebox\myboxB
\newlength\mylenA

\newcommand*\xoverline[2][0.75]{
    \sbox{\myboxA}{$\m@th#2$}%
    \setbox\myboxB\null
    \ht\myboxB=\ht\myboxA%
    \dp\myboxB=\dp\myboxA%
    \wd\myboxB=#1\wd\myboxA
    \sbox\myboxB{$\m@th\overline{\copy\myboxB}$}
    \setlength\mylenA{\the\wd\myboxA}
    \addtolength\mylenA{-\the\wd\myboxB}%
    \ifdim\wd\myboxB<\wd\myboxA%
       \rlap{\hskip 0.5\mylenA\usebox\myboxB}{\usebox\myboxA}%
    \else
        \hskip -0.5\mylenA\rlap{\usebox\myboxA}{\hskip 0.5\mylenA\usebox\myboxB}%
    \fi}
\makeatother

\numberwithin{equation}{section}

\newcommand{\ud}{\mathrm{d}}
\newcommand{\lab}[1]{{\mathrm{#1}}}
\newcommand{\mb}[1]{{\mathbf{#1}}}
\newcommand{\minus}{{\scalebox {0.8}[1.0]{$-$}}}
\newcommand{\sminus}{{\scalebox {0.6}[0.85]{$-$}}}

\theoremstyle{definition}

\DeclareRobustCommand{\SkipTocEntry}[4]{}

\newcommand{\es}{\hspace{0.5pt}}

\definecolor{pyBlue}{RGB}{31, 119, 180}
\definecolor{pyRed}{RGB}{214, 39, 40}
\definecolor{pyGreen}{RGB}{44, 160, 44}
\definecolor{pyBlue2}{RGB}{0, 111, 237}
\definecolor{pyRed2}{RGB}{224, 52, 36}

\newcommand{\slab}[1]{{\textsc{#1}}}

\tikzstyle{intSty}=[draw=white, thick, line width=0.24mm]
\tikzstyle{inflSty}=[cornellRed]

\begin{document}

\pagenumbering{roman}
\begin{titlepage}
\baselineskip=15.5pt \thispagestyle{empty}

\bigskip\

\vspace{1cm}
\begin{center}
{\fontsize{18}{24}\selectfont  {\bfseries Light Scalars at the Cosmological Collider}}
\end{center}
\vspace{0.1cm}
\begin{center}
{\fontsize{12}{18}\selectfont Priyesh Chakraborty and John Stout} 
\end{center}

\begin{center}
\vskip8pt
\textit{Department of Physics, Harvard University, Cambridge, MA 02138, USA}

\end{center}

\vspace{1.2cm}
\hrule \vspace{0.3cm}
\noindent {\bf Abstract}\\[0.1cm]
	We study the self-energies of weakly interacting scalar fields in de Sitter space with one field much lighter than the Hubble scale.  We argue that self-energies drastically simplify in this light limit. We illustrate this in theories with two scalar fields, one heavy and one light, interacting with one another through either cubic or quartic interactions. To regulate infrared divergences, we compute these self-energies in Euclidean de Sitter space and then carefully analytically continue to Lorentzian signature. In particular, we do this for the most general renormalizable theory of two scalar fields with even interactions to leading order in the coupling and the mass of the light field. These self-energies are determined by de Sitter sunset diagrams, whose analytic structure and UV divergences we derive. Even at very weak couplings, the light field can substantially change how the heavy field propagates over long distances. The light field's existence may then be inferred from how it modifies the heavy field's oscillatory contribution to the primordial bispectrum in the squeezed limit, i.e. its cosmological collider signal.
\vskip10pt
\hrule
\vskip10pt

\end{titlepage}

\newcommand\emd{\xi}
\newcommand\imd{\zeta}

\thispagestyle{empty}
\setcounter{page}{2}
\begin{spacing}{1.03}
\tableofcontents
\end{spacing}

\clearpage
\pagenumbering{arabic}
\setcounter{page}{1}

\newpage

\newpage

\section{Introduction}
	
	Light, weakly-coupled scalar fields are ubiquitous in modern high energy physics. Not only do they often appear in solutions to various theoretical puzzles---for example, in the axion solution to the Strong CP problem~\cite{Peccei:1977ur,Peccei:1977hh,Weinberg:1977ma,Wilczek:1977pj,Essig:2013lka}, as viable dark matter candidates~\cite{Preskill:1982cy,Dine:1982ah,Abbott:1982af,Hu:2000ke,Cooley:2022ufh}, in solutions to the electroweak hierarchy problem~\cite{Graham:2015cka,Arkani-Hamed:2016rle,Hook:2018jle,TitoDAgnolo:2021pjo,Craig:2022uua}---they also arise naturally, and often in great numbers, in low energy theories consistent with quantum gravity~\cite{Arvanitaki:2009fg,Demirtas:2018akl,Demirtas:2021gsq}. If such light scalars exist, they must be coupled so weakly to the Standard Model to have so far evaded detection. In light of the substantial theoretical pressure towards their existence, it is important to find new ways of detecting them and to understand their impact on other more directly observable fields. The goal, then, of this paper is to study the impact such light, weakly-coupled scalars can have on primordial cosmological signals, specifically in the very light and very weakly-coupled limit.

	It is widely believed that our universe once enjoyed a phase of rapid cosmic inflation at energy scales far beyond those accessible in terrestrial experiments, with an associated Hubble scale that could have been as high as $H \sim \mathcal{O}(10^{14}\, \lab{GeV})$. Information about extremely high-energy processes could then be imprinted upon inflationary correlators which, in turn, would affect future observations of the cosmic microwave background and our universe's large-scale structure~\cite{Meerburg:2019qqi}. In particular, they would change the inferred primordial bispectrum, or equivalently the three-point function of the co-moving curvature perturbation $\langle \zeta_{\mb{k}_1} \zeta_{\mb{k}_2} \zeta_{\mb{k}_3} \rangle$. This may be characterized by a dimensionless ``shape'' function $S(k_1, k_2, k_3)$,  
	\begin{equation}
		\langle \zeta_{\mb{k}_1} \zeta_{\mb{k}_2} \zeta_{\mb{k}_3} \rangle \equiv \frac{(2 \pi)^4 P^2_\zeta}{(k_1 k_2 k_3)^2}  (2 \pi)^3 \delta^{(3)}\big(\mb{k}_1 + \mb{k}_2 + \mb{k}_3) S(k_1, k_2, k_3)\,,
	\end{equation}
	where $P_\zeta \simeq 2 \times 10^{\sminus 9}$ is the amplitude of the scalar power spectrum and $k_i = |\mb{k}_i|$. Unfortunately, because their energy is relatively small during inflation, light and weakly-coupled scalars often have a very small effect on density fluctuations~\cite{Marsh:2015xka}. Similarly, their effect on the shape $S(k_1, k_2, k_3)$ is difficult to distinguish\footnote{Recent work, however, has shown that the shape function can receive important contributions from isocurvature modes excited by such light scalars~\cite{Lu:2021gso,Chen:2023txq}}
 from the so-called local shape \cite{Chen:2009zp,Chen:2010xka}, making their impact on inflationary correlation functions either ambiguous or unobservable.

	In contrast, a heavy scalar $\sigma$, with $m_\sigma > \frac{3}{2} H$, imparts an unambiguous  ``cosmological collider'' signal \cite{Chen:2009we,Chen:2009zp,Chen:2010xka, Baumann:2011nk,Chen:2012ge,Arkani-Hamed:2015bza,Assassi:2012zq,Noumi:2012vr,Lee:2016vti,Flauger:2016idt,Wang:2022eop} onto the bispectrum in its so-called squeezed limit, $k_1 \approx k_2 \gg k_3$. Assuming that $\sigma$ couples directly to the curvature perturbation $\zeta$, it will contribute to the bispectrum via the tree-level exchange that can be depicted diagrammatically as
	\begin{equation}
			\def\circSize{0.6}
			\begin{tikzpicture}[thick, baseline=-28pt]
				\coordinate (c1) at (-1.0, -1.75);
				\coordinate (c2) at (1.0, -1.75);
				\coordinate (c3) at (-1.75, 0);
				\coordinate (c4) at (-0.25, 0);
				\coordinate (c5) at (1.75, 0);

				\begin{scope}[shift={(0, -1.75)}]
					\draw (c1) -- (c2) node[midway, below] {$\sigma$};
				\end{scope}

				\draw[inflSty] (c1) -- (c3) node[midway, shift={(-0.3, 0)}] {$\zeta_{\mb{k}_1}$};
				\draw[inflSty] (c4) -- (c1) node[midway, shift={(0.4, 0)}] {$\zeta_{\mb{k}_3}$};
				\draw[inflSty] (c2) -- (c5) node[midway, shift={(0.4, 0)}] {$\zeta_{\mb{k}_2}$};
				\draw[line width=0.6mm, gray] (-2.75, 0) -- (2.75, 0);
				\fill[cornellRed, intSty] (c1) circle (0.07);
				\fill[cornellRed, intSty] (c2) circle (0.07);
				\fill[cornellRed, intSty] (c3) circle (0.07);
				\fill[cornellRed, intSty] (c4) circle (0.07);
				\fill[cornellRed, intSty] (c5) circle (0.07);
			
				\end{tikzpicture} \label{eq:treeLevelSig}
			\end{equation}
			where the top-most grey line denotes the time at which inflation ends. This depicts a process in which two $\zeta$ curvature perturbations (in [{\color{cornellRed}\emph{red}}]) and a $\sigma$ particle (in [\emph{black}]) are spontaneously created from the vacuum. The $\sigma$ particle then evolves freely until it eventually decays into a third $\zeta$, ultimately correlating the fluctuations of $\zeta$ at three distinct points. In the squeezed limit, this $\sigma$ particle is long-lived and freely propagates over large distances, oscillating at a frequency determined by its rest mass. It thus acquires a phase proportional to this frequency and the distance it propagates, which in turn imparts a characteristic oscillatory signature onto the shape function of the form
			\begin{equation}\label{eq:cosmo_coll}
				S(k_1 \approx k_2 \gg k_3) \sim \mathcal{A} \left(\frac{k_3}{k_1}\right)^{\beta}\! \sin \!\left[\omega \log\left(\frac{k_3}{k_1}\right) + \delta\right]\,,
			\end{equation}
			where the parameters $\mathcal{A}$, $\beta$, $\omega$ and $\delta$ correspond to the amplitude, rate of decay, frequency, and phase of these oscillations and are all calculable given a specific model. In particular, for the tree-level exchange of a heavy scalar $\sigma$ shown above, the decay rate $\beta$ and frequency $\omega$ are completely controlled by how $\sigma$ propagates freely in de Sitter space. The decay rate $\beta = \frac{1}{2}$ is determined by how quickly fluctuations in $\sigma$ dilute due to Hubble expansion and is the same for all free scalar fields, while $\omega = \sqrt{m_\sigma^2/H^2 - 9/4}$ is its frequency at ``rest'' in de Sitter.

			Unfortunately, a light particle $\varphi$ with mass $m_\varphi < \frac{3}{2} H$ that directly couples to the curvature perturbation does not impart such an oscillatory signature in bispectrum. However, even in the absence of a direct coupling to the inflaton, a light scalar $\varphi$ will modify how the heavy scalar $\sigma$ propagates over large distances~\cite{Lu:2021wxu}, imbuing it with non-trivial self-energy. Diagrammatically, in the presence of $\varphi$ the process (\ref{eq:treeLevelSig}) becomes
			\begin{equation}
				\def\circSize{0.6}
				\begin{tikzpicture}[thick, baseline=-28pt]
					\coordinate (c1) at (-1.5, -1.75);
					\coordinate (c2) at (1.5, -1.75);
					\coordinate (c3) at (-2.25, 0);
					\coordinate (c4) at (-0.75, 0);
					\coordinate (c5) at (2.25, 0);

					\begin{scope}[shift={(0, -1.75)}]
						
						\draw (c1) -- (-\circSize, 0) node[midway, below] {$\sigma$};
						\draw (c2) -- (\circSize, 0) node[midway, below] {$\sigma$};
						\begin{scope}
							\fill[white] (0, 0) circle (\circSize);
							\clip[draw] (0, 0) circle (\circSize);
							\foreach \x in {-0.575, -0.5, ..., 0.575} 
							{	
								\draw[rotate=45, thin] (-\circSize, \x) -- (\circSize, \x);
							}
						\end{scope}
						\fill[black, intSty] (-\circSize, 0) circle (0.07);
						\fill[black, intSty] (\circSize, 0) circle (0.07);
					\end{scope}

					\draw[inflSty] (c1) -- (c3) node[midway, shift={(-0.3, 0)}] {$\zeta_{\mb{k}_1}$};
					\draw[inflSty] (c4) -- (c1) node[midway, shift={(0.4, 0)}] {$\zeta_{\mb{k}_3}$};
					\draw[inflSty] (c2) -- (c5) node[midway, shift={(0.4, 0)}] {$\zeta_{\mb{k}_2}$};
					\draw[line width=0.6mm, gray] (-2.75, 0) -- (2.75, 0);
					\fill[cornellRed, intSty] (c1) circle (0.07);
					\fill[cornellRed, intSty] (c2) circle (0.07);
					\fill[cornellRed, intSty] (c3) circle (0.07);
					\fill[cornellRed, intSty] (c4) circle (0.07);
					\fill[cornellRed, intSty] (c5) circle (0.07);
				\end{tikzpicture}\,,
			\end{equation}
			where we use a hatched blob to denote the exact $\sigma$ propagator. Specifically, interactions with $\varphi$  will cause $\sigma$ to decay \emph{faster} than any free field~\cite{Marolf:2010zp}, and so $\beta > \frac{1}{2}$. Interestingly, these effects are seemingly \emph{enhanced} as the field becomes lighter and lighter, causing a massive suppression of the cosmological collider signal as $m_\varphi \to 0$. This enhancement, as we show in the main text, is directly tied to the behavior of the light scalar in the infrared, which is known to fluctuate violently in the limit of light mass~\cite{Allen:1985ux,Starobinsky:1986fx,Rajaraman:2010xd,Hollands:2011we,Beneke:2012kn,Burgess:2009bs,Burgess:2010dd,Chen:2016nrs,LopezNacir:2016gzi,Gorbenko:2019rza,Mirbabayi:2019qtx,Mirbabayi:2020vyt,Cohen:2020php,Baumgart:2020oby,Cohen:2021fzf}.

			To leading order in the slow roll parameter, we can extract this induced decay by studying $\sigma$'s self-energy in pure de Sitter space. The main goal of this paper is to understand how to compute the self-energies of both the heavy scalar $\sigma$ and light scalar $\varphi$ in the limit $m_\varphi/H \to 0$. Interacting quantum field theory in de Sitter is a notoriously rich\footnote{This may be interpreted as a euphemism for ``complicated.''} subject, and many treatments are plagued by infrared divergences. We choose to work first in Euclidean de Sitter~\cite{Marolf:2010zp,Marolf:2010nz,Higuchi:2010xt,Hollands:2010pr,Hollands:2011we,Higuchi:2021sxn} in which correlations functions are free of infrared divergences and then analytically continue to Lorentzian signature. This analytic continuation is subtle~\cite{Miao:2013isa} but can be done via the Froissart-Gribov inversion formula~\cite{Correia:2020xtr,Hogervorst:2021uvp,DiPietro:2021sjt,Loparco:2023rug,Gribov:2003nw,Newton:2002stw} and we explain how to take advantage of the small parameter $m_\varphi/H$ in making cosmological predictions. We explain how to efficiently extract physical predictions from this formula, and find that the self-energies induced by cubic and quartic interactions drastically simplify in the limit $m_\varphi \to 0$, though our methods are straightforwardly applicable to any two-vertex loop diagram. Furthermore, we explain how to efficiently extract the singularity structure of, and thus physically meaningful information from, these diagrams. This is especially useful as the complete analytic expressions can be exceedingly complex (see, e.g.~\cite{Marolf:2010zp}). We will argue that, in this light limit, only a few of these singularities are needed to determine the infrared behavior of two-point functions in de Sitter. We also explain how to extract the ultraviolet divergences of these contributions, and do so for the two-loop ``sunset'' diagram. Our perturbative approximations can only be trusted at suitably weak couplings, in a sense that we make precise in the main text, and so our results only apply to light fields that are also very weakly coupled.

			\vspace{8pt}
			\noindent \textbf{Outline} In Section~\ref{sec:review}, we review the basic properties of free scalar fields in de Sitter space. Specifically, in \S\ref{sec:geometry}, we first describe the geometry of $D$-dimensional de Sitter space in Euclidean and Lorentzian signatures. Free fields in Euclidean de Sitter space permit a useful momentum space representation in terms of the hyperspherical harmonics, which we review in \S\ref{sec:freeFields}. There, we use this representation to derive the free field propagator and discuss its analytic continuation from Euclidean signature to Lorentzian signature, via the Watson-Sommerfeld transformation. This transformation requires that one use the ``correct'' momentum space representation of a correlator, and we explain in \S\ref{sec:linv} how this is provided by the Froissart-Gribov or Lorentzian inversion formula, illustrating its use by applying it to the free field propagator. 

			Our main results appear in Section~\ref{sec:loop}. After reviewing general aspects of two-point functions of interacting quantum fields in de Sitter, we derive the self-energy of a heavy scalar $\sigma$ induced by a cubic interaction with a light scalar $\varphi$ in \S\ref{sec:bubble} and explain how it simplifies in the limit $m_\varphi/H \to 0$. This section is meant to illustrate the techniques we use, and so we only study a particular contribution to the self-energy, the so-called bubble diagram. In \S\ref{sec:sunset}, we study the most general (renormalizable) theory of a heavy scalar $\sigma$ interacting with a light scalar $\varphi$, with a $\mathbb{Z}_2 \times \mathbb{Z}_2$ global symmetry $\sigma \to \minus \sigma$ and $\varphi \to \minus \varphi$. We derive the self-energies of both $\sigma$ and $\varphi$ to leading order in perturbation theory. These are determined by the so-called sunset diagram which we study in the limit $m_\varphi \to 0$. Finally, we present our conclusions and mention some future directions in Section~\ref{sec:conclusions}.

			In Appendix~\ref{app:formulary}, we have collected various definitions and conventions for the special functions we use in the main text. In Appendix~\ref{app:unique}, we discuss the uniqueness of the interpolations used in the analytic continuation of two-point functions from Euclidean to Lorentzian signature. Appendix~\ref{app:uvDiv} computes the UV divergences for both the bubble and sunset diagrams in dimensional regularization via the Mellin-Barnes representation of the self-energy and contains several new results. These are used in \S\ref{sec:bubble} and \S\ref{sec:sunset} to set the kinetic and mass counterterms for $\sigma$ and $\varphi$. Finally, in Appendix~\ref{app:UVLight}, we analyze the ultraviolet contributions to the self-energies and argue that they are subleading to the infrared contributions as $m_\varphi \to 0$.

	\newpage
	\section{Free Fields in de Sitter Space} \label{sec:review}

		We will study interacting quantum fields in $D$-dimensional de Sitter space and, in particular, how interactions with a light field $\varphi$ affect the long-distance propagation of a heavy field $\sigma$. These theories we consider will have Euclidean actions of the form
		\begin{equation}
			S_\slab{e} = \int \!\ud^D x\, \sqrt{g} \left[\tfrac{1}{2}(\partial \sigma)^2 + \tfrac{1}{2} m_\sigma^2 \sigma^2 + \tfrac{1}{2}(\partial \varphi)^2 + \tfrac{1}{2} m_\varphi^2 \varphi^2\right] + S_\lab{int}
		\end{equation}
		with $(\partial \sigma)^2 = g^{\mu \nu} \partial_\mu \sigma \es\es \partial_\nu \sigma$ while $S_\lab{int}$ encodes the interactions between the two fields. As discussed in~\cite{Marolf:2010zp,Marolf:2010nz,Higuchi:2010xt}, we may define the de Sitter correlation functions of these fields by first computing them in Euclidean signature and then appropriately analytically continuing them to Lorentzian signature. The goal of this section is to review how free fields in de Sitter behave and explain how this analytic continuation works as a way of setting the stage for our loop calculations. We first review the geometry of de Sitter space in both signatures in \S\ref{sec:geometry}, and then discuss basic properties of de Sitter free fields in \S\ref{sec:freeFields}. There exists a useful ``momentum space'' representation of observables in terms of the hyperspherical harmonics---their analytic continuation to Lorentzian signature requires the use of the Froissart-Gribov inversion formula, which we discuss in \S\ref{sec:linv}.

		\subsection{The Geometry of de Sitter Space} \label{sec:geometry}

			In Lorentzian signature, $D$-dimensional de Sitter space is defined as the maximally symmetric space with positive constant curvature, and has isometry group $\lab{SO}(D, 1)$. In global coordinates, its metric is given by
			\begin{equation}
				\ud s^2 = \ell^2 \left[\minus \ud t^2 + \cosh^2 t \, \ud \Omega_d\right]
			\end{equation}
			where $\ud \Omega_d^2 = \ud \theta_1^2 + \sin^2 \theta_1 \, \ud \Omega_{d-1}^2$ is the standard round metric on the $d$-dimensional sphere, with $\theta_1, \ldots, \theta_{d-2}\in [0, \pi]$ and $\theta_{d-1} \in [0, 2 \pi)$, while $t \in \mathbb{R}$~\cite{Marolf:2010zp,Spradlin:2001pw,Anninos:2012qw}. We will find it extremely convenient to introduce the shorthand $\alpha = d/2$, as it will simplify many expressions. The radius of curvature $\ell$ is determined by the Hubble constant $\ell = H^{\sminus 1}$. It will also be convenient to measure all quantities in terms of this radius of curvature, and so we thus set $\ell = 1$.\footnote{This also has the effect of setting our renormalization group scale (often denoted $\mu$) to the Hubble scale $H = \ell^{\sminus 1}$, as we only analytically continue dimensionless quantities that are made so by multiplying by appropriate factors of $\ell$.} Of course, dimensions can be restored in any expression by restoring appropriate powers of $\ell$.

			The main goal of this paper is to compute loop corrections induced by a light scalar $\varphi$ on a heavy scalar $\sigma$. However, such corrections are plagued by IR divergences in Lorentzian signature which make it difficult to extract physical predictions. As discussed in \cite{Marolf:2010zp,Marolf:2010nz}, we can instead compute these loop corrections in Euclidean de Sitter space, which is simply the sphere $\lab{S}^D$ equipped with the standard round metric
			\begin{equation}
				\ud s^2 = \ud \Omega_D^2 = \ud \tau^2 + \sin^2 \tau\, \ud \Omega_d^2\,. \label{eq:edsMetric}
			\end{equation} 
			Generally, we will parameterize a point $x$ in this $(d+1)$-dimensional sphere in terms of its $\tau$ coordinate and a unit vector $\mb{x}$ on the $d$-dimensional sub-sphere, $x = (\tau, \mb{x})$, with $|\mb{x}|^2 = 1$. All potential IR divergences are automatically regulated by the finite volume of the sphere, and so---up to UV divergences that can be renormalized away---these loop corrections are finite and physical. We may then analytically continue the loop-corrected Euclidean correlation functions to a Lorentzian correlation function\footnote{Euclidean-signature Feynman diagrams will converge to define an interacting $\lab{SO}(D+1)$-invariant state on the sphere as long as the linearized field theory admits an $\lab{SO}(D+1)$-invariant propagator, which will be the case as long as our scalar fields both have non-zero mass. The correlators in this state will thus be invariant under the isometries of Euclidean de Sitter space and will satisfy the Euclidean Schwinger-Dyson equations. Upon analytic continuation to Lorentzian signature, these correlators will automatically be $\lab{SO}(D, 1)$-invariant and satisfy the Lorentzian Schwinger-Dyson equations, thus defining a consistent de Sitter-invariant state and correlators.} by taking $\tau \to i t + \frac{\pi}{2}$ with an appropriate $i\varepsilon$-prescription to specify an operator ordering~\cite{Marolf:2010zp}. It is particularly natural to use Euclidean de Sitter to study the dynamics of light fields since it explicitly isolates the mode that causes physical results to diverge in the massless limit.  

			Any function of two points on the sphere that is invariant under the $\lab{SO}(D+1)$ isometry group, and thus any de Sitter-invariant two-point function, may be written in terms of the so-called  \emph{embedding distance},\footnote{This embedding distance is also called $Z$ by \cite{Marolf:2010zp}, $\sigma$ by \cite{Hogervorst:2021uvp}, and $s$ by \cite{DiPietro:2021sjt}. }
			\begin{equation}
				\emd_{12} \equiv \emd(x_1, x_2) = \cos \tau_1 \cos \tau_2 + \sin \tau_1  \sin \tau_2 \, (\mb{x}_1 \cdot \mb{x}_2)\,,
			\end{equation}
			where $\mb{x}_1 \cdot \mb{x}_2$ is the standard dot product in $\mathbb{R}^D$.
			Specifically, if we embed the $\lab{S}^D$ into $\mathbb{R}^{D+1}$, the embedding distance $\emd$ is the cosine of the angle subtended by the great arc connecting the two points. We will drop the subscripts $\emd_{12} \to \emd$ when there is no chance of ambiguity.

			In Euclidean signature, this embedding distance is constrained to the interval $\emd\in [-1, 1]$ while upon analytic continuation to Lorentzian signature, $\tau_i \to i t_i + \frac{\pi}{2}$, it takes values on the entire real line $\emd \in \mathbb{R}$. The coincident limit $x_1 \to x_2$ corresponds to $\emd \to 1$, and this is also true for points connected by a null geodesic in Lorentzian signature. Furthermore, $|\emd| < 1$ or $\emd > 1$ if the two points are connected by a spacelike or timelike geodesic, respectively. Finally, points with $\emd < \minus 1$ are not connected by a geodesic. This limit is particularly important for cosmological observations, as taking two points to future infinity with fixed spatial separation corresponds to the limit $\emd \to \minus \infty$.

			Finally, we will find it convenient to simplify many expressions like (\ref{eq:gegQ}) by instead working in terms of the variable\footnote{Not to be confused with the comoving curvature perturbation $\zeta_{\mb{k}}$.}
			\begin{equation}
				\imd = \frac{2}{\emd - 1}\,, \label{eq:imdDef}
			\end{equation}
			in which the limits $\emd \to \pm \infty$ correspond to $\imd \to 0^{\pm}$. Because of its relationship to the long-time or long-distance limit, we will refer to the region $|\imd| \in [0, 1)$ as the ``infrared.'' Likewise, we will call $|\imd| \in [1, \infty)$ the ``ultraviolet.''

		\subsection{Free Fields from Euclidean to Lorentzian} \label{sec:freeFields}

			One very nice feature of Euclidean de Sitter space is that there exists a useful momentum space representation of scalar fields in terms of the hyperspherical harmonics,
			\begin{equation}
				\sigma(x) = \sum_{\mb{J}} \sigma_{\mb{J}} Y_{\mb{J}}(x)\,.
			\end{equation} 
			These harmonics are the $D$-dimensional analogs of the familiar spherical harmonics. They are labeled by an integer vector $\mb{J} = (J, m_1, \cdots, m_d)$, where ${J \in \mathbb{N}}$ is a non-negative integer ($J = 0, 1, \ldots$) and $\mb{m} = (m_1, m_2, \ldots, m_d) \in \mathbb{Z}^d$ are a set of integers such that $J \geq m_1 \geq m_2 \geq \cdots \geq |m_d|$. Most importantly, they diagonalize the Laplacian on the sphere $\lab{S}^{D}$,
			\begin{equation}
				\nabla^2 Y_{\mb{J}}(x) = -J(J+d) Y_{\mb{J}}(x)\,, \label{eq:shCasimir}
			\end{equation}
			where we call $J$ and $\mb{m}$ the \emph{total angular momentum quantum number} and \emph{magnetic quantum numbers}, respectively. We will not need the explicit forms for these harmonics, though a special role will be played by the \emph{zero mode} with $\mb{J} = 0$,
			\begin{equation}
				Y_0(x) = \sqrt{\frac{\Gamma(\alpha + 1)}{2 \pi^{\alpha + 1}}} = \frac{1}{\sqrt{\lab{vol}\, \lab{S}^D}} \label{eq:zmProf}
			\end{equation}
			which is just a constant profile on the sphere with amplitude determined by the volume of $\lab{S}^D$.

			These hyperspherical harmonics are orthonormal and complete,
			\begin{equation}
				\int_{\lab{S}^D} \!\ud \Omega_D \, Y_{\mb{J}}(x) \es \SHc{\mb{K}}(x) = \delta_{\mb{J}\mb{K}}\quad \text{and} \quad \sum_{\mb{J}} Y_{\mb{J}}(x) \SHc{\mb{J}}(y) = \delta^{(D)}(x - y)/\sqrt{g}\,, \label{eq:shOrthog}
			\end{equation}
			and so the momentum space representation of a function of a single point on the sphere can be easily found by computing, for example,  $\sigma_{\mb{J}} = \int\!\ud \Omega_D \, \sigma(x) \SHc{\mb{J}}(x)$. Usefully, the sum over magnetic quantum numbers 
			\begin{equation}
				\sum_{\mb{m}} Y_{J\mb{m}}(x) \SHc{J \mb{m}}(y) = \frac{\Gamma(\alpha)}{2 \pi^{\alpha + 1}} (J + \alpha) \, C_{J}^{\alpha}(\emd)\,,
			\end{equation}
			Since any de Sitter-invariant two-point function $H(x, y)$ is necessarily a function of the embedding distance $\emd(x, y)$, by completeness its harmonic decomposition will only depend on the total angular momentum $J$,
			\begin{equation}
				H(x, y) = \sum_{\mb{J}} [H]_J Y_{\mb{J}}(x) \SHc{\mb{J}}(y) = \frac{\Gamma(\alpha)}{2 \pi^{\alpha + 1}}\sum_{J = 0}^{\infty} (J+ \alpha) [H]_J C_{J}^{\alpha}(\emd)\,. \label{eq:hJDef}
			\end{equation}
			For integer $J$, the coefficients $[H]_J$ may be extracted by the ``Euclidean'' inversion formula,
			\begin{equation}
				[H]_J =  \frac{(4 \pi)^\alpha \Gamma(\alpha) \Gamma(J+1)}{\Gamma(J + 2 \alpha)} \int_{\sminus 1}^{1}\!\ud \emd \, \big(1 - \emd^2\big)^{\alpha - \frac{1}{2}} C_{J}^{\alpha}(\emd) H(\emd)\,. \label{eq:euclideanInv}
			\end{equation}
			Strictly speaking, the expansion (\ref{eq:hJDef}) only converges when the inversion formula (\ref{eq:euclideanInv}) does. Analogously, since $C_{J}^{\alpha}(\minus 1) \sim J^{2\alpha - 1}$ as $J \to \infty$, these expansions only converge if $[H]_J$ decays faster than $1/J^{2 \alpha+1}$ as $J \to \infty$. Generally, these conditions will not be satisfied for most of the functions we work with in four dimensions, $\alpha = \frac{3}{2}$. We will instead keep $\alpha = \frac{1}{2}(3 - \epsilon)$ arbitrary throughout our calculations and use dimensional regularization to remove any $\epsilon$-divergences, \emph{defining} these functions via analytic continuation in $\alpha$.

			The hyperspherical harmonics can allow us to easily determine the propagator $G(x, y)$ for a free field with mass $m$. This propagator obeys the Klein-Gordon equation
			\begin{equation}
				(-\nabla^2 + m^2) G(x, y) = \delta^{(D)}(x - y)/\sqrt{g}\,,
			\end{equation}
			which, using (\ref{eq:shCasimir}) and (\ref{eq:shOrthog}), can be solved to find
			\begin{equation}
				G(x, y) = \sum_{\mb{J}} \frac{Y_{\mb{J}}(x) \SHc{\mb{J}}(y)}{J(J+2 \alpha) + m^2} = \frac{\Gamma(\alpha)}{2 \pi^{\alpha + 1}} \sum_{J = 0}^{\infty} \frac{J + \alpha}{J(J+2 \alpha) + m^2} C_{J}^{\alpha}(\emd_{xy})\,. \label{eq:propSH}
			\end{equation}
			The propagator in momentum space is thus
			\begin{equation}
				[G]_J = \frac{1}{J(J+2 \alpha) + m^2} = \frac{1}{(J + \Delta)(J + \bar{\Delta})}\,, \label{eq:propMom}
			\end{equation}
			the poles of which are determined by the so-called scaling dimension of the field,
			\begin{equation}
				\Delta = \begin{dcases}
							\alpha + i \sqrt{m^2 - \alpha^2}\,, & m \geq \alpha \\
							\alpha - \sqrt{\alpha^2 - m^2}\,, & m < \alpha
						\end{dcases}
			\end{equation}
			and its conjugate or \emph{shadow} dimension $\bar{\Delta} \equiv d - \Delta$. A ``heavy'' scalar field, with mass $m \geq \alpha$ is said to belong to the \emph{principal series} with dimension $\Delta = \alpha + i \nu$, $ \nu \in \mathbb{R}$. Likewise, a ``light'' scalar field, with $m < \alpha$, is said to belong to the \emph{complementary series} with $\Delta \in (0, \alpha)$. Specifically, the single-particle states created by these fields fall into irreducible representations of de Sitter's $\lab{SO}(D, 1)$ isometry group, of which there are two continuous families called the principal and complementary series.\footnote{There is a third series, the \emph{discrete series}, with non-negative integer dimension $\Delta \in \mathbb{N}$, which includes exactly massless and tachyonic fields. Since we always work with fields of finite (albeit potentially small) and non-negative~$m^2$, we will not consider discrete series fields here.} 

			Unfortunately, this representation is useless if we are interested in the propagator in Lorentzian signature, since each term in the sum diverges as $\emd^{J}$ as $|\emd| \to \infty$. To analytically continue this expression to $\emd \in \mathbb{R}$, we can make use of the Watson-Sommerfeld transform, in which we rewrite a sum $\sum_{J = 0}^{\infty} s(J)$ as a contour integral over the product of a ``kernel'' $k(J) = \minus \e^{i \pi J}\Gamma(\minus J) \Gamma(J+1)$  with unit residue poles at the non-negative integers and a meromorphic ``interpolation'' $\tilde{s}(J) \propto (J+\alpha) [G]_J$ which agrees with $s(J)$ at the integers and continues it to complex $J$. For (\ref{eq:propSH}), this takes the form
			\begin{equation}
				G(\emd) = \frac{(\minus 1)}{2 \pi^{\alpha + 1}} \frac{\Gamma(\alpha)}{\Gamma(2 \alpha)} \int_{\mathcal{C}} \frac{\ud J}{2 \pi i}\, \Gamma(\minus J) \Gamma(J+2 \alpha) (J+\alpha)\, [G]_J  \,  \tFo{\minus J}{J+2 \alpha}{\alpha + \tfrac{1}{2}}{\frac{1 + \emd}{2}}\,. \label{eq:propInt}
			\end{equation}
			The contour $\mathcal{C}$ sandwiches the positive real axis, enclosing the poles at the non-negative integers $J \in \mathbb{N}$ in a counterclockwise fashion. This integral representation again only converges for $\emd \in [\minus 1, 1]$, but we can derive a convergent integral expression for $\emd \in \mathbb{C}/[\minus 1, \infty)$ by deforming the contour $\mathcal{C}$ to lie parallel to the imaginary axis, cf. Appendix~\ref{app:unique}. Crucially, the integrand (\ref{eq:propInt}) is well-behaved as $|J| \to \infty$, decaying as $\exp(\minus \pi |\es \es \lab{Im}\, J\es|)$ as $|\lab{Im}\, J| \to \infty$, and analytic away from the real axis, so we do not pick up any additional contributions as we deform the contour in this way as long as we do not shift it too far to the left. Thus, (\ref{eq:propInt}) defines an analytic continuation of the series (\ref{eq:propSH}) to Lorentzian signature.

			The integrand in (\ref{eq:propInt}) has poles at $J \in \mathbb{N}$, $J \in - 2 \alpha - \mathbb{N}$, and at the poles of (\ref{eq:propMom}). It is also odd under $J \to \minus (J + 2 \alpha)$, so by deforming $\mathcal{C}$ so that it passes through the fixed point of this transformation $J = \minus \alpha$~\cite{Marolf:2010zp}, we can make use of this antisymmetry to force the integral to vanish. However, to do this we must always encircle one of the poles of (\ref{eq:propMom}), and so the free field propagator reduces to the residue of this pole,
			\begin{equation}
				G(\emd) = \frac{\Gamma(\Delta) \Gamma(\bar{\Delta})}{(4 \pi)^{\alpha + \frac{1}{2}}\Gamma\big(\alpha + \frac{1}{2}\big)}\,  {}_2 F_1\big(\Delta, \bar{\Delta}; {\alpha + \tfrac{1}{2}}; \tfrac{1}{2}(1 + \emd)\big)\,. \label{eq:freeProp}
			\end{equation}
			The free field propagator (\ref{eq:freeProp}) is analytic for all $\emd$ aside from a branch cut along $\emd \in [1, \infty)$ or, in terms of our $\imd$ variable (\ref{eq:imdDef}), $\imd \in [0, \infty)$. It will be helpful to rewrite the propagator in the form
			\begin{equation}
				G(\imd) = \mathcal{G}_{\Delta}(\imd) + \mathcal{G}_{\bar{\Delta}}(\imd) = \mathcal{A}(\Delta) (\minus 1/\imd)^{\sminus \Delta}  \tFo{\Delta}{\Delta - \alpha + \frac{1}{2}}{2 \Delta - 2 \alpha + 1}{\minus \imd } + (\Delta \to \bar{\Delta})\,, \label{eq:propAsymp}
			\end{equation}
			with coefficient
			\begin{equation}
				\mathcal{A}(\Delta) = \frac{1}{(4 \pi)^{\alpha + \frac{1}{2}}} \frac{\Gamma(\Delta) \Gamma(2 \alpha - 2 \Delta)}{\Gamma\big(\alpha + \frac{1}{2} - \Delta\big)}\,. \label{eq:propCoeff}
			\end{equation}
			It is clear from this expression that the dimension $\Delta$, and thus the field's mass, controls the asymptotic behavior of the free propagator for very large separations $\imd \to 0$ or as $\emd \to \pm \infty$.

	\subsection{The Lorentzian Inversion Formula} \label{sec:linv}
		
		While the momentum space representation in Euclidean signature is an extremely useful tool, it can be nontrivial to extract physics in Lorentzian signature from it. As we saw for the propagator~(\ref{eq:propSH}), the core issue is that these momentum space expressions cannot be analytically continued term-by-term, but we must instead rely on the Watson-Sommerfeld transformation to continue the entire sum away from $\emd \in [\minus 1, 1]$. As we discussed, this transformation proceeds by first identifying a meromorphic interpolation $\tilde{s}(J)$ that agrees with the summand $s(J)$ at all positive integers and extends it to arbitrary complex values. We then recast the sum as a contour integral over $J$, deforming the contour so that the resulting expression is absolutely convergent for Lorentzian separations. There are infinitely many such interpolations but, fortunately, there is only one well-behaved enough as $|J| \to \infty$ to enable this analytic continuation. Given a function $H(\emd)$, how do we determine this \emph{correct} momentum space representation $[H]_J$? How do we invert expressions like (\ref{eq:propInt})? This is the role of the Lorentzian inversion formula, which we describe and illustrate in this section.

		The ambiguity in the interpolation is easy to see---given any interpolation $\tilde{s}(J)$ of the summand $s(J)$, we can add to it an \emph{arbitrary} analytic function multiplied by $\sin \pi J$ and it will still agree with the summand $s(J)$ at the integers. However, this will always mess up the behavior of $\tilde{s}(J)$ as $|J| \to \infty$, causing it to diverge at least as fast as $\propto \e^{\pi |J|}$. As we describe in more detail in Appendix~\ref{app:unique}, this would make the interpolation completely useless for analytic continuation from Euclidean to Lorentzian signature. So, if we also require that the interpolation $\tilde{s}(J)$ does not diverge as $|J| \to \infty$, Carlson's theorem\footnote{Carlson's theorem states that, if $f(z)$ is regular in the right half plane $\lab{Re}\, z > 0$ and $|f(z)| \leq C \e^{k|z|}$ with $C > 0$ and $k < \pi$, and if $f(z) = 0$ for $z = 0, 1, 2, \ldots$, then $f(z) \equiv 0$ is identically zero.} guarantees that this interpolation is \emph{unique}, see e.g. \cite{Newton:2002stw}. Thus, while there may be infinitely many interpolations of the summand $s(J)$, there is only one which should be used to analytically continue the sum beyond its domain of convergence.

		Thus, we should understand how to compute this correct interpolation given a function $H(\emd)$ defined, aside from possible singularities and branch cuts, on $\emd \in \mathbb{C}$. The Euclidean inversion formula (\ref{eq:euclideanInv}) cannot work because the Gegenbauer-$C$ functions behave very poorly away from the real $J$-axis, growing exponentially as $|\es \es \lab{Im}\, J \es | \to \infty$. For instance, $C_J^{\alpha}(\minus 1) \propto \exp(\pi \es | \es\es  \lab{Im}\, J \es|)$ as $\lab{Im}\, J \to \infty$, cf. (\ref{eq:gegEndpoints}), and so the $[H]_J$ provided by (\ref{eq:euclideanInv}) behave very poorly for complex $J$ and thus yields one of the wrong interpolations. Instead, the appropriate interpolation is provided by the Froissart-Gribov formula~\cite{Hogervorst:2021uvp,Loparco:2023rug,Correia:2020xtr,Newton:2002stw}, which defines $[H]_J$~as 
		\begin{equation}
			[H]_J = \frac{(4 \pi)^\alpha \Gamma(\alpha) \Gamma(J+1)}{\Gamma(J+2\alpha)} \oint_{\mathcal{C}} \frac{\ud \emd}{2 \pi i} \big(\emd^2 - 1\big)^{\alpha - \frac{1}{2}} Q_J^\alpha (\emd) H(\emd)\,. \label{eq:lorentzianInversion}
		\end{equation}
		The contour $\mathcal{C}$ is taken to wrap the interval $\emd \in [\minus 1, 1]$ counterclockwise, while the $Q_J^{\alpha}(\xi)$ are the Gegenbauer $Q$-functions defined in (\ref{eq:gegQ}).
		
		Let us justify this expression. The Gegenbauer $Q$-functions $Q_J^{\alpha}(\xi)$ satisfy the same differential equation (\ref{eq:gegDeq}) as the Gegenbauer $C$-functions~$C_J^{\alpha}(\emd)$ and have a branch cut along $\emd \in [\minus 1, 1]$. While we could have chosen any linear combination of $C_{J}^{\alpha}(\emd)$ and $Q_{J}^{\alpha}(\xi)$ and written a formula analogous to (\ref{eq:lorentzianInversion}) that agrees with (\ref{eq:euclideanInv}) for non-negative integer $J$, the trick is that $Q_{J}^{\alpha}(\emd)$ are the \emph{unique} solutions to the Gegenbauer differential equation which \emph{decay} as $\emd^{\es -J-2\alpha}$ as $\xi \to \infty$, and so the interpolation defined by (\ref{eq:lorentzianInversion}) necessarily decays as $\lab{Re}\, J \to \infty$. By Carlson's theorem, (\ref{eq:lorentzianInversion}) is then the unique well-behaved extension of (\ref{eq:euclideanInv}) from the integers to complex $J$, and is the one relevant for physics in Lorentzian signature.

		The functions we work with will all have a discontinuity along $\emd \in [1, \infty)$, and so we may deform the contour\footnote{Technically, (\ref{eq:lorentzianInversion}) only applies to functions which are analytic in a region around the interval $\emd \in [\minus 1, 1]$. This is not the case for the functions we consider, which have a discontinuity along $\emd \in [1,\infty)$, and is related to the fact that (products of) the propagator diverge in the coincident limit $\emd \to 1$, causing their spectral representations to converge poorly. As is clear from (\ref{eq:gegEndpoints}), the convergence of these series is strongly dependent on $\alpha$, and so we will work at small enough $\alpha$ so that (\ref{eq:lInvForm}) applies and then \emph{define} the $[H]_J$ at $\alpha =\frac{3}{2}$ by analytic continuation. Of course, these will often diverge as $\alpha \to \frac{3}{2}$, but these are the typical divergences one encounters in any loop calculation and may be absorbed by local counterterms.} in (\ref{eq:lorentzianInversion}) so that it becomes an integral over the discontinuity of $H(z)$,
		\begin{equation}
			[H]_J = - \frac{2 \pi^{\alpha+1}\Gamma(J+1)}{4^J\es \Gamma(J+\alpha+1)} \int_{0}^{\infty}\!\frac{\ud \imd}{2 \pi i} \, \imd^{J-1}\,  \tFo{J+\alpha + \frac{1}{2}}{J+1}{2 J + 2 \alpha + 1}{\minus \imd} \, \lab{disc}\, H(\imd)\,, \label{eq:lInvForm}
		\end{equation}
		with the discontinuity defined as
		\begin{equation}
			\lab{disc} \, H(\imd) = \lim_{\epsilon \to 0^+} H(\imd + i \epsilon) - H(\imd - i \epsilon)\,.
		\end{equation}
		The inversion formula (\ref{eq:lInvForm}) is the main tool we will use in this work---it defines the momentum space representation of $H(\emd)$ as long as $\lab{Re} \, J$ is large enough and $\alpha$ is small enough so that the integral converges at its endpoints $\imd \to 0$ and $\imd \to \infty$, respectively.

		It will be useful to illustrate the inversion formula (\ref{eq:lInvForm}) by applying it to the propagator (\ref{eq:propAsymp}). The discontinuity of the propagator along $\imd \in [0, \infty)$ is given by
		\begin{equation}
			\lab{disc} \, G(\imd) = -\frac{2 \pi i}{(4 \pi)^{\alpha + \frac{1}{2}}} \frac{\imd^{\alpha - \frac{1}{2}}}{\Gamma\big(\frac{3}{2}-\alpha\big)}\,  \tFo{\alpha + \tfrac{1}{2} - \Delta}{\alpha+\tfrac{1}{2}-\bar{\Delta}}{\tfrac{3}{2} - \alpha}{\minus \frac{1}{\imd}}\,. \label{eq:propDisc}
		\end{equation}
		The integral (\ref{eq:lInvForm}) can be explicitly evaluated by a computer algebra system to again yield (\ref{eq:propMom}),
		\begin{equation}
			[G]_{J} = \frac{1}{(J+\Delta)(J + \bar{\Delta})}\,, \label{eq:propMom2}
		\end{equation}
		extending (\ref{eq:lInvForm}) to arbitrary complex $J$ and arbitrary $\alpha$. However, it will be more helpful to evaluate (\ref{eq:lInvForm}) in a way more readily applicable to cases in which an exact answer is not known, or is too complicated to be useful.

		For example, throughout this paper we will be interested in expanding integrals like (\ref{eq:lInvForm}) order-by-order in the mass of a field or, analogously, order-by-order in $\Delta$. Already from (\ref{eq:propMom2}), it is clear that this expansion can be complicated by the fact that singularities of $[H]_J$ may depend on the dimension $\Delta$, and an expansion in $\Delta$ may depend sensitively on $J$. It will thus be helpful to rewrite (\ref{eq:lInvForm}) in a way that both analytically continues it to arbitrary $J$ and identifies any potential singularities.

		By use of a Kummer relation, the discontinuity (\ref{eq:propDisc}) can be written in a form that evokes~(\ref{eq:propAsymp}),
		\begin{equation}
			\lab{disc}\, G(\imd) = -\frac{2 \pi i}{(4 \pi)^{\alpha + \frac{1}{2}}} \frac{\Gamma(2 \alpha - 2 \Delta)}{\Gamma(1 - \Delta) \Gamma\big(\alpha + \tfrac{1}{2} - \Delta\big)}\,  \imd^{\Delta} \tFo{\Delta}{\Delta - \alpha + \frac{1}{2}}{2 \Delta - 2 \alpha +1}{\minus \imd} + (\Delta \to \bar{\Delta})\,. \label{eq:propDisc2}
		\end{equation}
		The benefit of this form is that it makes it clear \emph{why} (\ref{eq:lInvForm}) generates a pole at both $J = \minus \Delta$ and $J = \minus \bar{\Delta}$. We have
		\begin{equation}
			\begin{aligned}
				&\quad [G]_J = \mathcal{N}_{J, \Delta}\int_{0}^{\infty}\!\ud \imd\, \imd^{J + \Delta - 1}\,  \tFo{J+\alpha + \frac{1}{2}}{J+1}{2J + 2 \alpha +1 }{\minus \imd} \tFo{\Delta}{\Delta - \alpha + \frac{1}{2}}{2 \Delta - 2 \alpha + 1}{\minus \imd} + (\Delta \to \bar{\Delta}) \label{eq:propSplit}
			\end{aligned}
		\end{equation}
		where we have defined the coefficient
		\begin{equation}
			\mathcal{N}_{J, \Delta} = \frac{1}{2^{2J+2 \Delta+1}} \G{J+1,\, \alpha - \Delta}{J+ \alpha + 1,\, 1-\Delta} \,.\label{eq:propLInvCoeff}
		\end{equation}
		When $J = -\Delta$, the integrand in (\ref{eq:propSplit}) behaves as $\imd^{\es \sminus 1}$ as $\imd \to 0$ since ${}_2 F_1(a, b;c;\minus \imd) \sim 1 + \mathcal{O}(\imd)$, and so the integral diverges in the infrared. Indeed, we can analytically continue (\ref{eq:lInvForm}) to arbitrary $J$ by splitting the integral into an infrared contribution sensitive to the long-distance $|\emd| \to \infty$ ($\imd \to 0$) behavior of the propagator and an ``ultraviolet'' contribution sensitive to the short-distance $\imd \to \infty$ behavior, 
		\begin{equation}
			\begin{aligned}
				[G]_{J}^{\slab{ir}} &= \mathcal{N}_{J, \Delta} \int_{0}^{1}\!\ud \imd\, \imd^{J + \Delta - 1}\,  \tFo{J+\alpha + \frac{1}{2}}{J+1}{2J + 2 \alpha +1 }{\minus \imd} \tFo{\Delta}{\Delta - \alpha + \frac{1}{2}}{2 \Delta - 2 \alpha + 1}{\minus \imd} + \cdots 
			\end{aligned} \label{eq:irPropInt}
		\end{equation}
		where the $\cdots$ denotes the contribution from the shadow $\Delta \to \bar{\Delta}$, while the UV contribution $[G]_J^{\slab{uv}} = [G]_J - [G]_J^{\slab{ir}}$ has the same integrand but is instead integrated over $\imd \in [1, \infty)$. 

		Since the integrand is a regular function for all $J$ and $\imd \in (0, \infty)$, the only way the integral could develop a singularity in $J$ is if the integrand diverges in a $J$-dependent way at one of its endpoints, $\imd = 0$ or $\imd = \infty$. As we also discuss in Appendix~\ref{app:UVLight}, the integrand is regular in $J$ as $\imd \to \infty$ if we include the coefficient (\ref{eq:propLInvCoeff}), and so the only way a singularity can develop is if the integrand diverges as $\imd \to 0$. It is then trivial to isolate the singularities of $[G]_J$ and compute their residues by series expanding the integrand in $[G]_J^\slab{ir}$ about $\imd = 0$ and then integrating term-by-term. This is a technique that we will rely on throughout this work.

		Since the integral (\ref{eq:irPropInt}) is over $\imd \in [0, 1]$, we have no problem in replacing the integrand with its series expansion around $\imd = 0$ since it has radius of convergence $|\imd| = 1$, 
		\begin{equation}
			\imd^{J+\Delta - 1}\tFo{J+\alpha + \frac{1}{2}}{J+1}{2J + 2 \alpha +1 }{\minus \imd} \tFo{\Delta}{\Delta - \alpha + \frac{1}{2}}{2 \Delta - 2 \alpha + 1}{\minus \imd} = \sum_{k = 0}^{\infty} c_k(J, \Delta)\es\es \imd^{J+\Delta + k - 1}\,. \label{eq:propagatorExpCoeff}
		\end{equation}
		The series coefficients are explicitly given by
		\begin{equation}
			\begin{aligned}
				c_k(J, \Delta)
				&= \frac{(\minus 1)^k 4^{\Delta - \alpha}}{\sqrt{\pi}} \G{\Delta + k,\, \Delta - \alpha + 1,\, \Delta - \alpha + \frac{1}{2} + k }{k+1,\, \Delta,\, 2 \Delta - 2 \alpha + 1 + k} \\ 
				&\qquad \times {}_4 F_3\!\left[\genfrac..{0pt}{0}{J+1,\, \minus k,\, J+ \alpha + \frac{1}{2},\, 2 \alpha - 2 \Delta - k}{2 J + 2 \alpha + 1,\, 1 - \Delta - k,\, \alpha + \frac{1}{2} - \Delta - k} \, \Big| \, 1\, \right]\,.
			\end{aligned}
		\end{equation}
		Integrating term-by-term we find that $[G]_J^{\slab{ir}}$ has a infinite number of potential poles at $J = -\Delta - k$ and $J = -\bar{\Delta} - k$, for positive integer $k \in \mathbb{N}$,
		\begin{equation}
			[G]_J^\slab{ir} = \mathcal{N}_{J, \Delta} \sum_{k = 0}^{\infty} \frac{c_k(J, \Delta)}{J + \Delta + k} + (\Delta \to \bar{\Delta})\,, 
		\end{equation}
		but since $c_k\big(\minus [\Delta + k], \Delta\big) = \delta_{k,0}$, the only poles with non-vanishing residue are those at $J = -\Delta$ and $J= -\bar{\Delta}$, with residues $\mathcal{N}_{\sminus \Delta, \Delta} = 1/(2 \alpha - \Delta)$ and $\mathcal{N}_{\sminus \bar{\Delta}, \bar{\Delta}} = 1/(2 \alpha - \bar{\Delta})$ respectively. The rest are then ``spurious'' in the language of \cite{Loparco:2023rug}. Since $[G]_J^\slab{uv}$ is necessarily regular in $J$, we may then write
		\begin{equation}
			[G]_J = \frac{1}{(J +\Delta)(J + \bar{\Delta})} + f(J)
		\end{equation}
		with $f(J)$ is some entire function. But since $[G]_J \to 0$ as $\lab{Re}\, J \to +\infty$, this entire function must also vanish as $|J| \to \infty$ and thus $f(J) = 0$. $[G]_J$ is thus completely determined by its singularities, which can be easily extracted from (\ref{eq:irPropInt}).

		In the next section, we will find that a similar structure appears when analyzing the self-energy of a heavy field $\sigma$ in the presence of a light field $\varphi$. Fortunately, corrections to the long-distance behavior of the two-point function $\langle \sigma(x) \sigma(y)\rangle$ will be dominated by only one singularity (and its shadow) and thus the strategy outlined above provides an extremely efficient way of extracting corrections in the limit $m_\varphi \to 0$.

\section{Loop Corrections from a Light Scalar} \label{sec:loop}

	Having reviewed the basics of free de Sitter quantum field theory in both Euclidean and Lorenztian signatures, we are now ready to include interactions. The main goal of this paper is to determine how an interaction with a light scalar field $\varphi$, with mass $m_\varphi$, corrects the two-point function $\langle \sigma(x)\sigma(y)\rangle$ of a heavy scalar field $\sigma$, with mass $m_\sigma$, in the limit that $m_\varphi \to 0$. Specifically, we will work with actions of the form
	\begin{equation}
		S_\slab{e} = \!\int \!\ud^D x\, \sqrt{g} \left[\tfrac{1}{2}(\partial \sigma)^2 + \tfrac{1}{2} m_\sigma^2 \sigma^2 + \tfrac{1}{2}(\partial \varphi)^2 + \tfrac{1}{2} m_\varphi^2 \varphi^2 + \mathcal{L}_\lab{int} + \mathcal{L}_\lab{ct}\right]\,,
	\end{equation}
	where $\mathcal{L}_\lab{int}$ is the interaction Lagrangian and $\mathcal{L}_\lab{ct}$ is the counterterm Lagrangian. We consider cubic and quartic interactions in \S\ref{sec:bubble} and \S\ref{sec:sunset}, respectively, though it will be helpful to discuss general aspects of perturbation theory first.

	We organize corrections to the two-point functions $\langle \sigma(x) \sigma(y) \rangle$ and $\langle \varphi(x) \varphi(y) \rangle$ diagrammatically. We denote the free-field propagators of $\sigma$ and $\varphi$ by $G^{\sigma}(\emd)$ and $G^{\varphi}(\emd)$, respectively, and their dimensions as $\Delta_\sigma$ and $\Delta_\varphi$. We represent $\sigma$ by black lines and $\varphi$ by blue lines,
	\begin{equation}
		\begin{aligned}
			\begin{tikzpicture}[thick, baseline=-3pt]
				\draw (-1, 0) node[below] {$x$} -- (1, 0) node[below] {$y$};
				\draw[fill=white] (-1, 0) circle (0.05);
				\draw[fill=white] (1, 0) circle (0.05);
			\end{tikzpicture} = G^{\sigma}(\emd) \\
			\begin{tikzpicture}[thick, baseline=-3pt]
				\draw[cornellBlue] (-1, 0) node[below, black] {$x$} -- (1, 0) node[below, black] {$y$};
				\draw[fill=white] (-1, 0) circle (0.05);
				\draw[fill=white] (1, 0) circle (0.05);
			\end{tikzpicture} = G^{\varphi}(\emd)
		\end{aligned}  \qquad\quad \begin{aligned}
			\begin{tikzpicture}[thick, baseline=-3pt]
				\def\circSize{0.1}
				\draw[] (-1, 0)  -- (1, 0) node[below] {$\vphantom{y}$};
				\fill[white, draw=black, line width=0.3mm] (0, 0) circle (\circSize);
				\draw[rotate=45, line width=0.2mm] (-\circSize, 0) -- (\circSize, 0);
				\draw[rotate=-45, line width=0.2mm] (-\circSize, 0) -- (\circSize, 0);
			\end{tikzpicture} &= -J(J+2 \alpha) \delta_{Z_\sigma} \!- \delta_{m_\sigma}\\
			\begin{tikzpicture}[thick, baseline=-3pt]
				\def\circSize{0.1}
				\draw[cornellBlue] (-1, 0) -- (1., 0) node[below] {$\vphantom{y}$};
				\fill[white, draw=cornellBlue, line width=0.3mm] (0, 0) circle (\circSize);
				\draw[cornellBlue, rotate=45, line width=0.25mm] (-\circSize, 0) -- (\circSize, 0);
				\draw[cornellBlue, rotate=-45, line width=0.25mm] (-\circSize, 0) -- (\circSize, 0);
			\end{tikzpicture}  &= -J(J+2\alpha) \delta_{Z_\varphi} \! - \delta_{m_\varphi} 
		\end{aligned} \label{eq:propDiagram}
	\end{equation}
	with associated counterterm vertices from $\mathcal{L}_\lab{ct} \supset \tfrac{1}{2} \delta_{Z_\sigma} (\partial \sigma)^2 + \tfrac{1}{2} \delta_{Z_\varphi} (\partial \varphi)^2 + \tfrac{1}{2} \delta_{m_\sigma} \sigma^2 + \tfrac{1}{2} \delta_{m_\varphi} \varphi^2$, which account for mass and wavefunction renormalization and whose Feynman rules we display in momentum space.
	In position space, each vertex is associated with an integral over the $D$-dimensional unit sphere which we write in shorthand as $\int_{\lab{S}^D} \ud^D z_i\, \sqrt{g(z_i)} \equiv \int\!\ud^D z_i$ suppressing the metric factor, with $z_i$ always representing integration dummy variables. Finally, open circles like \es\es \begin{tikzpicture}[baseline=-3pt] \draw[fill=white, thick] (0, 0) circle (0.05); \end{tikzpicture} \es\es represent external legs at coordinates labeled by $x_i$ or, when there are only two, $x$ and $y$.
	
	As usual, corrections to the two-point function can be organized in terms of the self-energy~$\Pi_\sigma(J)$, which is defined by the sum over all one-particle irreducible (1PI) diagrams. We denote the self-energy as
	\begin{equation}
		\def\lSize{0.65}
		\def\lSizeA{0.8}
		\def\circSize{0.5}
		\def\circSizeCT{0.08}
		\begin{tikzpicture}[baseline=-3pt, thick]
						\draw (-\lSizeA, 0) -- (-\circSize, 0);
						\draw (\circSize, 0) -- (\lSizeA, 0);
						\fill (-\circSize, 0) circle (0.05);
						\fill (\circSize, 0) circle (0.05);
						\begin{scope}
							\clip[draw] (0, 0) circle (\circSize);
							\foreach \x in {-0.5, -0.425, ..., 0.5} 
							{	
								\draw[rotate=45, thin] (-\circSize, \x) -- (\circSize, \x);
							}
						\end{scope}
					\end{tikzpicture} = \Pi_{\sigma}(z_1, z_2) = \sum_{\mb{J}} \Pi_{\sigma}(J) Y_{\mb{J}}(z_1) \SHc{\mb{J}}(z_2)\,,
	\end{equation}
	and it can be computed order-by-order in perturbation theory in Euclidean signature and then appropriately analytically continued to non-integer $J$, or Lorentzian signature, by using (\ref{eq:lInvForm}).\footnote{Strictly speaking, we only require that inversion of the full two-point function $[\sigma \sigma]_J$ be well-behaved as $J \to \infty$ to be able to perform the Watson-Sommerfeld transform and analytically continue the Euclidean de Sitter result to Lorentzian signature. However, the same should be true for the self-energy $\Pi_\sigma(J)$, which should also be well-behaved as $J \to \infty$, since there is an order-by-order equivalence between in-in perturbation theory and Euclidean perturbation theory~\cite{Higuchi:2010xt}.}  Diagrammatically, the exact two-point function is then given by
	\begin{equation}
		\def\lSize{0.65}
		\def\lSizeA{0.8}
		\def\circSize{0.4}
		\def\circSizeCT{0.08}
		 \langle \sigma(x) \sigma(y) \rangle =  
		 			\begin{tikzpicture}[baseline=-3pt, thick]
						\draw (-\lSize, 0) -- (\lSize, 0);
						\draw[fill=white] (-\lSize, 0) circle (0.05);
						\draw[fill=white] (\lSize, 0) circle (0.05);
					\end{tikzpicture} + 
					\begin{tikzpicture}[baseline=-3pt, thick]
						\draw (-\lSizeA, 0) -- (-\lSizeA+\circSize, 0);
						\draw (\lSizeA-\circSize, 0) -- (\lSizeA, 0);
						\fill (-\lSizeA + \circSize, 0) circle (0.05);
						\fill (\lSizeA - \circSize, 0) circle (0.05);
						\draw[fill=white] (-\lSizeA, 0) circle (0.05);
						\draw[fill=white] (\lSizeA, 0) circle (0.05);
						\begin{scope}
							\clip[draw] (0, 0) circle (\circSize);
							\foreach \x in {-0.5, -0.425, ..., 0.5} 
							{	
								\draw[rotate=45, thin] (-\circSize, \x) -- (\circSize, \x);
							}
						\end{scope}
					\end{tikzpicture}
					+ 
					\begin{tikzpicture}[baseline=-3pt, thick]
					\draw (-\lSizeA, 0) -- (-\lSizeA+\circSize, 0);
					\draw (\lSizeA-\circSize, 0) -- (\lSizeA, 0);
					\fill (-\lSizeA + \circSize, 0) circle (0.05);
					\fill (\lSizeA - \circSize, 0) circle (0.05);
					\draw[fill=white] (-\lSizeA, 0) circle (0.05);
					\fill (\lSizeA, 0) circle (0.05);
					\begin{scope}
						\clip[draw] (0, 0) circle (\circSize);
						\foreach \x in {-0.5, -0.425, ..., 0.5} 
						{	
							\draw[rotate=45, thin] (-\circSize, \x) -- (\circSize, \x);
						}
					\end{scope}
					\begin{scope}[shift={(\lSizeA + \circSize, 0)}]
						\clip[draw] (0, 0) circle (\circSize);
						\foreach \x in {-0.5, -0.425, ..., 0.5} 
						{	
							\draw[rotate=45, thin] (-\circSize, \x) -- (\circSize, \x);
						}
					\end{scope}
					\draw (\lSizeA + 2*\circSize, 0) -- (2*\lSizeA+ \circSize, 0);
					\fill (\lSizeA + 2*\circSize, 0) circle (0.05);
					\draw[fill=white] (2*\lSizeA + \circSize, 0) circle (0.05);
					\end{tikzpicture} + \cdots\,,
	\end{equation}
	which forms a geometric series that can be subsequently summed to yield
	\begin{equation}
		\langle \sigma(x) \sigma(y) \rangle = \frac{(\minus 1)}{2 \pi^{\alpha + 1}} \frac{\Gamma(\alpha)}{\Gamma(2 \alpha)} \oint_{\mathcal{C}} \frac{\ud J}{2 \pi i}\, \frac{\Gamma(\minus J) \Gamma(J+2 \alpha) (J+\alpha)}{(J+\Delta_\sigma)(J+\bar{\Delta}_\sigma) - \Pi_\sigma(J)}\, \tFo{\minus J}{J+2 \alpha}{\alpha + \tfrac{1}{2}}{\frac{1 + \emd}{2}}\,. \label{eq:correctedPropInt}
	\end{equation}
	Following~\cite{Marolf:2010zp}, the asymptotic behavior of (\ref{eq:correctedPropInt}) at future infinity and fixed spatial separation, $\xi \to \minus \infty$, and thus the physical mass of the field,\footnote{This is familiar from quantum field theory in flat space, where the poles of the momentum space propagator $G(k) = 1/(k^2 +m^2 - \Pi(k^2))$, with $\Pi(k^2)$ the self-energy, control the long-distance behavior of the position space propagator, and thus provides a physical notion of mass in the interacting theory.} is controlled by the pole $J_*$ of the integrand, 
	\begin{equation}
		(J_*+\Delta_\sigma)(J_* + \bar{\Delta}_{\sigma}) - \Pi_\sigma(J_*) = 0\,,
	\end{equation}
	with maximal real part.
	
	For a weakly interacting heavy field $\sigma$, there is a pair of poles with largest real part, one the complex conjugate of the other, which we denote as $J_*$ and $\bar{J}_*$. At future infinity, the heavy field propagator thus behaves as $\langle \sigma(x) \sigma(y) \rangle \sim \mathcal{C}_1 \es (\minus 2 \xi)^{J_*} + \mathcal{C}_2 \es (\minus 2 \xi)^{\bar{J}_*}$ as $\xi \to \minus \infty$ for some constants $\mathcal{C}_{1}$ and $\mathcal{C}_2$. For free fields, $\Pi_\sigma(J) = 0$ and so $J_* = \minus \Delta_\sigma$ and $\bar{J}_*= \minus \bar{\Delta}_\sigma$, while at leading order in perturbation theory the non-zero self-energy shifts the pole to
	\begin{equation}
		J_* \approx -\Delta_\sigma - \frac{\Pi_{\sigma}(\minus \Delta_\sigma)}{\Delta_\sigma - \bar{\Delta}_\sigma}\,,
	\end{equation}
	with $\bar{J}_*$ given by the conjugate of this expression, or by taking $\Delta_\sigma \to \bar{\Delta}_\sigma$. Specifically, decomposing the pole into its real and imaginary parts,
	\begin{equation}
		J_* \approx - \!\! \left[\alpha + \frac{\lab{Im}\, \Pi_\sigma(\minus \Delta_\sigma)}{2 \sqrt{m_\sigma^2 - \alpha^2}}\right] - i\left[ \sqrt{m_\sigma^2 - \alpha^2} + \frac{\lab{Re}\, \Pi_\sigma(\minus \Delta_\sigma)}{2 \sqrt{m_\sigma^2 - \alpha^2}}\right]\,. \label{eq:poleShift}
	\end{equation}
	we find that the real part of the self-energy affects the physically-measured mass of $\sigma$, while its imaginary part changes the decay of the correlator at long distances. We will work in an ``on-shell'' mass renormalization scheme in which $m_\sigma$ is $\sigma$'s physically-measured mass, and so we adjust our counterterms so that
	\begin{equation}
		\lab{Im}\, J_* = \minus \sqrt{m_\sigma^2 - \alpha^2}\,, \label{eq:renormCond}
	\end{equation}
	or equivalently, to first order in perturbation theory, $\lab{Re}\, \Pi_\sigma(\minus \Delta_\sigma) = 0$. The shift in $J_*$'s real part is a purely physical effect that cannot be mimicked in free field theory. Furthermore, to ensure that the two-point function is properly normalized in the long-distance limit, we require that the residue of this pole is unchanged and thus $\Pi_\sigma'(\minus \Delta_\sigma) = 0$.

	A weakly interacting light field $\varphi$ only has one pole with largest real part, which we again denote as $J_*$. As for the heavy field, $J_* = -\Delta_\varphi$ in the free theory, while at leading order in perturbation theory a non-zero self-energy $\Pi_\varphi(J)$ corrects this to
	\begin{equation}
		J_* \approx - \Delta_\varphi - \frac{\Pi_\varphi(\minus \Delta_\varphi)}{\Delta_\varphi -\bar{\Delta}_\varphi}\,.
	\end{equation}
	We will find that $\Pi_\varphi(\minus \Delta_\varphi)$ is purely real. If we define the physically-measured mass of a light field by $J_* \equiv -\alpha + \sqrt{\alpha^2 - m_{\smash{\varphi}}^2}$,  thus working in an on-shell renormalization scheme, this requires that we choose our counterterms such that $\Pi_\varphi(\minus \Delta_\varphi) = 0$. So, unlike the heavy field, a non-zero self-energy $\Pi_\sigma(J)$ does not result in a measurable effect as $\emd \to \minus \infty$.

	Finally, before we move on to our specific examples, it will be useful to discuss the general strategy we will use for simplifying self-energy corrections in the limit $m_\varphi \to 0$. The self-energies we study will be expressed in terms of the inversion formula (\ref{eq:lInvForm}) applied to products of $\sigma$ and $\varphi$ propagators. As discussed in \S\ref{sec:freeFields}, the free field propagator can be decomposed into a sum~(\ref{eq:propAsymp}) of terms with definite scaling behavior in the infrared,
	\begin{equation}
		G(\imd) = \mathcal{G}_{\Delta}(\imd) + \mathcal{G}_{\bar{\Delta}}(\imd) \sim \mathcal{A}(\Delta) (\minus 1 /\imd)^{\sminus \Delta} + \mathcal{A}(\bar{\Delta}) (\minus 1/\imd)^{\sminus \bar{\Delta}}\,, \mathrlap{\quad \imd \to 0\,,}
	\end{equation}
	where $\mathcal{A}(\Delta) \propto \Gamma(\Delta) \sim \Delta^{\sminus 1}$ as $\Delta \to 0$ is given by (\ref{eq:propCoeff}). Our self-energies will then be sums of terms like $[\mathcal{G}_{\Delta_\varphi} \mathcal{G}_{\Delta_\sigma}]_{J}$ and $[\mathcal{G}_{\smash{\bar{\Delta}_{\varphi}}} \mathcal{G}_{\Delta_{\sigma}}]_{J}$. As we will show in this section, and as we might expect from our analysis of the propagator in \S\ref{sec:linv}, a term like $[\mathcal{G}_{\Delta_\varphi} \mathcal{G}_{\Delta_\sigma}]_{J}$ contributes ``potential'' poles to the self-energy at $J = \minus (\Delta_{\varphi} + \Delta_\sigma + k)$, with $k\in \mathbb{N}$,  whose residues may be zero (i.e. the so-called ``spurious poles'' of \cite{Loparco:2023rug}) but are proportional to $\mathcal{A}(\Delta_\varphi) \mathcal{A}(\Delta_{\sigma}) \propto \Gamma(\Delta_\varphi) \Gamma(\Delta_\sigma)$. 

	Since we are interested in how the light scalar $\varphi$ affects the long-distance behavior of the heavy scalar $\sigma$, we want to approximate $\Pi_\sigma(J)$ near the free-field pole $J_* = \minus \Delta_\sigma$. As $m_\varphi \to 0$, the dimension of $\varphi$ also vanishes, $\Delta_\varphi \sim m_\varphi^2/(2 \alpha) \to 0$, and so we find that there are two simplifications of the self-energy in this limit. The first is that the terms with factors of $\mathcal{G}_{\Delta_\varphi} \propto \Gamma(\Delta_\varphi) \propto m_\varphi^{\sminus 2}$ are \emph{enhanced} compared to those with~$\bar{\mathcal{G}}_{\Delta_\varphi}$, whose amplitude does not diverge as $m_\varphi \to 0$. The second is that only terms with factors of $\mathcal{G}_{\Delta_\varphi}$ can contribute poles that approach the free-field pole at $J_* = \minus \Delta_\sigma$ as $\Delta_\varphi \to 0$, and there are only finitely many that do. Such poles are further enhanced by a factor of $\Delta_\varphi^{\sminus 1} \propto m_\varphi^{\sminus 2}$ compared to other poles or regular terms in the self-energy. We thus find that the self-energy near the free-field pole, and thus the correction to $\sigma$'s long-distance behavior, is governed by a single easily-calculated term as $m_\varphi \to 0$.  

	We will first illustrate this procedure in the so-called ``bubble'' diagram in \S\ref{sec:bubble}, which is the leading correction in a theory with a cubic interaction $\mathcal{L}_\lab{int} = \tfrac{1}{2}g \varphi \sigma^2$. Here, there are exact analytic results we can use to check our approximations. Then, in \S\ref{sec:sunset}, we study the general theory with quartic interactions with $\mathcal{L}_\lab{int} = \tfrac{1}{2} g \sigma^2 \varphi^2 + \tfrac{1}{4!} g_\varphi \varphi^4 + \tfrac{1}{4!} g_\sigma \sigma^4$, computing the leading order physical corrections to the two-point functions in the limit $m_\varphi \to 0$. Throughout, we will regulate UV divergences by first working with $\alpha = \tfrac{1}{2}(3-\epsilon)$ and then taking $\epsilon \to 0$.

	\subsection{The Bubble} \label{sec:bubble}

		We begin by studying the theory with a cubic interaction, $\mathcal{L}_\lab{int} = \frac{1}{2}g \varphi \sigma^2$, which we denote diagrammatically as
		\begin{equation}
			\begin{tikzpicture}[baseline=-3pt, thick]
				\def\lSize{0.7}
				\draw (0, 0) --+(120:\lSize);
				\draw (0, 0) --+(-120:\lSize);
				\draw[cornellBlue] (0, 0) --+(0:\lSize);
				\fill[black, draw=white] (0,0) circle (0.07);
			\end{tikzpicture} = -g\,.
		\end{equation}
		We should also include the relevant vertex counterterms, i.e. $\mathcal{L}_\lab{ct} = \frac{1}{2} \delta_{g} \varphi \sigma^2 + \cdots$, but since the goal of this section is to compute $\sigma$'s self-energy $\Pi_\sigma(J)$ to leading order in perturbation theory, these counterterms will not contribute to the final answer and so we will not include them.
		
\begin{figure}
			\centering
			\includegraphics[scale=0.85]{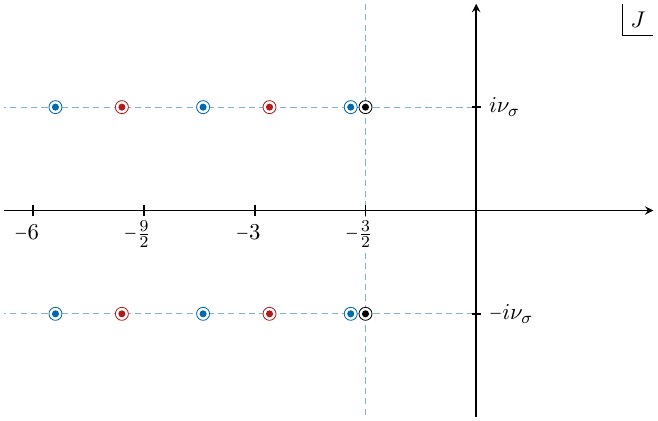}
			\caption{The analytic structure of the bubble diagram $[G^{\sigma} G^{\varphi}]_J$ for $\alpha = \tfrac{3}{2}$. In [\emph{black}], we denote the free-field poles at $J = \minus \Delta_\sigma = \minus (\alpha + i \nu_\sigma)$ and $J = \minus \bar{\Delta}_\sigma = \minus(\alpha - i \nu_\sigma)$ at which $[G^{\sigma} G^{\varphi}]_J$ is regular. It instead has singularities at $J = \minus (\Delta_\sigma + \Delta_\varphi + 2k)$ and $J = \minus (\bar{\Delta}_\sigma + \Delta_\varphi + 2k)$, with $k \in \mathbb{N}$, in [\emph{\color{cornellBlue}blue\es}]. These singularities encroach upon the free-field poles and dominate the self-energy as $\Delta_\varphi \to 0$. The bubble also has singularities at $J = \minus(\Delta_\sigma + \bar{\Delta}_\varphi + 2k)$ and $J = \minus(\bar{\Delta}_\sigma + \bar{\Delta}_\varphi + 2k)$, with $k \in \mathbb{N}$, pictured in [\emph{\color{cornellRed}red\es\es}], which remain well-separated from the free-field poles as $\Delta_\varphi \to 0$.\label{fig:bubbleSings}}
		\end{figure}

		The first correction to $\sigma$'s self-energy appears at $\mathcal{O}(g^2)$, and is generated by the bubble diagram
		\begin{equation}
			\def\lSize{0.65}
			\def\lSizeA{0.9}
			\def\circSize{0.5}
			\def\circSizeCT{0.1}
			\begin{tikzpicture}[baseline=-3pt, thick]
					\draw (-\lSizeA, 0) -- (-0.5, 0);
					\draw (0.5, 0) -- (\lSizeA, 0);
					\begin{scope}
							\clip[draw] (0, 0) circle (\circSize);
							\foreach \x in {-0.5, -0.425, ..., 0.5} 
							{	
								\draw[rotate=45, thin] (-\circSize, \x) -- (\circSize, \x);
							}
					\end{scope}
					\fill[black, draw=white] (-\circSize,0) circle (0.06);
					\fill[black, draw=white] (\circSize,0) circle (0.06);
			\end{tikzpicture} = 
			\begin{tikzpicture}[baseline=-3pt, thick]
					\draw (-\lSizeA, 0) -- (-0.5, 0);
					\draw (0.5, 0) -- (\lSizeA, 0);
					\draw (0, 0)+(\circSize, 0) arc (0:180:\circSize);
					\draw[cornellBlue] (0, 0)+(\circSize, 0) arc (0:-180:\circSize);
					\fill[black, draw=white] (-\circSize,0) circle (0.06);
					\fill[black, draw=white] (\circSize,0) circle (0.06);
			\end{tikzpicture} + 
			\begin{tikzpicture}[baseline=-3pt, thick]
					\draw (-\lSize, 0) -- (\lSize, 0);
					\fill[white, draw=black, line width=0.3mm] (0, 0) circle (\circSizeCT);
					\draw[rotate=45, line width=0.2mm] (-\circSizeCT, 0) -- (\circSizeCT, 0);
					\draw[rotate=-45, line width=0.2mm] (-\circSizeCT, 0) -- (\circSizeCT, 0);
			\end{tikzpicture}\,,
		\end{equation}
		or equivalently,
		\begin{equation}
			\Pi_{\sigma}(J) = (\minus g)^2 [G^{\sigma} G^{\varphi}]_{J} - J(J+2 \alpha) \delta_{Z_\sigma} - \delta_{m_\sigma}\,,
		\end{equation}
		where $[G^{\sigma} G^{\varphi}]_{J}$ is determined by the inversion formula (\ref{eq:lInvForm}) applied to the collection of propagators appearing in the loop, $H(\imd) = G^{\sigma}(\imd) G^{\varphi}(\imd)$. An exact expression for $[G^{\sigma} G^{\varphi}]_{J}$ was found in~\cite{Marolf:2010zp}, which we present in (\ref{eq:mmResult}). As we illustrate in Figure~\ref{fig:bubbleSings}, they found that the bubble has poles at $J = -\Delta_\sigma - \Delta_\varphi - 2k$, $J = -\Delta_\sigma - \bar{\Delta}_\varphi - 2k$, $J = -\bar{\Delta}_\sigma - \Delta_\varphi - 2k$ and $J = -\bar{\Delta}_{\sigma} - \bar{\Delta}_{\varphi} - 2k$, for all non-negative integer $k \in \mathbb{N}$. Physically, these correspond to the appearance of long-lived two-particle states with non-zero momenta. Our goal is to see this structure directly from the inversion formula (\ref{eq:lInvForm}) and to extract the self-energy's $m_\varphi \to 0$ behavior.

		Using (\ref{eq:propAsymp}), we can expand the bubble into its constituents with definite scaling as $\imd \to 0$,
		\begin{equation}
			[G^{\sigma} G^{\varphi}]_J = [\mathcal{G}_{\Delta_\sigma} \mathcal{G}_{\Delta_\varphi}]_J + [\mathcal{G}_{\smash{\bar{\Delta}_\sigma}} \mathcal{G}_{\Delta_\varphi}]_J + [\mathcal{G}_{\Delta_\sigma} \mathcal{G}_{\smash{\bar{\Delta}_\varphi}}]_J + [\mathcal{G}_{\smash{\bar{\Delta}_\sigma}} \mathcal{G}_{\smash{\bar{\Delta}_\varphi}}]_J \,.\,\label{eq:bubbleDecomp}
		\end{equation}
		We will focus on the first term, $[\mathcal{G}_{\Delta_\sigma} \mathcal{G}_{\Delta_{\varphi}}]_J$, and recover the behavior of the others by simply interchanging the dimensions $\Delta_\sigma$ and $\Delta_\varphi$ with their shadows $\bar{\Delta}_\sigma$ and $\bar{\Delta}_\varphi$. The relevant discontinuity is then given by
		\begin{equation}
			\begin{aligned}
				\lab{disc}\, \mathcal{G}_{\Delta_\sigma} \mathcal{G}_{\Delta_\varphi}(\imd) &= -\frac{2 \pi i \es\es \mathcal{A}(\Delta_\sigma) \mathcal{A}(\Delta_\varphi)}{\Gamma(\Delta_\sigma + \Delta_{\varphi}) \Gamma(1 - \Delta_\sigma - \Delta_\varphi)} \\
				&\times  \imd^{\Delta_{\varphi} + \Delta_{\varphi}}  \tFo{\Delta_{\sigma}}{\Delta_\sigma - \alpha + \frac{1}{2}}{2 \Delta_\sigma - 2 \alpha + 1}{\minus \imd} \tFo{\Delta_{\varphi}}{\Delta_\varphi - \alpha + \frac{1}{2}}{2 \Delta_\varphi - 2 \alpha + 1}{\minus \imd}
			\end{aligned}
		\end{equation}
		and so the inversion formula reads
		\begin{equation}
			\begin{aligned}
				[\mathcal{G}_{\Delta_\sigma} \mathcal{G}_{\Delta_\varphi}]_{J} &= \mathcal{N}_{J, \Delta_{\sigma} \Delta_{\varphi}} \int_{0}^\infty \!\ud \imd \, \imd^{J + \Delta_\varphi + \Delta_\sigma - 1} \, \tFo{J+\alpha + \frac{1}{2}}{J+1}{2 J + 2 \alpha + 1}{\minus \imd} \\  & \qquad \qquad \qquad \times \tFo{\Delta_{\sigma}}{\Delta_\sigma - \alpha + \frac{1}{2}}{2 \Delta_\sigma - 2 \alpha + 1}{\minus \imd}\tFo{\Delta_{\varphi}}{\Delta_\varphi - \alpha + \frac{1}{2}}{2 \Delta_\varphi - 2 \alpha + 1}{\minus \imd}  \,, \label{eq:cyclopsLinv}
			\end{aligned}
		\end{equation}
		with 
		\begin{equation}
			\mathcal{N}_{J, \Delta_\sigma \Delta_\varphi} =  \frac{2 \pi^{\alpha+1}\mathcal{A}(\Delta_\sigma) \mathcal{A}(\Delta_\varphi) \Gamma(J+1)}{4^{J} \Gamma(J+\alpha+1) \Gamma(\Delta_\sigma + \Delta_\varphi) \Gamma(1 - \Delta_\sigma - \Delta_\varphi)}\,. \label{eq:bubbleCoeff}
		\end{equation}
		The coefficients $\mathcal{A}(\Delta)$ are defined in (\ref{eq:propCoeff}), and crucially diverge $\mathcal{A}(\Delta) \propto \Delta^{\sminus 1}$ as $\Delta \to 0$. This implies that, barring any enhancements coming from singularities in $J$, the terms that depend on $\mathcal{G}_{\Delta_\varphi}$ will be enhanced over those that depend on $\mathcal{G}_{\bar{\Delta}_{\varphi}}$ in the limit that $\Delta_\varphi \to 0$ and $\bar{\Delta}_\varphi \to \alpha$.

		Following our analysis of the propagator in \S\ref{sec:linv}, it is convenient to treat the infrared and ultraviolet regions of (\ref{eq:cyclopsLinv}) separately, writing $[\mathcal{G}_{\Delta_\sigma} \mathcal{G}_{\Delta_\varphi}]_J = [\mathcal{G}_{\Delta_\sigma} \mathcal{G}_{\Delta_\varphi}]_J^\slab{ir} + [\mathcal{G}_{\Delta_\sigma} \mathcal{G}_{\Delta_\varphi}]_J^{\slab{uv}}$ with
		\begin{equation}
			\begin{aligned}
				[\mathcal{G}_{\Delta_\sigma} \mathcal{G}_{\Delta_\varphi}]_{J}^{\slab{ir}} = \mathcal{N}_{J, \Delta_{\sigma} \Delta_{\varphi}} \int_{0}^1\!\ud \imd &\, \imd^{J + \Delta_\varphi + \Delta_\sigma - 1} \, \tFo{J+\alpha + \frac{1}{2}}{J+1}{2 J + 2 \alpha + 1}{\minus \imd} \\  & \times \tFo{\Delta_{\sigma}}{\Delta_\sigma - \alpha + \frac{1}{2}}{2 \Delta_\sigma - 2 \alpha + 1}{\minus \imd} \tFo{\Delta_{\varphi}}{\Delta_\varphi - \alpha + \frac{1}{2}}{2 \Delta_\varphi - 2 \alpha + 1}{\minus \imd}  \,. \label{eq:cyclopsIR}
			\end{aligned}
		\end{equation}
		We will focus on this infrared contribution first since we can use it to immediately read off where $[G^{\sigma} G^{\varphi}]_J$'s singularities are. As $\imd \to 0$, the integrand behaves as $\imd^{J + \Delta_\varphi + \Delta_\sigma - 1}$ and so it diverges when $J = -\Delta_\varphi - \Delta_\sigma$. Technically, this integral representation is only well-defined for $\lab{Re}\, J > - \lab{Re} \, \Delta_\varphi - \lab{Re}\, \Delta_\sigma$, but we may derive an analytic continuation by series expanding the integrand about $\imd = 0$,
		\begin{equation}
			\begin{aligned}
				&\tFo{J+\alpha + \frac{1}{2}}{J+1}{2 J + 2 \alpha + 1}{\minus \imd} \tFo{\Delta_{\sigma}}{\Delta_\sigma - \alpha + \frac{1}{2}}{2 \Delta_\sigma - 2 \alpha + 1}{\minus \imd}   \\
				&\qquad\qquad \qquad \times  \tFo{\Delta_{\varphi}}{\Delta_\varphi - \alpha + \frac{1}{2}}{2 \Delta_\varphi - 2 \alpha + 1}{\minus \imd}
			\end{aligned}\, =\,  \sum_{k = 0}^{\infty} c_{k}(J, \Delta_\sigma, \Delta_\varphi)\es\es \imd^{k}\,, \label{eq:bubbleExpCoeff}
		\end{equation}
		and then integrating (\ref{eq:cyclopsIR}) term-by-term to find
		\begin{equation}
			[\mathcal{G}_{\Delta_\sigma} \mathcal{G}_{\Delta_\varphi}]_{J}^\slab{ir} = \mathcal{N}_{J, \Delta_\sigma \Delta_\varphi}\sum_{k = 0}^{\infty} \frac{c_{k}(J, \Delta_\sigma, \Delta_\varphi)}{J + \Delta_\sigma + \Delta_\varphi + k}\,.
		\end{equation}
		Immediately, we see that the deep infrared $\imd \to 0$ of the integral (\ref{eq:cyclopsIR}) generates a family of potential singularities at $J = -\Delta_\sigma - \Delta_\varphi -k$, with $k \in \mathbb{N}$. However, some of their residues may vanish and so some of these poles may be spurious. 
		
		Ignoring the coefficient $\mathcal{N}_{J, \Delta_\sigma \Delta_\varphi}$ for the moment, the residue around each pole $c_k({\Delta_\sigma, \Delta_\varphi}) \equiv c_{k}(\minus[\Delta_\sigma +\Delta_\varphi + k], \Delta_\sigma, \Delta_\varphi)$ takes a relatively simple closed form, with non-zero even coefficients
		\begin{equation}
			\begin{aligned}
				c_{2k}({\Delta_\sigma,\Delta_\varphi}) = \frac{1}{2^{4k} k!} \frac{\Gamma(\Delta_\varphi + k)}{\Gamma(\Delta_\varphi)}\,&  \G{\alpha + k\,,\,\Delta_\sigma+k\,,\, \alpha - \Delta_\sigma - k\,,\,\alpha - \Delta_\varphi - k}{\alpha\,,\,\Delta_\sigma\,,\, \alpha - \Delta_\sigma\,,\,\alpha - \Delta_\varphi}\\ 
				\times & \G{\Delta_\sigma + \Delta_\varphi - 2 \alpha + 1 + 2k \,,\, \Delta_\sigma + \Delta_\varphi - \alpha + k}{\Delta_\sigma + \Delta_\varphi - 2 \alpha + 1 +k\,,\, \Delta_\sigma + \Delta_\varphi - \alpha + 2k}
			\end{aligned}
		\end{equation}
		and vanishing odd coefficients, $c_{2k+1}(\Delta_\varphi, \Delta_\sigma) = 0$. From this, we find that the bubble indeed has poles at $J = -\Delta_\sigma - \Delta_\varphi - 2k$, with $k \in \mathbb{N}$. Furthermore, there is a hierarchy in residues as $\Delta_\varphi \to 0$, with $c_{0}(\Delta_\sigma, \Delta_\varphi)= 1$ while the rest are $\mathcal{O}(\Delta_\varphi)$. We can understand this hierarchy as follows. Since the integral (\ref{eq:cyclopsIR}) is constrained to $\imd \in [0, 1]$, we may approximate the last hypergeometric function as
		\begin{equation}
			\tFo{\Delta_\varphi}{\Delta_\varphi - \alpha + \frac{1}{2}}{2 \Delta_\varphi - 2 \alpha + 1}{\minus \imd} = 1 +\frac{1}{2} \Delta_\varphi \, (\minus \imd) \, \pFq{3}{2}{1\,,\, 1\,,\, \frac{3}{2}- \alpha}{2\,,\, 2 - 2 \alpha}{\minus \imd} + \mathcal{O}(\Delta_\varphi^2)\,
		\end{equation}
		and so, to leading order in $\Delta_\varphi$, the integral in (\ref{eq:cyclopsIR}) reduces to that of (\ref{eq:irPropInt}), except that the factor of $\imd^{J + \Delta - 1}$ there is now a factor of $\imd^{J + \Delta_\sigma + \Delta_\varphi - 1}$, which as we argued there only has a single non-vanishing residue. 

		Since we are primarily interested in the behavior of the bubble near the free-field pole $J = \minus\Delta_\sigma$, to leading order in $\Delta_\varphi$ we can approximate 
		\begin{equation}
			\mathcal{N}_{J, \Delta_\sigma \Delta_\varphi} \sim \frac{\Gamma(\alpha)}{4 \pi^{\alpha+1}}\frac{\mathcal{N}_{J, \Delta_\sigma}}{ \Delta_\varphi}\,,\mathrlap{\qquad \Delta_\varphi \to 0\,,}
		\end{equation}
		with $\mathcal{N}_{J, \Delta_\sigma}$ the coefficient appearing in the inversion of the propagator (\ref{eq:propLInvCoeff}), and thus (\ref{eq:cyclopsIR}) as
		\begin{equation}
			\begin{aligned}
				[\mathcal{G}_{\Delta_\sigma} \mathcal{G}_{\Delta_\varphi}]_{J}^{\slab{ir}}
				&\approx \frac{\Gamma(\alpha)}{4 \pi^{\alpha+1}}\frac{1}{\Delta_\varphi} \frac{\mathcal{N}_{J, \Delta_\sigma}}{J + \Delta_\sigma + \Delta_\varphi}\,. \label{eq:cyclopsIRAns}
			\end{aligned}
		\end{equation}
		Near the free-field pole $J = \minus \Delta_\sigma$, the terms we have dropped are suppressed by a factor of $\Delta_\varphi^2$ compared to the one we have kept.

		The analysis of the UV contribution $[\mathcal{G}_{\Delta_\sigma} \mathcal{G}_{\Delta_\varphi}]_J^\slab{uv}$ is more involved because it diverges as $\alpha \to \frac{3}{2}$. We leave its analysis in Appendix~\ref{app:UVLight}, but the upshot is that $[\mathcal{G}_{\Delta_\sigma} \mathcal{G}_{\Delta_\varphi}]_{J}^{\slab{uv}}$ is regular in $J$ and is well-approximated by
		\begin{equation}
			\begin{aligned}
				[\mathcal{G}_{\Delta_\sigma} \mathcal{G}_{\Delta_\varphi}]_{J}^\slab{uv} \approx \frac{\Gamma(\alpha)}{4 \pi^{\alpha+1}} \frac{\mathcal{N}_{J, \Delta_\sigma}}{\Delta_{\varphi}} \int_{1}^\infty\!\ud \imd\, &\imd^{J + \Delta_\sigma - 1} \, \tFo{J+\alpha + \frac{1}{2}}{J+1}{2 J + 2 \alpha + 1}{\minus \imd} \\ &\times  \tFo{\Delta_{\sigma}}{\Delta_\sigma - \alpha + \frac{1}{2}}{2 \Delta_\sigma - 2 \alpha + 1}{\minus \imd} + \cdots 
			\end{aligned}\label{eq:cyclopsUV}
		\end{equation}
		as $\Delta_\varphi \to 0$, 
		where the $\cdots$ denote terms that diverge as $\epsilon \to 0$. These divergences simplify drastically only once we sum over dimensions and their shadows, i.e. only in $[G^{\sigma} G^{\varphi}]_J$ and not in expressions like $[\mathcal{G}_{\Delta_\sigma} \mathcal{G}_{\Delta_\varphi}]_J$. Regardless, aside from the UV divergence, (\ref{eq:cyclopsUV}) is subleading in $\Delta_\varphi$ compared to (\ref{eq:cyclopsIRAns}) near $J = -\Delta_\sigma$ and so we may drop it. 

\begin{figure}
			\centering
			\includegraphics{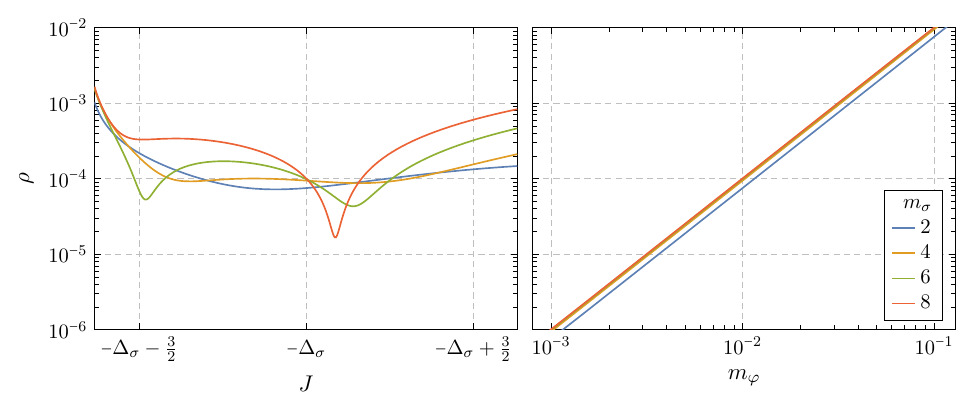}
			\caption{Plots of the relative error $\rho \equiv | \lab{exact} - \lab{approx}|/|\lab{exact}|$ between the approximation~(\ref{eq:bubbleAns}) to the exact result (\ref{eq:mmResult}), with $\alpha = \frac{1}{2}(3 - 10^{\sminus 8})$. On the \emph{left}, we plot this relative error as a function of $J$, along a line in the complex plane that is parallel to the real axis and passes through the free-field pole $J = \minus \Delta_\sigma$, for $m_\varphi^2 = 10^{\sminus 2}$ and various values of $m_\sigma$. On the \emph{right}, we plot this relative error at the free-field pole $J = \minus \Delta_\sigma$ as a function of $m_\varphi$. It can be seen that the relative error between the approximate (\ref{eq:bubbleAns}) and exact (\ref{eq:mmResult}) results is~$\mathcal{O}(m_\varphi^2)$. \label{fig:mmCompare}}
		\end{figure}

		A similar story applies to $[\mathcal{G}_{\smash{\bar{\Delta}_\sigma}} \mathcal{G}_{\Delta_\varphi}]_J$, which dominates near the free-field pole at $J = \minus \bar{\Delta}_\sigma$, and to compute it we only need to replace $\Delta_\sigma$ with $\bar{\Delta}_\sigma$ in (\ref{eq:cyclopsIRAns}). The other terms in (\ref{eq:bubbleDecomp}) are neither singular near the free-field poles $J = \minus \Delta_\sigma$ or $J = -\bar{\Delta}_\sigma$, nor are they enhanced by diverging factors of $\mathcal{A}(\Delta_\varphi)$. Thus, at leading order in $\Delta_\varphi$, we can approximate the bubble as
		\begin{equation}
			[G^{\sigma} G^{\varphi}]_J \approx [G^{\sigma} \mathcal{G}_{\Delta_\varphi}]_J^\slab{ir} + \cdots\approx \frac{\Gamma(\alpha)}{4 \pi^{\alpha+1}} \frac{1}{\Delta_\varphi} \frac{1}{(J + \Delta_\sigma + \Delta_\varphi)(J + \bar{\Delta}_\sigma + \Delta_\varphi)} + \cdots\,,
		\end{equation}
		where the $\cdots$ denote the UV divergences from $[G^{\sigma} G^{\varphi}]_J^\slab{uv}$. 
		Specifically, as $\alpha \to \frac{3}{2}$, this becomes 
		\begin{equation}
			[G^{\sigma} G^{\varphi}]_J \approx \frac{3}{8\pi^2} \frac{1}{m_\varphi^2} \frac{1}{(J + \Delta_\sigma + \Delta_\varphi)(J + \bar{\Delta}_\sigma + \Delta_\varphi)} + \frac{1}{8 \pi^2 \epsilon}\,, \label{eq:bubbleAns}
		\end{equation}
		where the $1/(8 \pi^2 \epsilon)$ is the divergence of the bubble in dimensional regularization (\ref{eq:bubbleUVDiv}). We compare this approximation to the exact result (\ref{eq:mmResult}) in Figure~\ref{fig:mmCompare}, where we find excellent agreement as $m_\varphi \to 0$. To leading order in $\Delta_\varphi$, we thus find that the self-energy reduces to
		\begin{equation}
			\Pi_{\sigma}(J)  \approx \frac{ g^2}{8 \pi^2 \Delta_\varphi} \frac{1}{(J+ \Delta_\sigma + \Delta_\varphi) (J + \bar{\Delta}_{\sigma} + \Delta_\varphi)} + \frac{g^2}{8 \pi^2 \epsilon} - J(J+2 \alpha) \delta_{Z_\sigma} - \delta_{m_\sigma}\,
		\end{equation}
		in the regions of $J$ which are relevant for $\sigma$'s propagation over long distances. This allows us to determine our counterterms as
		\begin{equation}
			\delta_{Z_\sigma} = \frac{g^2}{32 \pi^2 \nu_\sigma^2 \Delta_\varphi^3} \quad\text{and}\qquad \delta_{m_\sigma} = \frac{g^2}{8 \pi^2 \epsilon} + \frac{g^2 m_\sigma^2}{32 \pi^2 \nu_\sigma^2 \Delta_\varphi^3}\,,
		\end{equation}
		where $\nu_\sigma \equiv \sqrt{m_\sigma^2-\alpha^2}$. Furthermore, the free-field pole at $J_* = \minus \Delta_\sigma = \minus (\alpha + i \nu_\sigma)$ is shifted to
		\begin{equation}
			J_* = -\frac{3}{2}\left(1 + \frac{3}{16 \pi^2} \frac{g^2}{\nu_\sigma^2 m_\varphi^4}\right)  - i \nu_\sigma\,.
		\end{equation}
		As noted in \cite{Marolf:2010zp}, the heavy field $\sigma$'s interaction with the light field $\varphi$ causes it to decay faster than any free field since $\sigma$'s coupling to $\varphi$ provides it with another decay channel in addition to the standard dilution due to Hubble expansion. On the other hand, as explained in \cite{Lu:2021wxu}, the enhancement of this decay of $\sigma$ with respect to $m_\varphi$ can be interpreted as the washing out of $\sigma$ correlations at long distances due to the enhanced fluctuations of $\varphi$ in the infrared. In this way, we can make contact with in-in perturbation theory wherein the enhanced fluctuations of $\varphi$ are tied to the dynamics of the super-horizon modes \cite{Burgess:2009bs,Burgess:2010dd}.

		We have argued that $[G^{\sigma} G^{\varphi}]_J$'s singularity structure in $J$ can be extracted from the IR contribution $[G^{\sigma} G^{\varphi}]_J$ alone since the UV contribution is regular in $J$. We may then write
		\begin{equation}
			[G^{\sigma} G^{\varphi}]_J = \sum_{k = 0}^{\infty} \left[\frac{\mathcal{R}(\Delta_\sigma + \Delta_\varphi + 2k)}{J + \Delta_\sigma + \Delta_\varphi + 2 k} + \frac{\mathcal{R}(\Delta_\sigma + \bar{\Delta}_\varphi + 2k)}{J + \Delta_\sigma + \bar{\Delta}_\varphi + 2 k} + \big(\Delta_\sigma \to \bar{\Delta}_\sigma\big)\right] + f(J)\,, \label{eq:bubbleResSum}
		\end{equation}
		where $\mathcal{R}(J_*)$ denotes the residue of this function at $J = \minus J_*$ and $f(J)$ is an analytic function. Unlike the propagator in \S\ref{sec:linv}, the bubble does not necessarily decay as $|J| \to \infty$ because the product $G^{\sigma}(\emd) G^{\varphi}(\emd)$ is too singular as $\emd \to 1$, ultimately leading to the UV divergence $[G^{\sigma} G^{\varphi}]_J \sim 1/(8 \pi^2 \epsilon)$. Analytically continuing our Euclidean results defined on $\emd \in [-1, 1]$ to Lorentzian signature with $\emd \in \mathbb{R}$ requires that these coefficients decay sufficiently rapidly to use the Watson-Sommerfeld transform, and we ensure this is possible by adjusting our counterterms, subtracting off the constant---or the $J(J+3)$ terms in cases where the kinetic counterterm is needed---as $J \to \infty$. Once this is done, the self-energy $\Pi_\sigma(J)$ vanishes as $J \to \infty$ and thus we can conclude that $f(J) = 0$. The full self-energy can thus be recovered from the IR contribution $[G^{\sigma} G^{\varphi}]_J$ alone. This agrees with (\ref{eq:bubbleAns}), where we have truncated the sum in (\ref{eq:bubbleResSum}) to only its dominant poles at $J = \minus(\Delta_\sigma + \Delta_\varphi)$ and $ -(\bar{\Delta}_\sigma + \Delta_\varphi)$ and approximated their residues in the $\Delta_\varphi \to 0$ limit.    

		Before we proceed to study quartic interactions, it will be helpful to understand how the approximation (\ref{eq:bubbleAns}) arises in Euclidean signature as in~\cite{Lu:2021wxu}. Here, the analysis is much more straightforward since we can use the Euclidean inversion formula (\ref{eq:euclideanInv}) for non-negative integer~$J$,
		\begin{equation}
			[G^{\sigma} G^{\varphi}]_J =  \frac{(4 \pi)^\alpha \Gamma(\alpha) \Gamma(J+1)}{\Gamma(J + 2 \alpha)} \int_{\sminus 1}^{1}\!\ud \emd \, \big(1 - \emd^2\big)^{\alpha - \frac{1}{2}} C_{J}^{\alpha}(\emd) \, G^{\sigma}(\emd) G^{\varphi}(\emd)\,,{\quad J \in\mathbb{N}\,.} \label{eq:bubbleEuclidean}
		\end{equation}
		Since this integral is restricted to the finite interval $\emd \in [-1, 1]$, we may expand the light field propagator into its harmonics (\ref{eq:propSH}),
		\begin{equation}
			G^\varphi(\emd) = \frac{\Gamma(\alpha + 1)}{2 \pi^{\alpha + 1}} \frac{1}{m_\varphi^2} + \frac{\Gamma(\alpha)}{2 \pi^{\alpha + 1}} \sum_{J = 1}^{\infty} \frac{J+\alpha}{J(J+ 2\alpha) + m_\varphi^2} C_{J}^{\alpha}(\emd)\,,\label{eq:zmDecomp}
		\end{equation}
		where the first term is the Euclidean zero mode (\ref{eq:zmProf}) propagator. As $m_\varphi \to 0$, we find that the Euclidean zero mode dominates the bubble $[G^{\sigma} G^{\varphi}]_J$, and (\ref{eq:bubbleEuclidean}) may be approximated as
		\begin{equation}
			[G^{\sigma} G^{\varphi}]_J \sim  \frac{\Gamma(\alpha+1)}{2 \pi^{\alpha + 1}} \frac{[G^{\sigma}]_J }{m_\varphi^2} \sim\frac{\Gamma(\alpha+1)}{2 \pi^{\alpha + 1}} \frac{1}{m_\varphi^2} \frac{1}{(J+\Delta_\sigma)(J + \bar{\Delta}_\sigma)}\,.
		\end{equation}
		This agrees with (\ref{eq:bubbleAns}) as $\Delta_\varphi \to 0$ when $J \in \mathbb{N}$, and so we can interpret the enhancement factor as the Euclidean zero mode of the light field $\varphi$ which becomes strongly coupled~\cite{Rajaraman:2010xd} as $m_\varphi \to 0$.

\subsection{The Sunset} \label{sec:sunset}

	Having studied the bubble diagram, we now focus on sunset diagrams. Specifically, we will study the theory with interaction and counterterm Lagrangians
	\begin{equation}
		\begin{aligned}
			\mathcal{L}_\lab{int} + \mathcal{L}_\lab{ct} &= \tfrac{1}{2} g \sigma^2 \varphi^2 + \tfrac{1}{4!} g_\varphi \varphi^4 + \tfrac{1}{4!} g_\sigma \sigma^4  + \tfrac{1}{2} \delta_{Z_\sigma} (\partial \sigma)^2 + \tfrac{1}{2} \delta_{Z_\varphi} (\partial \varphi)^2 + \tfrac{1}{2} \delta_{m_\sigma} \sigma^2 + \tfrac{1}{2} \delta_{m_\varphi} \varphi^2 \\
			&+ \tfrac{1}{2} \delta_{g} \sigma^2 \varphi^2 + \tfrac{1}{4!} \delta_{g_{\varphi}} \varphi^4 + \tfrac{1}{4!} \delta_{g_{\sigma}} \sigma^4 
		\end{aligned}
	\end{equation}
	and associated interaction vertices
	\begin{equation}
		\begin{tikzpicture}[baseline=-3pt, thick]
			\def\lSize{0.7}
			\draw (-\lSize, \lSize) -- (\lSize, -\lSize);
			\draw[cornellBlue] (-\lSize, -\lSize) -- (\lSize, \lSize);
			\fill[cornellBlue, draw=white, very thick] (0,0) circle (0.09);
		\end{tikzpicture} = \minus 2 g\,, \qquad\qquad \begin{tikzpicture}[baseline=-3pt, thick]
			\def\lSize{0.7}
			\draw[cornellBlue] (-\lSize, \lSize) -- (\lSize, -\lSize);
			\draw[cornellBlue] (-\lSize, -\lSize) -- (\lSize, \lSize);
			\fill[cornellBlue, draw=white, very thick] (0,0) circle (0.09);
		\end{tikzpicture} = \minus g_\varphi\,, \qquad\qquad \begin{tikzpicture}[baseline=-3pt, thick]
			\def\lSize{0.7}
			\draw (-\lSize, \lSize) -- (\lSize, -\lSize);
			\draw (-\lSize, -\lSize) -- (\lSize, \lSize);
			\fill[black, draw=white, very thick] (0,0) circle (0.09);
		\end{tikzpicture} = \minus g_\sigma\,, \label{eq:vertexDiagrams}
	\end{equation}
	To these, we add the analogous counterterm vertices, with $g \to \delta_g$, $g_\varphi \to \delta_{g_{\varphi}}$ and $g_\sigma \to \delta_{g_\sigma}$, which we distinguish with a crossed circle, e.g. \begin{tikzpicture}[ baseline=-3pt] \draw[fill=white, line width=0.3mm] (0, 0) circle (0.125); \draw[rotate=45, line width=0.25mm] (-0.125, 0) -- (0.125, 0); \draw[rotate=45, line width=0.25mm] (0, -0.125)--(0, 0.125); \end{tikzpicture}.
	Specifically, we will compute the self-energies of $\sigma$ and $\varphi$, $\Pi_\sigma(J)$ and $\Pi_\varphi(J)$ respectively, in the limit $m_\varphi \to 0$ with $g$, $g_\varphi$, and $g_\sigma$ all $\ll 1$. We will work perturbatively in each of these couplings to second order, which we will collectively denote $\mathcal{O}(g^2)$ as shorthand---this includes terms that are $\mathcal{O}(g g_\sigma)$, $\mathcal{O}(g_\sigma^2)$. This is the first order at which physical corrections arise and they will be generated by so-called ``sunset'' diagrams.\footnote{The sunset diagram in de Sitter has also been studied using the Schwinger-Dyson equations for the $\lab{O}(N)$ scalar field theory in~\cite{Gautier:2013aoa}. }

	At first order in the couplings, $\sigma$'s self-energy is given by
	\begin{equation}
			\def\lSize{0.65}
			\def\lSizeA{0.9}
			\def\circSize{0.35}
			\def\circSizeA{0.5}
			\def\circSizeCT{0.1}
			\begin{tikzpicture}[baseline=-3pt, thick]
					\draw (-\lSizeA, 0) -- (-0.5, 0);
					\draw (0.5, 0) -- (\lSizeA, 0);
					\begin{scope}
							\clip[draw] (0, 0) circle (\circSizeA);
							\foreach \x in {-0.5, -0.425, ..., 0.5} 
							{	
								\draw[rotate=45, thin] (-\circSizeA, \x) -- (\circSizeA, \x);
							}
					\end{scope}
					\fill[black, draw=white] (-\circSizeA,0) circle (0.06);
					\fill[black, draw=white] (\circSizeA,0) circle (0.06);
			\end{tikzpicture} = 
			\begin{tikzpicture}[baseline=-3pt, thick]
			\draw (-\lSize, 0) -- (\lSize, 0);
			\draw[cornellBlue] (0, \circSize) circle (\circSize);
			\fill[cornellBlue, draw=white] (0,0) circle (0.06);
			\end{tikzpicture} + \begin{tikzpicture}[baseline=-3pt, thick]
			\draw (-\lSize, 0) -- (\lSize, 0);
			\draw[black] (0, \circSize) circle (\circSize);
			\fill[black, draw=white] (0,0) circle (0.06);
			\end{tikzpicture} + 
			\begin{tikzpicture}[baseline=-3pt, thick]
					\draw (-\lSize, 0) -- (\lSize, 0);
					\fill[white, draw=black, line width=0.3mm] (0, 0) circle (\circSizeCT);
					\draw[rotate=45, line width=0.2mm] (-\circSizeCT, 0) -- (\circSizeCT, 0);
					\draw[rotate=-45, line width=0.2mm] (-\circSizeCT, 0) -- (\circSizeCT, 0);
			\end{tikzpicture}\,,
		\end{equation}
		where we exclude terms associated with vertex renormalization that are not needed to this order in perturbation theory. Specifically, we have
		\begin{equation}
			\Pi_{\sigma}(J) = -g \es \es G^{\varphi}(1) - \tfrac{1}{2} g_\sigma \es G^{\sigma}(1) - J(J+2\alpha)\delta_{Z_\sigma}\!- \delta_{m_\sigma}\,,
		\end{equation}
		where $G(1) \equiv \lim_{\emd \to 1} G(\emd)\equiv \lim_{y\to x} G(x, y) $ represents the coincident limit of the propagator. These diagrams are UV divergent and, as before, we regularize them via dimensional regularization by working in $\alpha = \tfrac{1}{2}(3-\epsilon)$ and taking $\epsilon \to 0$ with masses held fixed. We define the coincident limit of the free-field propagator in general dimensions as
		\begin{equation}
			\begin{aligned}
				G(1) &= \frac{1}{(4 \pi)^{\alpha + \frac{1}{2}}} \, \G{\frac{1}{2} - \alpha\,,\, \Delta\,,\, 2\alpha - \Delta}{\frac{1}{2} +\alpha - \Delta\,,\, \frac{1}{2} -\alpha + \Delta}
			\end{aligned}
		\end{equation}
		which, as $\epsilon \to 0$, behaves as
		\begin{equation}
			G(1) \sim -\frac{(\Delta - 1)(\bar{\Delta}-1)}{16 \pi^2}\left[\frac{2}{\epsilon}- \psi\big(\Delta-1\big) - \psi\big(\bar{\Delta} - 1\big) + \log 4 \pi \e^{\sminus \gamma_\slab{e}} + 1\right] - \frac{1}{8 \pi^2}\,,
		\end{equation}
		with $\gamma_\slab{e} \equiv -\psi(1)$ the Euler-Mascheroni constant and $\psi(z) = \Gamma'(z)/\Gamma(z)$ the digamma function. In our renormalization scheme, this contribution is completely absorbed by the mass counterterm, such that $\delta_{Z_\sigma} = \mathcal{O}(g^2)$ and $\delta_{m_\sigma} = -g \es\es G^{\varphi}(1) - \frac{1}{2} g_\sigma\es\es G^{\sigma}(1) + \mathcal{O}(g^2)$. The same is true for the self-energy of $\varphi$ to first order, which is given by
		\begin{equation}
			\def\lSize{0.65}
			\def\lSizeA{0.9}
			\def\circSize{0.35}
			\def\circSizeA{0.5}
			\def\circSizeCT{0.1}
			\begin{tikzpicture}[baseline=-3pt, thick]
					\draw[cornellBlue] (-\lSizeA, 0) -- (-0.5, 0);
					\draw[cornellBlue] (0.5, 0) -- (\lSizeA, 0);
					\begin{scope}[cornellBlue]
							\clip[draw] (0, 0) circle (\circSizeA);
							\foreach \x in {-0.5, -0.425, ..., 0.5} 
							{	
								\draw[rotate=45, thin, cornellBlue] (-\circSizeA, \x) -- (\circSizeA, \x);
							}
					\end{scope}
					\fill[cornellBlue, draw=white] (-\circSizeA,0) circle (0.06);
					\fill[cornellBlue, draw=white] (\circSizeA,0) circle (0.06);
			\end{tikzpicture} = 
			\begin{tikzpicture}[baseline=-3pt, thick]
			\draw[cornellBlue](-\lSize, 0) -- (\lSize, 0);
			\draw[] (0, \circSize) circle (\circSize);
			\fill[cornellBlue, draw=white] (0,0) circle (0.06);
			\end{tikzpicture} + \begin{tikzpicture}[baseline=-3pt, thick]
			\draw[cornellBlue] (-\lSize, 0) -- (\lSize, 0);
			\draw[cornellBlue] (0, \circSize) circle (\circSize);
			\fill[cornellBlue, draw=white] (0,0) circle (0.06);
			\end{tikzpicture} + 
			\begin{tikzpicture}[baseline=-3pt, thick]
					\draw[cornellBlue] (-\lSize, 0) -- (\lSize, 0);
					\fill[white, draw=cornellBlue, line width=0.3mm] (0, 0) circle (\circSizeCT);
					\draw[rotate=45, line width=0.2mm, cornellBlue] (-\circSizeCT, 0) -- (\circSizeCT, 0);
					\draw[rotate=-45, line width=0.2mm, cornellBlue] (-\circSizeCT, 0) -- (\circSizeCT, 0);
			\end{tikzpicture}\,,
		\end{equation}
		or equivalently
		\begin{equation}
			\Pi_{\varphi}(J) = -g\es\es G^\sigma(1) - \tfrac{1}{2} g_\varphi \es G^\varphi(1) - J(J+2\alpha) \delta_{Z_\varphi} - \delta_{m_\varphi}\,.
		\end{equation}
		For a light field, there is only one pole with maximal real part, and ensuring this pole yields the physically measured mass forces us to choose $\delta_{Z_\varphi} = \mathcal{O}(g^2)$ and $\delta_{m_\varphi} = -g \es\es G^\sigma(1) - \tfrac{1}{2} g_{\varphi} \es G^{\varphi}(1) + \mathcal{O}(g^2)$. Physical corrections to the propagation of both $\sigma$ and $\varphi$ instead occur at $\mathcal{O}(g^2)$.
		
		At second order in perturbation theory, $\sigma$'s self-energy $\Pi_\sigma(J)$ is given by
				\begin{equation}
			\def\lSize{0.65}
			\def\lSizeA{0.9}
			\def\circSize{0.3}
			\def\circSizeA{0.5}
			\def\circSizeCT{0.08}
			\begin{aligned}
				\begin{tikzpicture}[baseline=-3pt, thick]
					\draw (-\lSizeA, 0) -- (-0.5, 0);
					\draw (0.5, 0) -- (\lSizeA, 0);
					\begin{scope}
							\clip[draw] (0, 0) circle (\circSizeA);
							\foreach \x in {-0.5, -0.425, ..., 0.5} 
							{	
								\draw[rotate=45, thin] (-\circSizeA, \x) -- (\circSizeA, \x);
							}
					\end{scope}
					\fill[black, draw=white] (-\circSizeA,0) circle (0.06);
					\fill[black, draw=white] (\circSizeA,0) circle (0.06);
			\end{tikzpicture} &= \begin{tikzpicture}[baseline=-3pt, thick]
					\draw (-\lSizeA, 0) -- (\lSizeA, 0);
					\draw[cornellBlue] (0, 0) circle (0.5);
					\fill[cornellBlue, draw=white] (-0.5,0) circle (0.06);
					\fill[cornellBlue, draw=white] (0.5,0) circle (0.06);
					\end{tikzpicture} + \begin{tikzpicture}[baseline=-3pt, thick]
					\draw (-\lSizeA, 0) -- (\lSizeA, 0);
					\draw[] (0, 0) circle (0.5);
					\fill[black, draw=white] (-0.5,0) circle (0.06);
					\fill[black, draw=white] (0.5,0) circle (0.06);
					\end{tikzpicture} + \begin{tikzpicture}[baseline=-3pt, thick]
					\def\circSize{0.35}
					\def\circSizeCT{0.08}
					\draw (-\lSize, 0) -- (\lSize, 0);
					\draw (0, \circSize) circle (\circSize);
					\fill[white, draw=black, line width=0.3mm] (0, 0) circle (\circSizeCT);
					\draw[rotate=45, line width=0.25mm] (-\circSizeCT, 0) -- (\circSizeCT, 0);
					\draw[rotate=-45, line width=0.25mm] (-\circSizeCT, 0) -- (\circSizeCT, 0);
				\end{tikzpicture} + \begin{tikzpicture}[baseline=-3pt, thick]
					\def\circSize{0.35}
					\def\circSizeCT{0.08}
					\draw (-\lSize, 0) -- (\lSize, 0);
					\draw[cornellBlue] (0, \circSize) circle (\circSize);
					\fill[white, draw=black, line width=0.3mm] (0, 0) circle (\circSizeCT);
					\draw[rotate=45, line width=0.25mm] (-\circSizeCT, 0) -- (\circSizeCT, 0);
					\draw[rotate=-45, line width=0.25mm] (-\circSizeCT, 0) -- (\circSizeCT, 0);
				\end{tikzpicture} +  \begin{tikzpicture}[baseline=-3pt, thick]
					\def\circSize{0.35}
					\def\circSizeCT{0.09}
					\def\lSize{0.8}
					\draw (-\lSize, 0) -- (\lSize, 0);
					\fill[white, draw=black, line width=0.3mm] (0, 0) circle (\circSizeCT);
					\draw[rotate=45, line width=0.25mm] (-\circSizeCT, 0) -- (\circSizeCT, 0);
					\draw[rotate=-45, line width=0.25mm] (-\circSizeCT, 0) -- (\circSizeCT, 0);
				\end{tikzpicture}\\
					& + \begin{tikzpicture}[baseline=-3pt, thick]
					\draw (-\lSize, 0) -- (\lSize, 0);
					\draw[cornellBlue] (0, \circSize) circle (\circSize);
					\draw[cornellBlue] (0, 3*\circSize) circle (\circSize);
					\fill[cornellBlue, draw=white] (0,0) circle (0.06);
					\fill[cornellBlue, draw=white] (0,2*\circSize) circle (0.06);
					\end{tikzpicture} + \begin{tikzpicture}[baseline=-3pt, thick]
					\draw (-\lSize, 0) -- (\lSize, 0);
					\draw[cornellBlue] (0, \circSize) circle (\circSize);
					\draw[black] (0, 3*\circSize) circle (\circSize);
					\fill[cornellBlue, draw=white] (0,0) circle (0.06);
					\fill[cornellBlue, draw=white] (0,2*\circSize) circle (0.06);
					\end{tikzpicture} + \begin{tikzpicture}[baseline=-3pt, thick]
					\draw (-\lSize, 0) -- (\lSize, 0);
					\draw[cornellBlue] (0, \circSize) circle (\circSize);
					\fill[cornellBlue, draw=white] (0,0) circle (0.06);
					\begin{scope}[shift={(0, 2*\circSize)}]
						\fill[white, draw=cornellBlue, line width=0.3mm] (0, 0) circle (\circSizeCT);
						\draw[rotate=45, line width=0.25mm, cornellBlue] (-\circSizeCT, 0) -- (\circSizeCT, 0);
						\draw[rotate=-45, line width=0.25mm, cornellBlue] (-\circSizeCT, 0) -- (\circSizeCT, 0);
					\end{scope}
					\end{tikzpicture} + \begin{tikzpicture}[baseline=-3pt, thick]
					\draw (-\lSize, 0) -- (\lSize, 0);
					\draw[black] (0, \circSize) circle (\circSize);
					\draw[cornellBlue] (0, 3*\circSize) circle (\circSize);
					\fill[cornellBlue, draw=white] (0,0) circle (0.06);
					\fill[cornellBlue, draw=white] (0,2*\circSize) circle (0.06);
					\end{tikzpicture}  + \begin{tikzpicture}[baseline=-3pt, thick]
					\draw (-\lSize, 0) -- (\lSize, 0);
					\draw[black] (0, \circSize) circle (\circSize);
					\draw[black] (0, 3*\circSize) circle (\circSize);
					\fill[cornellBlue, draw=white] (0,0) circle (0.06);
					\fill[cornellBlue, draw=white] (0,2*\circSize) circle (0.06);
					\end{tikzpicture}  + \begin{tikzpicture}[baseline=-3pt, thick]
					\draw (-\lSize, 0) -- (\lSize, 0);
					\draw[black] (0, \circSize) circle (\circSize);
					\fill[black, draw=white] (0,0) circle (0.06);
					\begin{scope}[shift={(0, 2*\circSize)}]
						\fill[white, draw=black, line width=0.3mm] (0, 0) circle (\circSizeCT);
						\draw[rotate=45, line width=0.25mm, black] (-\circSizeCT, 0) -- (\circSizeCT, 0);
						\draw[rotate=-45, line width=0.25mm, black] (-\circSizeCT, 0) -- (\circSizeCT, 0);
					\end{scope}
					\end{tikzpicture} \,\,.
			\end{aligned}\label{eq:sigSecondOrder}
		\end{equation}
		It is clear that the diagrams on the second line of (\ref{eq:sigSecondOrder}) all cancel in our renormalization scheme, while the two diagrams associated with the $\mathcal{O}(g^2)$ vertex counterterms can be absorbed into the $\mathcal{O}(g^2)$ part of the $\delta_{m_\sigma}$ counterterm, as they yield just a constant factor $-\frac{1}{2} \delta_{g_\sigma} \es\es G^{\sigma}(1) - \frac{1}{2} \delta_{g} \es \es G^{\varphi}(1)$. So, to $\mathcal{O}(g^2)$ $\sigma$'s self-energy is given by the sum of sunset diagrams,
		\begin{equation}
			\def\lSize{0.65}
			\def\lSizeA{0.9}
			\def\circSize{0.3}
			\def\circSizeA{0.5}
			\def\circSizeCT{0.08}
				\begin{tikzpicture}[baseline=-3pt, thick]
					\draw (-\lSizeA, 0) -- (-0.5, 0);
					\draw (0.5, 0) -- (\lSizeA, 0);
					\begin{scope}
							\clip[draw] (0, 0) circle (\circSizeA);
							\foreach \x in {-0.5, -0.425, ..., 0.5} 
							{	
								\draw[rotate=45, thin] (-\circSizeA, \x) -- (\circSizeA, \x);
							}
					\end{scope}
					\fill[black, draw=white] (-\circSizeA,0) circle (0.06);
					\fill[black, draw=white] (\circSizeA,0) circle (0.06);
			\end{tikzpicture} = \begin{tikzpicture}[baseline=-3pt, thick]
					\draw (-\lSizeA, 0) -- (\lSizeA, 0);
					\draw[cornellBlue] (0, 0) circle (0.5);
					\fill[cornellBlue, draw=white] (-0.5,0) circle (0.06);
					\fill[cornellBlue, draw=white] (0.5,0) circle (0.06);
					\end{tikzpicture} + \begin{tikzpicture}[baseline=-3pt, thick]
					\draw (-\lSizeA, 0) -- (\lSizeA, 0);
					\draw[] (0, 0) circle (0.5);
					\fill[black, draw=white] (-0.5,0) circle (0.06);
					\fill[black, draw=white] (0.5,0) circle (0.06);
					\end{tikzpicture} +  \begin{tikzpicture}[baseline=-3pt, thick]
					\def\circSize{0.35}
					\def\circSizeCT{0.09}
					\def\lSize{0.8}
					\draw (-\lSize, 0) -- (\lSize, 0);
					\fill[white, draw=black, line width=0.3mm] (0, 0) circle (\circSizeCT);
					\draw[rotate=45, line width=0.25mm] (-\circSizeCT, 0) -- (\circSizeCT, 0);
					\draw[rotate=-45, line width=0.25mm] (-\circSizeCT, 0) -- (\circSizeCT, 0);
				\end{tikzpicture}
		\end{equation}
		or
		\begin{equation}
			\Pi_\sigma(J) = \frac{1}{2}(\minus 2 g)^2 [G^{\varphi} G^{\varphi} G^{\sigma}]_{J} + \frac{1}{3!}(\minus g_{\sigma})^2 [G^{\sigma} G^{\sigma} G^{\sigma}]_{J} - J(J+2\alpha)\es \delta^{\scriptscriptstyle (2)}_{Z_\sigma} - \delta_{m_\sigma}^{\scriptscriptstyle (2)}
		\end{equation}
		where we use $\delta^{\scriptscriptstyle (2)}_{Z_\sigma}$ and $\delta^{\scriptscriptstyle (2)}_{m_\sigma}$ to denote the counterterms at $\mathcal{O}(g^2)$, having absorbed the vertex counterterms. The logic is identical for the $\mathcal{O}(g^2)$ corrections to $\Pi_\varphi(J)$, and so we have
		\begin{equation}
			\def\lSize{0.65}
			\def\lSizeA{0.9}
			\def\circSize{0.3}
			\def\circSizeA{0.5}
			\def\circSizeCT{0.08}
				\begin{tikzpicture}[baseline=-3pt, thick, cornellBlue]
					\draw (-\lSizeA, 0) -- (-0.5, 0);
					\draw (0.5, 0) -- (\lSizeA, 0);
					\begin{scope}
							\clip[draw] (0, 0) circle (\circSizeA);
							\foreach \x in {-0.5, -0.425, ..., 0.5} 
							{	
								\draw[rotate=45, thin] (-\circSizeA, \x) -- (\circSizeA, \x);
							}
					\end{scope}
					\fill[cornellBlue, draw=white] (-\circSizeA,0) circle (0.06);
					\fill[cornellBlue, draw=white] (\circSizeA,0) circle (0.06);
			\end{tikzpicture} = \begin{tikzpicture}[baseline=-3pt, thick]
					\draw[cornellBlue] (-\lSizeA, 0) -- (\lSizeA, 0);
					\draw[] (0, 0) circle (0.5);
					\fill[cornellBlue, draw=white] (-0.5,0) circle (0.06);
					\fill[cornellBlue, draw=white] (0.5,0) circle (0.06);
					\end{tikzpicture} + \begin{tikzpicture}[baseline=-3pt, thick, cornellBlue]
					\draw (-\lSizeA, 0) -- (\lSizeA, 0);
					\draw[] (0, 0) circle (0.5);
					\fill[cornellBlue, draw=white] (-0.5,0) circle (0.06);
					\fill[cornellBlue, draw=white] (0.5,0) circle (0.06);
					\end{tikzpicture} +  \begin{tikzpicture}[baseline=-3pt, thick, cornellBlue]
					\def\circSize{0.35}
					\def\circSizeCT{0.09}
					\def\lSize{0.8}
					\draw (-\lSize, 0) -- (\lSize, 0);
					\fill[white, draw=cornellBlue, line width=0.3mm] (0, 0) circle (\circSizeCT);
					\draw[rotate=45, line width=0.25mm] (-\circSizeCT, 0) -- (\circSizeCT, 0);
					\draw[rotate=-45, line width=0.25mm] (-\circSizeCT, 0) -- (\circSizeCT, 0);
				\end{tikzpicture}
		\end{equation}
		or, equivalently,
		\begin{equation}
			\Pi_\varphi(J) = \frac{1}{2}(\minus 2 g)^2 [G^{\varphi} G^{\sigma} G^{\sigma}]_{J} + \frac{1}{3!}(\minus g_{\varphi})^2 [G^{\varphi} G^{\varphi} G^{\varphi}]_{J} - J(J+2\alpha)\es \delta^{\scriptscriptstyle (2)}_{Z_\varphi} - \delta_{m_\varphi}^{\scriptscriptstyle (2)}\,. \label{eq:phiSE}
		\end{equation}
		We see that the leading order corrections to the self-energies of $\sigma$ and $\varphi$ are determined by sums of sunset diagrams, and our goal is to approximate how they correct the long-distance behavior of the propagators in the limit $m_\varphi \to 0$.
		
		We begin by considering $\sigma$'s mixed sunset diagram, which we may decompose (\ref{eq:propAsymp}) as
		\begin{equation}
			\begin{aligned}
				\begin{tikzpicture}[baseline=-3pt, thick]
					\def\lSize{0.9}
					\def\lSizeA{1}
					\def\circSize{0.3}
					\def\circSizeCT{0.08}
					\draw (-\lSize, 0) -- (\lSize, 0);
					\draw[cornellBlue] (0, 0)+(0.5, 0) arc (0:180:0.5);
					\draw[cornellBlue] (0, 0)+(0.5, 0) arc (0:-180:0.5);
					\fill[cornellBlue, draw=white] (-0.5,0) circle (0.06);
					\fill[cornellBlue, draw=white] (0.5,0) circle (0.06);
				\end{tikzpicture} \propto [ G^{\sigma} G^{\varphi} G^{\varphi}]_{J} =  [G^{\sigma}\mathcal{G}_{\Delta_\varphi} \mathcal{G}_{\Delta_{\varphi}} ]_{J} + 2 \times [G^{\sigma}\mathcal{G}_{\Delta_{\varphi}} \mathcal{G}_{\smash{\bar{\Delta}_{\varphi}}} ]_{J} + [ G^{\sigma}\mathcal{G}_{\smash{\bar{\Delta}_{\varphi}}} \mathcal{G}_{\smash{\bar{\Delta}_{\varphi}}}]_{J}\,. \label{eq:sigmaMixedSunset}
			\end{aligned}
		\end{equation}
		Given our experience with the bubble diagram in \S\ref{sec:bubble}, we already know which terms in (\ref{eq:sigmaMixedSunset}) will dominate as $\Delta_\varphi \to 0$. The first term, $[G^{\sigma}\mathcal{G}_{\Delta_\sigma} \mathcal{G}_{\Delta_\sigma}]_J$, will contain two factors of $\mathcal{A}(\Delta_\varphi) \propto \Delta_\varphi^{\sminus 2}$ and have two families of integer-spaced poles at $J = \minus (\Delta_\sigma + 2 \Delta_\varphi + 2k)$ and $J = \minus( \bar{\Delta}_\sigma + 2 \Delta_\varphi + 2k)$, for $k \in \mathbb{N}$.\footnote{From the representation (\ref{eq:sunsetIR}), it is trivial to see that $[G^{\sigma} \mathcal{G}_{\Delta_\varphi} \mathcal{G}_{\Delta_\varphi}]_J^\slab{ir}$ \emph{potentially} has poles at $J = \minus (\Delta_\sigma + 2 \Delta_\varphi + k)$ and $J = \minus(\bar{\Delta}_\sigma + 2 \Delta_\varphi + k)$, for $k \in \mathbb{N}$. Unfortunately, it is not so easy to see from this form that the residues of the poles vanish for all odd $k$, and we do not have a proof of this. These residues are trivial to compute---one simply sets $J = \minus(\Delta_\sigma + 2 \Delta_\varphi + k)$ and series expands the integrand about $\imd=0$ to find the coefficient of the $\imd^k$ term. In practice, we have confirmed that residues of the odd $k$ poles vanish up to $k = 15$, though whether or not this is true does not affect our results. Physically, that these odd poles vanish follows from momentum conservation---the pole at $J = -(\Delta_\sigma + 2 \Delta_\varphi + k)$ is generated by a long-lived state of a $\Delta_\sigma$ ``particle'' and two $\Delta_\varphi$ ``particles,'' with $k$ units of integer-quantized momentum distributed among them. Since momentum must be conserved, we cannot add momentum along one leg without adding the opposite momentum along the other, and so $k$ must be even to correspond to a long-lived physical state.} As $\Delta_\varphi \to 0$, the right-most poles with $k = 0$ will encroach on the free-field poles at $J = \minus \Delta_\sigma$ and $J = \minus \bar{\Delta}_\sigma$ and dominate the self-energy. 

		This does not happen with the other two terms in (\ref{eq:sigmaMixedSunset}). By the same logic, $[ G^\sigma\mathcal{G}_{\Delta_\varphi} \mathcal{G}_{\smash{\bar{\Delta}_\varphi}}]_J$ will only contain one factor of $\mathcal{A}(\Delta_\varphi)$ and have two families of poles at $J = \minus(\Delta_\sigma + \Delta_\varphi + \bar{\Delta}_\varphi + 2k) = \minus(\Delta_\sigma + 2 \alpha + 2k)$ and $J = \minus(\bar{\Delta}_\sigma + 2 \alpha + 2k)$, neither of which encroach upon the free-field poles as $\Delta_\varphi \to 0$. Similarly, $[G^{\sigma}\mathcal{G}_{\smash{\bar{\Delta}_\varphi}} \mathcal{G}_{\smash{\bar{\Delta}_\varphi}}]_J$ has no factors of $\mathcal{A}(\Delta_\varphi)$ and poles at $J = \minus(\Delta_\sigma + 2 \bar{\Delta}_\varphi + 2k)$ and $J = \minus(\bar{\Delta}_\sigma + 2 \bar{\Delta}_\varphi + 2k)$. Thus, the terms $[G^{\sigma}\mathcal{G}_{\Delta_\varphi} \mathcal{G}_{\Delta_\varphi}]_J$, $[\mathcal{G}^{\sigma} \mathcal{G}_{\Delta_\varphi} \mathcal{G}_{\smash{\bar{\Delta}_{\varphi}}}]_J$ and $[\mathcal{G}^{\sigma} \mathcal{G}_{\smash{\bar{\Delta}_\varphi}} \mathcal{G}_{\smash{\bar{\Delta}_\varphi}}]_J$ will be of order $\Delta_\varphi^{\sminus 3}$, $\Delta_{\varphi}^{\sminus 1}$ and $\Delta_{\varphi}^{0}$, respectively, near the free-field poles as $\Delta_\varphi \to 0$ and so we will only need to keep the first. By similar logic, the other sunset diagram $\propto [G^{\sigma} G^{\sigma} G^{\sigma}]_J$ will be of $\mathcal{O}(\Delta^0_\varphi)$ as $\Delta_\varphi \to 0$, and so we need not consider it. 

		The dominant contribution to $\sigma$'s self-energy as $\Delta_\varphi \to 0$ is then given by
		\begin{align}
				[G^{\sigma} \mathcal{G}_{\Delta_{\varphi}} \mathcal{G}_{\Delta_{\varphi}}]_{J} &= \mathcal{N}_{J, \Delta_\sigma\es 2 \Delta_\varphi} \int_{0}^{\infty}\!\ud \imd \, \imd^{J + \Delta_\sigma + 2 \Delta_\varphi - 1} \tFo{J+\alpha + \frac{1}{2}}{J+1}{2 J + 2 \alpha + 1}{\minus \imd} \label{eq:sigmaSunsetDom} \\  &\qquad\times \tFo{\Delta_{\sigma}}{\Delta_\sigma - \alpha + \frac{1}{2}}{2 \Delta_\sigma - 2 \alpha + 1}{\minus \imd} \tFo{\Delta_{\varphi}}{\Delta_\varphi - \alpha + \frac{1}{2}}{2 \Delta_\varphi - 2 \alpha + 1}{\minus \imd}^2 + (\Delta_\sigma \to \bar{\Delta}_\sigma)\,, \nonumber
		\end{align}
		where we have introduced the coefficient 
		\begin{equation}
				\mathcal{N}_{J, \Delta_\sigma\es 2 \Delta_\varphi} = \frac{2 \pi^{\alpha+1}}{4^J} \frac{\mathcal{A}(\Delta_\sigma) \mathcal{A}(\Delta_\varphi) \mathcal{A}(\Delta_\varphi) \Gamma(J+1)}{\Gamma(J+ \alpha + 1) \Gamma(\Delta_\sigma + 2 \Delta_\varphi) \Gamma(1 - \Delta_\sigma - 2 \Delta_\varphi)}\, \underset{\Delta_\varphi \to 0}{\sim} \frac{\Gamma(\alpha)^2 \mathcal{N}_{J, \Delta_\sigma}}{16 \pi^{2 \alpha + 2} \Delta_\varphi^2}\,.
		\end{equation}
		Following the same strategy we used with the bubble diagram, the self-energy near the free-field pole $J = -\Delta_\sigma$ is dominated by the IR part of this integrand,
		\begin{equation}
			\begin{aligned}
				[G^{\sigma} \mathcal{G}_{\Delta_{\varphi}} \mathcal{G}_{\Delta_{\varphi}}]^{\slab{ir}}_{J} &\approx \frac{\Gamma(\alpha)^2}{16 \pi^{2\alpha + 2}}  \frac{\mathcal{N}_{J, \Delta_\sigma}}{\Delta_\varphi^2} \int_{0}^{1}\!\ud \imd \, \imd^{J + \Delta_\sigma + 2 \Delta_\varphi - 1}\, \tFo{J+\alpha + \frac{1}{2}}{J+1}{2 J + 2 \alpha + 1}{\minus \imd} \\
				&\qquad\qquad\qquad\qquad\qquad\qquad \times \tFo{\Delta_{\sigma}}{\Delta_\sigma - \alpha + \frac{1}{2}}{2 \Delta_\sigma - 2 \alpha + 1}{\minus \imd} +(\Delta_\sigma \to \bar{\Delta}_\sigma) 
			\end{aligned}\,, \label{eq:sunsetIR}
		\end{equation}
		which is well-approximated by the first term in the series expansion of the integrand about $\imd = 0$,
		\begin{equation}
			\begin{aligned}
				[G^{\sigma} \mathcal{G}_{\Delta_{\varphi}} \mathcal{G}_{\Delta_{\varphi}}]^{\slab{ir}}_{J} &\approx \left[\frac{\Gamma(\alpha)}{4 \pi^{\alpha + 1} \Delta_\varphi}\right]^2 \!\frac{1}{(J + \Delta_\sigma + 2 \Delta_\varphi)(J + \bar{\Delta}_\sigma + 2 \Delta_\varphi)}\,. \label{eq:sunsetIRAns}
			\end{aligned}
		\end{equation}
		To this, we must add the UV contribution $[G^{\sigma} \mathcal{G}_{\Delta_\varphi} \mathcal{G}_{\Delta_\varphi}]_J^\slab{uv}$. This is subleading as $\Delta_\varphi \to 0$, albeit UV divergent. We regularize the general sunset diagram in Appendix~\ref{app:uvDiv}, finding (\ref{eq:sunsetUVDiv}) as $\alpha \to \frac{3}{2}$. We find that as $\Delta_\varphi \to 0$, the sunset diagram (\ref{eq:sigmaMixedSunset}) is well-approximated by 
		\begin{equation}
			\begin{aligned}
				\begin{tikzpicture}[baseline=-3pt, thick]
					\def\lSize{0.9}
					\def\lSizeA{1}
					\def\circSize{0.3}
					\def\circSizeCT{0.08}
					\draw (-\lSize, 0) -- (\lSize, 0);
					\draw[cornellBlue] (0, 0)+(0.5, 0) arc (0:180:0.5);
					\draw[cornellBlue] (0, 0)+(0.5, 0) arc (0:-180:0.5);
					\fill[cornellBlue, draw=white] (-0.5,0) circle (0.06);
					\fill[cornellBlue, draw=white] (0.5,0) circle (0.06);
				\end{tikzpicture} \propto [G^{\sigma} G^{\varphi} G^{\varphi}]_{J} &\sim \frac{1}{64 \pi^4 \Delta_\varphi^2} \frac{1}{(J + \Delta_\sigma + 2 \Delta_\varphi)(J + \bar{\Delta}_\sigma + 2 \Delta_\varphi)} \\
				&  -\frac{J(J+3)}{2(4 \pi)^4 \epsilon} - \frac{1}{64 \pi^4 \epsilon} + \mathcal{U}(\Delta_\sigma) + 2 \,\mathcal{U}(\Delta_\varphi)
			\end{aligned} \label{eq:sigmaMixedSunsetAns}
		\end{equation}
		where $\mathcal{U}(\Delta)$ is defined in (\ref{eq:uDef}).
		Since $[G^{\sigma} G^{\sigma} G^{\sigma}]_J$ is subleading, (\ref{eq:sigmaMixedSunsetAns}) determines the self-energy $\Pi_\sigma(J)$ to leading order in $\epsilon$ and $\Delta_\varphi$. Imposing our renormalization conditions, $\lab{Re}\, \Pi_\sigma(\minus \Delta_\sigma) = 0$ and $\Pi'_{\sigma}(\minus \Delta_\sigma)$, we find second-order corrections to the kinetic and mass counterterms to be
		\begin{equation}
			\begin{aligned}
				\delta^{\scriptscriptstyle (2)}_{Z_\sigma} &= -\frac{g^2}{(4 \pi)^4} \left[\frac{1}{\epsilon} - \frac{1}{2 \es \nu_\sigma^2 \es \Delta_\varphi^4}\right]\\
				\delta^{\scriptscriptstyle (2)}_{m_\sigma} &= -\frac{8 g^2}{(4 \pi)^4 \epsilon} + \frac{g^2}{2(4\pi)^4} \frac{m_\sigma^2}{\nu_\sigma^2 \es \Delta_\varphi^4} + 2 g^2\,  \mathcal{U}(\Delta_\sigma)+ 4 g^2 \, \mathcal{U}(\Delta_\varphi)
			\end{aligned}
		\end{equation}
		at leading order in $\Delta_\varphi \to 0$. The imaginary part of the self-energy is unaffected by these real counterterms and is simply
		\begin{equation}
			\lab{Im}\, \Pi_\sigma(\minus \Delta_\sigma) \sim \frac{2 g^2}{(4 \pi)^4 \nu_\sigma \Delta_\varphi^3}\,,\mathrlap{\qquad \Delta_\varphi \to 0\,.}
		\end{equation}
		Using (\ref{eq:poleShift}), we find that the free-field pole is shifted by
		\begin{equation}
			J_* \approx \minus \frac{3}{2}\left(1 + \frac{9}{128 \pi^4 } \frac{g^2}{\nu_\sigma^2 m_\varphi^6}\right) - i \nu_\sigma\,.
		\end{equation}
		As with the bubble diagram of \S\ref{sec:bubble}, coupling to the light scalar $\varphi$ causes the heavy scalar $\sigma$ to decay faster than any free field in de Sitter. 

		This shift of the free-field pole translates into a clear observable in the context of the cosmological collider signal~\cite{Lu:2021wxu}. Ultimately, given any operator which couples to the scalar mode during inflation, it is the leading exponent of this operator in the infrared (i.e. in the late time limit) which fixes the decay of the three-point function $\VEV{\zeta_{\mb{k}_1}\zeta_{\mb{k}_3}\zeta_{\mb{k}_3}}$ in the squeezed limit~\cite{Arkani-Hamed:2015bza}. Therefore, the pole $J_*$ contains all the necessary information to characterize the scaling exponents in the shape function. Recalling (\ref{eq:cosmo_coll}), in the interacting theory the decay parameter corresponding to a $\sigma$ exchange in the cosmological collider precisely becomes 
		\begin{equation}
			\beta=\frac{1}{2}\!\left(1 + \frac{27}{128 \pi^4 } \frac{g^2}{\nu_\sigma^2 m_\varphi^6}\right),
		\end{equation} 
		thereby resulting in a clear suppression of the signal.

		We can also compute the leading-order corrections to the light scalar $\varphi$'s self-energy (\ref{eq:phiSE}). This is dominated by the sunset diagram involving three internal $\varphi$  legs,
		\begin{equation}
			\begin{aligned}
				\begin{tikzpicture}[baseline=-3pt, thick]
					\def\lSize{0.9}
					\def\lSizeA{1}
					\def\circSize{0.3}
					\def\circSizeCT{0.08}
					\draw[cornellBlue] (-\lSize, 0) -- (\lSize, 0);
					\draw[cornellBlue] (0, 0)+(0.5, 0) arc (0:180:0.5);
					\draw[cornellBlue] (0, 0)+(0.5, 0) arc (0:-180:0.5);
					\fill[cornellBlue, draw=white] (-0.5,0) circle (0.06);
					\fill[cornellBlue, draw=white] (0.5,0) circle (0.06);
				\end{tikzpicture} \propto [G^{\varphi} G^{\varphi} G^{\varphi}]_{J} &=  [\mathcal{G}_{\Delta_\varphi} \mathcal{G}_{\Delta_{\varphi}} \mathcal{G}_{\Delta_\varphi}]_{J} + 3 \times [\mathcal{G}_{\Delta_\varphi} \mathcal{G}_{\Delta_{\varphi}} \mathcal{G}_{\smash{\bar{\Delta}}_\varphi}]_{J}  \\
				& +  [\mathcal{G}_{\smash{\bar{\Delta}}_\varphi}\mathcal{G}_{\smash{\bar{\Delta}}_\varphi}\mathcal{G}_{\smash{\bar{\Delta}}_\varphi}]_{J} + 3 \times [\mathcal{G}_{\Delta_\varphi} \mathcal{G}_{\smash{\bar{\Delta}}_\varphi}\mathcal{G}_{\smash{\bar{\Delta}}_\varphi}]_{J}\,.
			\end{aligned} \label{eq:phiPureSunset}
		\end{equation}
		By the same logic we used for (\ref{eq:sigmaMixedSunset}), the first term $[\mathcal{G}_{\Delta_\varphi} \mathcal{G}_{\Delta_{\varphi}} \mathcal{G}_{\Delta_\varphi}]_{J}$ scales\footnote{From (\ref{eq:sunsetPhiCoeff}), we see that even though this term is proportional to $\mathcal{A}(\Delta_\varphi)^3 \sim \Delta_\varphi^{\sminus 3}$, the discontinuity in the inversion formula (\ref{eq:lInvForm}) introduces an extra $1/\Gamma(3 \Delta_\varphi)$ that causes this term to scale as $\Delta_\varphi^{\sminus 2}$ as $\Delta_\varphi \to 0$. This still matches with the Euclidean picture---if we replace every internal leg with a zero mode propagator as in (\ref{eq:zmDecomp}), we find that this diagram should scale as $m_\varphi^{\sminus 6} \propto \Delta_\varphi^{\sminus 3}$ when $J = 0$. At non-zero $J$, atleast one internal leg must be a non-zero mode, and so this diagram scales as $m_\varphi^{\sminus 4} \propto \Delta_\varphi^{\sminus 2}$ when $J \neq 0$.} as $\Delta_\varphi^{\sminus 2}$ and has poles at $J = \minus (3 \Delta_{\varphi} + 2 k)$, the second $[\mathcal{G}_{\Delta_\varphi} \mathcal{G}_{\Delta_{\varphi}} \mathcal{G}_{\smash{\bar{\Delta}}_\varphi}]_{J}$ scales as $\propto \Delta_\varphi^{\sminus 2}$ and has poles at $J = \minus (2 \Delta_\varphi + \smash{\bar{\Delta}}_{\varphi} + 2 k)$, the third $[\mathcal{G}_{\smash{\bar{\Delta}}_\varphi}\mathcal{G}_{\smash{\bar{\Delta}}_\varphi}\mathcal{G}_{\smash{\bar{\Delta}}_\varphi}]_{J}$ scales as $\propto \Delta_\varphi^{\sminus 1}$ and has poles at $J = \minus (\Delta_\varphi + 2 \bar{\Delta}_{\varphi} + 2 k)$, while the last term $[\mathcal{G}_{\Delta_\varphi} \mathcal{G}_{\smash{\bar{\Delta}}_\varphi}\mathcal{G}_{\smash{\bar{\Delta}}_\varphi}]_{J}$ scales as $\Delta_\varphi^0$ and has poles at $J = \minus (3 \smash{\bar{\Delta}}_{\varphi} + 2k)$, with $k \in \mathbb{N}$. We thus see that only the poles of the first term $[\mathcal{G}_{\Delta_\varphi} \mathcal{G}_{\Delta_{\varphi}} \mathcal{G}_{\Delta_\varphi}]_{J}$ encroach on the free-field pole at $J = \minus \Delta_\varphi$ and so dominates over the others---it is of order $\Delta_\varphi^{\sminus 3}$ while the others are of order $\Delta_\varphi^{\sminus 2}$, $\Delta_\varphi^{\sminus 1}$ and $\Delta_\varphi^{0}$, respectively. However, if we are interested in the behavior of the self-energy near the other free-field pole $J = \minus \bar{\Delta}_\varphi$, then $[\mathcal{G}_{\Delta_\varphi} \mathcal{G}_{\Delta_\varphi} \mathcal{G}_{\smash{\bar{\Delta}_\varphi}}]_J$ is of order $\Delta_\varphi^{\sminus 3}$ and dominates over the others.

		Focusing on $\Pi_\varphi(J)$ near $J = \minus \Delta_\varphi$, we argued that this is dominated by
		\begin{equation}
			[\mathcal{G}_{\Delta_\varphi} \mathcal{G}_{\Delta_\varphi} \mathcal{G}_{\Delta_{\varphi}}]_J = \mathcal{N}_{J, 3 \Delta_\varphi} \int_{0}^{\infty}\!\ud \imd\, \imd^{J + 3 \Delta_\varphi - 1}\tFoS{J+\alpha + \frac{1}{2}}{J+1}{2 J + 2 \alpha + 1}{\minus \imd} \tFoS{\Delta_{\varphi}}{\Delta_\varphi - \alpha + \frac{1}{2}}{2 \Delta_\varphi - 2 \alpha + 1}{\minus \imd}^3
		\end{equation}
		with
		\begin{equation}
			\mathcal{N}_{J, 3 \Delta_\varphi} = \frac{2 \pi^{\alpha +1}}{4^J} \frac{\mathcal{A}(\Delta_\varphi)^3 \Gamma(J+1)}{\Gamma(3 \Delta_\varphi) \Gamma(1 - 3 \Delta_\varphi) \Gamma(J+\alpha + 1)}\,. \label{eq:sunsetPhiCoeff}
		\end{equation}
		As with (\ref{eq:sigmaSunsetDom}), this integral is dominated by its IR contribution near $J = -\Delta_\varphi$, and we may approximate it by the closest pole,\footnote{When $J = 0$, we can easily evaluate (\ref{eq:phiPureSunset}) by decomposing the propagator into its zero mode and non-zero mode contributions, cf. (\ref{eq:zmDecomp}). Using the definition of $[G^{\varphi} G^{\varphi} G^{\varphi}]_J$ given in (\ref{eq:hJDef}), we have that (\ref{eq:phiPureSunset}) reduces to the product of three zero mode propagators multiplied the inverse square of the zero mode profiles (\ref{eq:zmProf}).  }
		\begin{equation}
			[\mathcal{G}_{\Delta_\varphi} \mathcal{G}_{\Delta_\varphi} \mathcal{G}_{\Delta_{\varphi}}]_J^\slab{ir} \approx \frac{3}{2 \alpha} \left[\frac{\Gamma(\alpha)}{4 \pi^{\alpha+1} \Delta_\varphi}\right]^{2} \frac{1}{J+3 \Delta_\varphi}\,.
		\end{equation}
		To this, we must add its UV divergence (\ref{eq:sunsetUVDiv}) and so we find that, as $\Delta_\varphi \to 0$,
		\begin{equation}
			\begin{tikzpicture}[baseline=-3pt, thick]
					\def\lSize{0.9}
					\def\lSizeA{1}
					\def\circSize{0.3}
					\def\circSizeCT{0.08}
					\draw[cornellBlue] (-\lSize, 0) -- (\lSize, 0);
					\draw[cornellBlue] (0, 0)+(0.5, 0) arc (0:180:0.5);
					\draw[cornellBlue] (0, 0)+(0.5, 0) arc (0:-180:0.5);
					\fill[cornellBlue, draw=white] (-0.5,0) circle (0.06);
					\fill[cornellBlue, draw=white] (0.5,0) circle (0.06);
				\end{tikzpicture} \propto [G^{\varphi} G^{\varphi} G^{\varphi}]_{J} \sim \frac{1}{64 \pi^4 \Delta_\varphi^2} \frac{1}{J+3 \Delta_\varphi}-\frac{J(J+3)}{2 (4 \pi)^4 \epsilon} - \frac{1}{64 \pi^4 \epsilon} + 3\, \mathcal{U}(\Delta_\varphi)
		\end{equation}
		near $J = \minus \Delta_\varphi$. Since this diagram also dominates over the other, $\varphi$'s self-energy is well-approximated by 
		\begin{equation}
			\Pi_\varphi(J) \approx \frac{g_\varphi^2}{3!} \left[\frac{1}{64 \pi^4 \Delta_\varphi^2} \frac{1}{J+3 \Delta_\varphi}-\frac{J(J+3)}{2 (4 \pi)^4 \epsilon} - \frac{1}{64 \pi^4 \epsilon} + 3\, \mathcal{U}(\Delta_\varphi)\right]- J(J+3)\es \delta^{\scriptscriptstyle (2)}_{Z_\varphi} - \delta_{m_\varphi}^{\scriptscriptstyle (2)}\,.
		\end{equation}
		This self-energy is completely real, and so its effect on the free-field pole $J = -\Delta_\varphi$ can be completely absorbed by the counterterms. Requiring $\lab{Re}\, \Pi(\minus \Delta_\varphi) = \Pi'(\minus \Delta_\varphi) = 0$, we find that
		\begin{equation}
			\begin{aligned}
				\delta_{Z_\varphi} &\sim -\frac{g_\varphi^2}{12 (4 \pi)^4 \epsilon} - \frac{g_\varphi^2}{18(4 \pi)^4 \Delta_\varphi^4} \\
				\delta_{m_\varphi} &\sim -\frac{2 g_\varphi^2}{3 (4 \pi)^4 \epsilon} + \frac{g_\varphi^2}{6(4 \pi)^4 \Delta_\varphi^3} + \frac{1}{2} g_\varphi^2 \,\mathcal{U}(\Delta_\varphi)\,,
			\end{aligned}
		\end{equation}
		as $\Delta_\varphi \to 0$.
		Of course, this is not to say that quantum fluctuations do not change the long-distance behavior of $\langle \varphi(x) \varphi(y)\rangle$, as the self-energy 
		\begin{equation}
			\Pi_\varphi(J) \sim \frac{ g_\varphi^2}{3 (4 \pi)^4 \Delta_\varphi^3}\left[ \frac{2 \Delta_\varphi}{J + 3\Delta_\varphi} + \frac{J(J+3)}{6 \Delta_\varphi} - \frac{1}{2}\right], \quad \Delta_\varphi \to 0\,,
		\end{equation}
		is still non-trivial in this limit.

		Before we conclude, it will be helpful to discuss the conditions under which these perturbative results are valid, concentrating on $\varphi$'s self-energy. There are a host of diagrams we can draw at $\mathcal{O}(g^3)$, though the only ones that yield physical effects are what we call the ``Homer'' diagrams,
		\begin{equation}
			\def\lSize{0.65}
			\def\lSizeA{0.9}
			\def\circSize{0.3}
			\def\circSizeA{0.5}
			\def\circSizeCT{0.08}
			\def\headWidth{0.65}
			\def\headHeight{0.5}
			\def\extLegs{0.4}
			\begin{tikzpicture}[baseline=-3pt, thick, cornellBlue]
					\draw (-\lSizeA, 0) -- (-0.5, 0);
					\draw (0.5, 0) -- (\lSizeA, 0);
					\begin{scope}
							\clip[draw] (0, 0) circle (\circSizeA);
							\foreach \x in {-0.5, -0.425, ..., 0.5} 
							{	
								\draw[rotate=45, thin] (-\circSizeA, \x) -- (\circSizeA, \x);
							}
					\end{scope}
					\fill[cornellBlue, draw=white] (-\circSizeA,0) circle (0.06);
					\fill[cornellBlue, draw=white] (\circSizeA,0) circle (0.06);
			\end{tikzpicture} \supset \begin{tikzpicture}[thick, cornellBlue, baseline=-3pt]
				\draw (0, 0) ++(\headWidth, 0) -- (\headWidth, \headHeight) arc(0:180:\headWidth) -- (-\headWidth, 0);
				\draw ({-\headWidth/2}, 0) circle ({\headWidth/2});
				\draw ({\headWidth/2}, 0) circle ({\headWidth/2});
				\draw ({-\headWidth}, 0) --+({-\extLegs}, 0);
				\draw ({\headWidth}, 0) --+({\extLegs}, 0);
				\fill[cornellBlue, draw=white] (-\headWidth,0) circle (0.06);
				\fill[cornellBlue, draw=white] (\headWidth,0) circle (0.06);
				\fill[cornellBlue, draw=white] (0,0) circle (0.06);
			\end{tikzpicture} + 
			\begin{tikzpicture}[thick, cornellBlue, baseline=-3pt]
				\draw[black] (0, 0) ++(\headWidth, 0) -- (\headWidth, \headHeight) arc(0:180:\headWidth) -- (-\headWidth, 0);
				\draw[black] ({\headWidth}, 0) arc (0:180:{\headWidth/2});
				\draw[black] ({-\headWidth}, 0) arc (180:0:{\headWidth/2});
				\draw[] ({\headWidth}, 0) arc (0:-180:{\headWidth/2});
				\draw[] ({-\headWidth}, 0) arc (-180:0:{\headWidth/2});
				\draw ({-\headWidth}, 0) --+({-\extLegs}, 0);
				\draw ({\headWidth}, 0) --+({\extLegs}, 0);
				\fill[cornellBlue, draw=white] (-\headWidth,0) circle (0.06);
				\fill[cornellBlue, draw=white] (\headWidth,0) circle (0.06);
				\fill[cornellBlue, draw=white] (0,0) circle (0.06);
			\end{tikzpicture} + \begin{tikzpicture}[thick, cornellBlue, baseline=-3pt]
				\draw[] (0, 0) ++(\headWidth, 0) -- (\headWidth, \headHeight) arc(0:180:\headWidth) -- (-\headWidth, 0);
				\draw[black] ({\headWidth}, 0) arc (0:180:{\headWidth/2});
				\draw[black] ({-\headWidth}, 0) arc (180:0:{\headWidth/2});
				\draw[black] ({\headWidth}, 0) arc (0:-180:{\headWidth/2});
				\draw[black] ({-\headWidth}, 0) arc (-180:0:{\headWidth/2});
				\draw ({-\headWidth}, 0) --+({-\extLegs}, 0);
				\draw ({\headWidth}, 0) --+({\extLegs}, 0);
				\fill[cornellBlue, draw=white] (-\headWidth,0) circle (0.06);
				\fill[cornellBlue, draw=white] (\headWidth,0) circle (0.06);
				\fill[black, draw=white] (0,0) circle (0.06);
			\end{tikzpicture} + \cdots\,.
		\end{equation}
		Specifically, the first dominates as $\Delta_\varphi \to 0$ and is given in position space by
		\begin{equation}
			\begin{tikzpicture}[thick, cornellBlue, baseline=8pt]
				\def\headWidth{0.65}
				\def\headHeight{0.5}
				\def\extLegs{0.4}
				
				\draw (0, 0) ++(\headWidth, 0) -- (\headWidth, \headHeight) arc(0:180:\headWidth) -- (-\headWidth, 0);
				\draw ({-\headWidth/2}, 0) circle ({\headWidth/2});
				\draw ({\headWidth/2}, 0) circle ({\headWidth/2});
				\draw ({-\headWidth}, 0) --+({-\extLegs}, 0);
				\draw ({\headWidth}, 0) --+({\extLegs}, 0);
				\fill[cornellBlue, draw=white] (-\headWidth,0) circle (0.06);
				\fill[cornellBlue, draw=white] (\headWidth,0) circle (0.06);
				\fill[cornellBlue, draw=white] (0,0) circle (0.06);
			\end{tikzpicture} = \mathcal{H}(z_1, z_2) \equiv \frac{1}{4}(\minus g_\varphi)^3 \int_{\lab{S}^D}\!\!\ud^D z_3 \,  \left[G^{\varphi}(z_1, z_3)\right]^2 \left[G^{\varphi}(z_3, z_2)\right]^2 G^{\varphi}(z_1, z_2)\,.
		\end{equation}
		We can easily compute this Homer's leading $\Delta_\varphi$-dependence at $J = 0$ by replacing each internal leg with a zero-mode propagator, cf. (\ref{eq:zmDecomp}), and then multiplying by the inverse-square of the zero mode profile (\ref{eq:zmProf}), yielding
		\begin{equation}
			[\mathcal{H}]_0 \sim \minus \frac{g^3_\varphi}{8  \alpha} \left[\frac{\Gamma(\alpha)}{4\pi^{\alpha+1} \Delta_\varphi}\right]^{4}\frac{1}{\Delta_\varphi}\,,\mathrlap{\qquad \Delta_\varphi \to 0\,.}
		\end{equation}
		Our perturbative expansion is valid as long as the Homer is subleading to the sunset (\ref{eq:phiPureSunset}), and their ratio at $J = 0$ is
		\begin{equation}
			\frac{[\mathcal{H}]_0}{[\mathcal{S}]_0} = \minus \frac{3 g_\varphi}{2} \left[\frac{\Gamma(\alpha)}{4 \pi^{\alpha+1} \Delta_\varphi}\right]^2 \underset{\alpha\to\frac{3}{2}}{\sim} \minus \frac{3 g_\varphi}{2} \left[\frac{3}{8 \pi^2 m_\varphi^2}\right]^2\,.
		\end{equation}
		We thus require that $g_\varphi \ll \frac{128}{27} \pi^4 m_\varphi^4$ to maintain perturbative control and ensure that $[\mathcal{H}]_0 \ll [\mathcal{S}]_0$. Equivalently, our results only apply to light and very weakly-coupled fields, and our perturbative expansions break down when $\varphi$ becomes very light with $m_\varphi \lesssim g_\varphi^{{1}/{4}}$.  Similar logic holds for the self-energy for $\sigma$ which has its own dominant Homer diagram with four internal $\varphi$ legs. Ensuring that this Homer is subleading to the sunset (\ref{eq:sigmaMixedSunset}) similarly requires that $g \ll \frac{128}{27} \pi^4 m_\varphi^4$. This is the familiar breakdown of perturbation theory for light fields in de Sitter \cite{Burgess:2009bs,Rajaraman:2010xd,Gorbenko:2019rza}. In Euclidean signature, this breakdown is tied to the fact that the zero mode becomes strongly coupled as $m_\varphi \to 0$ \cite{Rajaraman:2010xd}, and we will discuss how to restore control by rearranging perturbation theory around this strongly coupled zero mode in a follow-up work.

\section{Conclusions} \label{sec:conclusions}
	
	A heavy scalar $\sigma$, with $m_\sigma > \frac{3}{2} H$, that is excited during inflation will freely propagate, oscillating in phase at a frequency that depends on its mass, until it eventually decays. This, in turn, will impart an oscillatory signal onto the primordial bispectrum in the squeezed limit---the cosmological collider signal---which cannot be easily mimicked by other local processes, and so it serves as a ``smoking gun'' inflationary signature of the field $\sigma$, from which we can determine its mass (and spin). Unfortunately, a light field $\varphi$ does not generate a similar oscillatory signal in the bispectrum, but it \emph{will} modify how $\sigma$ propagates over long distances, causing it to decay more rapidly. We can thus infer the presence of such light fields via the cosmological collider signal by their impact on $\sigma$. These effects grow larger the lighter $\varphi$ is, and the goal of this paper was to study these corrections in the limit $m_\varphi/H \ll \frac{3}{2}$. 

	Specifically, we described how to compute the self-energies of $\sigma$ and $\varphi$ in de Sitter space, using the Froissart-Gribov or Lorentzian inversion formula~(\ref{eq:lInvForm}) to analytically continue from Euclidean to Lorentzian signature.  We found that these self-energies drastically simplified in the limit $m_\varphi/H \to 0$, and described how to appropriately leverage the small parameter $m_\varphi/H$ to make physical predictions. These techniques should apply straightforwardly to any diagram with only two vertices, regardless of the number of internal legs or loops. We first used them in \S\ref{sec:bubble} to analyze the bubble diagram, in which $\sigma$ and $\varphi$ interact via a cubic interaction~$g \varphi \sigma^2$. We did not consider all possible interactions in this theory, as we mainly used it as an illustrative test case to understand how self-energies behaved in the light limit. We then analyzed the fully interacting theory of $\sigma$ and $\varphi$ invariant under $\sigma \to \minus \sigma$ and $\varphi \to \minus \varphi$, whose physical corrections were determined by various sunset diagrams. We regulated these sunset diagrams in Appendix~\ref{app:uvDiv}, fully determining the mass and kinetic counterterms in this theory and computed how interactions affect the long-distance behavior of both fields in the limit $m_\varphi/H \to 0$. These results apply for light, weakly-coupled fields with couplings $g$, $g_\varphi \ll \frac{128}{27} \pi^4 m_\varphi^4$. 

	Given the ubiquity of such light scalar fields in modern high energy physics, there are several future directions worth pursuing:
	\begin{itemize}
		\item As mentioned previously, our techniques are easily and straightforwardly applicable to any diagram with only two vertices. It would be interesting to study diagrams with more internal vertices, like the Homer diagrams of \S\ref{sec:sunset}. Can we easily infer their analytic structure using (\ref{eq:lInvForm}), and do they also simplify in the $m_\varphi /H \to 0$ limit?
		\item Similarly, do other correlators like $\langle \sigma(x) \sigma(y) \sigma(z)\rangle$ also simplify in the limit $m_\varphi/H \to 0$?
		\item Given that de Sitter correlators are uniquely sensitive to the quantum fluctuations of light fields, are there other inflationary observables that can be used to detect them?
		\item Finally, how do these correlators behave away from very weak coupling, $g$, $g_\varphi \gtrsim \frac{128}{27} \pi^4 m_\varphi^4$?
	\end{itemize}
	We hope to return to some of these questions in the future.

\vspace{0.75cm}

\noindent \textbf{Acknowledgements}	

\noindent We would like to thank Arindam Bhattacharya, Michael Geller, Cody Long, Qianshu Lu, Austin Joyce, Rashmish Mishra, Matt Reece, and Lian-Tao Wang for very helpful discussions. We would especially like to thank Qianshu Lu and Matt Reece for comments on the draft. The work of both PC and JS is supported by the DOE grant \texttt{DE-SC0013607}, while the work of JS is also supported NASA Grant \texttt{80NSSC20K0506}. This work was performed in part at Aspen Center for Physics, which is supported by National Science Foundation grant \texttt{PHY-2210452}.

\newpage 
\appendix

\section{Formulary} \label{app:formulary}
	
	In this appendix, we collect various definitions and conventions of the special functions we use in the main text as well as several useful formulas.

	Many of our expressions will contain large ratios and products of gamma functions, and so we adopt the common shorthand
	\begin{equation}
		\G{a, b, c, \cdots}{d, e, f, \cdots} = \frac{\Gamma(a)\Gamma(b)\Gamma(c)\cdots}{\Gamma(d)\Gamma(e)\Gamma(f)\cdots} \quad \text{and} \quad \Gamma[a, b, c, \cdots] = \Gamma(a)\Gamma(b)\Gamma(c)\cdots\,,
	\end{equation}
	though we will also use actual products and ratios when they look better. For arguments with large imaginary parts, the magnitude of the gamma function is exponentially suppressed,
	\begin{equation}
			\big|\Gamma(x + i y)\big| = \sqrt{2 \pi}\es\es |y|^{x-\frac{1}{2}} \es\es \e^{-\pi |y|/2}\,,\mathrlap{\qquad y \to \pm \infty\,,} \label{eq:gammaAsymp}
	\end{equation} 
	with both $x$ and $y$ real.
	The digamma function $\psi(z)$ is defined as $\psi(z) = \frac{\ud}{\ud z} \log \Gamma(z)$ and usefully obeys the functional equation $\psi(z+1) = \psi(z) + 1/z$. Furthermore, $\minus \psi(1) = \gamma_\slab{e} \approx \minus 0.577216$ is the Euler-Mascheroni constant.

	The Gaussian or ordinary hypergeometric function is defined in the disc $|z|< 1$ by the series
	\begin{equation}
		{}_2 F_1(a, b; c; z) \equiv \tFo{a}{b}{c}{z} = \frac{\Gamma(c)}{\Gamma(a)\Gamma(b)} \sum_{n = 0}^{\infty} \frac{\Gamma(a+n)\Gamma(b+n)}{\Gamma(c+n)} \frac{z^n}{n!}\,. \label{eq:hypSum}
	\end{equation}
	This can also be defined through the Mellin-Barnes integral~\cite{Slater:1996}
	\begin{equation}
		{}_2 F_1(a, b, c; z) = \frac{\Gamma(c)}{\Gamma(a) \Gamma(b)} \int_{\gamma}\frac{\ud s}{2 \pi i} \frac{\Gamma(s) \Gamma(a-s) \Gamma(b-s)}{\Gamma(c-s)}(\minus z)^{\sminus s }\,, \label{eq:mellin2f1}
	\end{equation}
	in which the contour $\gamma$ runs along the imaginary $s$-axis and separates the so-called ``left poles'' generated by $\Gamma(s)$ from the ``right poles'' generated by $\Gamma(a-s)\Gamma(b-s)$.\footnote{This is typically written with $s \to \minus s$, but this form it will be slightly more convenient for our purposes.}  For certain values of $a$, $b$, and $c$, the contour $\gamma$ may need to be deformed to separate these families of poles and may not always be strictly parallel to the imaginary axis, as illustrated in Figure~\ref{fig:mellinProp}. This integral representation converges as long as $z$ is not on the positive real axis, $z \notin [0, \infty)$.

	Any function of two points on the Euclidean sphere that is invariant under its isometries must be a function of the embedding distance $\emd \in [\minus 1, 1]$, and so admits a decomposition into any complete set of orthogonal polynomials on the interval. An especially useful set for our purposes are the Gegenbauer polynomials. These can be defined in terms of the Gegenbauer-$C$ functions~\cite{Durand:1976efa}, which are solutions to the differential equations
	\begin{equation}
		\left[(\emd^2 - 1) \frac{\ud^2}{\ud \emd^2} + (2 J + 1) \, \emd \, \frac{\ud}{\ud \emd} - J(J+2 \alpha)\right] C_{J}^{\alpha}(\emd) = 0\,,\label{eq:gegDeq}
	\end{equation}
	and can be expressed in terms of the hypergeometric function as
	\begin{equation}
		C_{J}^{\alpha}(\emd) \equiv \frac{\Gamma(J+2\alpha)}{\Gamma(J+1)\Gamma(2\alpha)}\, \tFo{\minus J}{J+2 \alpha}{\alpha + \tfrac{1}{2}}{ \frac{1 - \emd}{2}}\,.\label{eq:gegC}
	\end{equation}
	The Gegenbauer polynomials are then defined as the Gegenbauer-$C$ functions restricted to non-negative integer $J$. Furthermore, for large argument, these behave as
	\begin{equation}
		C_{J}^{\alpha}(\emd) \sim \G{J+\alpha}{\alpha\,,\,J+1} (2\es  \emd)^{J} + \G{J+2 \alpha\,,\, -(J+\alpha)}{\alpha\,,\,\minus J\,,\, J+1} (2 \emd)^{-(J+2\alpha)}\,, \label{eq:gegAsymp}
	\end{equation}
	as $|\emd| \to \infty$.	
	In particular, at the endpoints of the interval
	\begin{equation}
		C_{J}^{\alpha}(1) = \frac{\Gamma(J+2 \alpha)}{\Gamma(J+1) \Gamma(2 \alpha)} \qquad \text{and} \qquad C_{J}^{\alpha}(\minus 1) = \frac{\cos \pi(J+\alpha)}{\cos \pi \alpha} \frac{\Gamma(J+2 \alpha)}{\Gamma(J+1) \Gamma(2 \alpha)}\,, \label{eq:gegEndpoints}
	\end{equation}
	and so both grow as $J^{2\alpha - 1}$ for large integer $J$. However, $C_{J}^{\alpha}(\minus 1)$ grows exponentially as $\lab{Im}\, J \to \pm \infty$. The Gegenbauer $C$-functions are thus very poorly behaved for non-integer $J$. 

	Any smooth function $H(\emd)$ can then be decomposed as
	\begin{equation}
		H(\emd) = \sum_{J = 0}^{\infty} (J+\alpha) [H]_J^{\phantom{\alpha}} \, C_{J}^{\alpha}(\emd)\,,
	\end{equation} 
	where the coefficients $[H]_J$ can be extracted via
	\begin{equation}
		[H]_J =  \frac{(4 \pi)^\alpha \Gamma(\alpha) \Gamma(J+1)}{\Gamma(J + 2 \alpha)} \int_{\sminus 1}^{1}\!\ud \emd \, \big(1 - \emd^2\big)^{\alpha - \frac{1}{2}} C_{J}^{\alpha}(\emd) H(\emd)\,
	\end{equation}
	for integer $J$.
	As discussed in the main text, they may also be extracted via a contour integral over the Gegenbauer $Q$-function
	\begin{equation}
		[H]_J = \frac{(4 \pi)^\alpha \Gamma(\alpha) \Gamma(J+1)}{\Gamma(J+2\alpha)} \oint_{\mathcal{C}} \frac{\ud \emd}{2 \pi i} \big(\emd^2 - 1\big)^{\alpha - \frac{1}{2}} Q_J^\alpha (\emd) H(\emd)\,, \label{eq:gegDiscInv}
	\end{equation}
	where $\mathcal{C}$ wraps the interval $\emd \in [-1, 1]$ in a counter-clockwise fashion. The Gegenbauer-$Q$ are the other solution to (\ref{eq:gegDeq}) and may be defined as 
	\begin{equation}
		Q_J^{\alpha}(\emd) \equiv \frac{2^{1 - J - 2 \alpha} \pi \Gamma(J+ 2 \alpha)}{\Gamma(\alpha) \Gamma(J+\alpha + 1)} (\emd - 1)^{-J - 2 \alpha} \tFo{J+\alpha + \frac{1}{2}}{J+2 \alpha}{2 J + 2 \alpha + 1}{\frac{2}{1 - \emd}}\,. \label{eq:gegQ}
	\end{equation}
	Importantly, $Q_{J}^{\alpha}(\emd)$ is the unique solution to (\ref{eq:gegDeq}) which \emph{decays} as $\xi^{-J-2\alpha}$ as $|\xi| \to \infty$. This implies that, as long as the integral (\ref{eq:gegDiscInv}) converges, it defines the appropriate analytic continuation of $[H]_J$ to non-integer $J$ that is well-behaved in the right-half $J$-plane.

	The functions we work with will have a discontinuity along $\emd \in [1, \infty)$, and so we may write
	\begin{equation}
		[H]_J = - \frac{2 \pi^{\alpha+1}\Gamma(J+1)}{4^J\es \Gamma(J+\alpha+1)} \int_{0}^{\infty}\!\frac{\ud \imd}{2 \pi i} \, \imd^{J-1}\,  \tFo{J+\alpha + \frac{1}{2}}{J+1}{2 J + 2 \alpha + 1}{\minus \imd} \, \lab{disc}\, H(\imd)\,, 
	\end{equation}
	where $\imd \equiv 2/(\emd -1)$. The discontinuity
	\begin{equation}
		\lab{disc}\,  (\minus 1/\imd)^{\sminus \Delta} = \minus 2 i \imd^{\Delta} \sin \pi \Delta = -\frac{2 \pi i \es \imd^{\Delta}}{\Gamma(\Delta) \Gamma(1 - \Delta)}\,,\mathrlap{\quad \imd > 0\,,}
	\end{equation}
	will be particularly useful throughout the text.

	In our notation, the inversion $[G^\sigma G^\varphi]_J$ of the bubble diagram was found in~\cite{Marolf:2010zp} to be
	\begin{align}
		&[G^{\sigma} G^{\varphi}]_{J} = \frac{\Gamma(\Delta_\varphi)}{16 \pi^{\alpha}} \frac{ \cos \pi \Delta_\varphi}{\sin \pi(\alpha - \Delta_\varphi)} \Gamma\big[2 - 2 \alpha\,,\,J+1\,,\, \tfrac{1}{2}(J+1-\alpha+\Delta_\varphi)\big] \times \nonumber \\ 
		& \qquad \qquad\quad \Gamma\big[ J+2 - \alpha + \Delta_\varphi\,,\, \tfrac{1}{2}(J+2 \alpha + \Delta_\varphi - \Delta_\sigma)\,,\,\tfrac{1}{2}(J+\Delta_\varphi + \Delta_\sigma)\big] \times \label{eq:mmResult} \\
		& {}_7 \!\es\es\es \xoverline{V}_{\!6}\Big[J+1 - \alpha + \Delta_\varphi\,;\, 1 - \alpha\,,\,1 - 2 \alpha + \Delta_\varphi\,,\, J+1\,,\, \tfrac{1}{2}(J+2 \alpha + \Delta_\varphi - \Delta_\sigma)\,,\,\tfrac{1}{2}(J+\Delta_\varphi + \Delta_\sigma)\Big] \nonumber \\
		&\qquad\qquad\quad  + (\Delta_\varphi, \Delta_\sigma) \to (\bar{\Delta}_{\varphi}, \Delta_\sigma) + (\Delta_\varphi, \Delta_\sigma) \to (\Delta_{\sigma}, \Delta_\varphi) + (\Delta_\varphi, \Delta_\sigma) \to (\bar{\Delta}_\sigma, \Delta_\varphi)\nonumber
	\end{align}
	where
	\begin{equation}
			{}_7 \!\es\es\es \xoverline{V}_{\!6}[a; b, c, d, e, f] = \frac{{}_7 V_{6}[a; b, c, d, e, f]}{\Gamma\big[\frac{1}{2}a, 1+a - b, 1+a-c, 1+a-d, 1+a-e, 1+a-f\big]}
	\end{equation}
	is a regularized ${}_7 V_6$, the so-called very well-poised ${}_7 F_6$ hypergeometric function~\cite{Slater:1996},
	\begin{equation}
		{}_7 V_{6}[a; b, c, d, e, f] = {}_7 F_6\!\left[\genfrac..{-1pt}{0}{\raisebox{-1pt}{$a\,,\, 1+\frac{1}{2}a\,,\, b \,,\, c \,,\, d\,,\, e\,,\, f$}}{\raisebox{1pt}{$\frac{1}{2}a\,,\, 1+a - b\,,\,1+a-c\,,\,1+a-d\,,\,1+a-e\,,\,1+a-f $}} \, \bigg| \, 1\, \right]\,,
	\end{equation}
	with
	\begin{equation}
		\pFq{p}{q}{a_1\,,\, a_2\,, \cdots\,,\, a_p}{b_1\,,\, b_2\,,\, \cdots \,,\, b_q}{z} = \G{b_1 \,,\, b_2\,,\, \cdots \,,\, b_q}{a_1\,,\, a_2 \,,\, \cdots\,,\, a_p} \sum_{n = 0}^{\infty}\, \G{a_1 + n\,,\, a_2 + n\,,\, \cdots\,,\, a_p + n}{b_1 + n \,,\, b_2 + n\,,\, \cdots \,,\, b_q + n} \frac{z^n}{n!} 
	\end{equation}
	for $|z| < 1$ is the generalized hypergeometric function. Such regularized hypergeometric functions are entire in all of their parameters, while a hypergeometric function is said to be well-poised if $p = q + 1$ and $1+a_1 = b_1 + a_2 = b_2 + a_3 = \cdots = b_q + a_{q+1}$. Such functions are very well-poised if they are well-poised and $a_2 = 1 + \frac{1}{2} a_1$.

	\section{Uniqueness of the Interpolation} \label{app:unique}

		In the main text, we defined Lorentzian de Sitter two-point functions on $\emd \in \mathbb{R}$ via the analytic continuation of their Euclidean counterparts defined on $\emd \in [\minus 1, 1]$. These Euclidean correlation functions were expressed as sums over integer momenta $J$ of weighted Gegenbauer polynomials, which only converge\footnote{Strictly speaking, $\alpha$ also needs to be small enough for these sums to converge, as the propagator is too singular when $\alpha \geq \frac{3}{2}$ to be faithfully represented by a Gegenbauer polynomial interpolation.} when $|\emd| < 1$. To analytically continue these expressions to $|\emd| \geq 1$, we rely on the Watson-Sommerfeld transformation to convert the discrete sum into a contour integral over the product of a meromorphic ``interpolation'' of the summand and a ``kernel'' with unit residue poles at non-negative integers. As we have discussed, there are infinitely many interpolations that match the summand at the integers, but Carlson's theorem guarantees that there can be only one with sub-exponential growth as $|J| \to \infty$. In this appendix, we give a pedagogical explanation for why this well-behaved interpolation, given by the Lorentzian inversion formula~(\ref{eq:lInvForm}), is the correct one to use for analytically continuing two-point functions to $\emd \in \mathbb{R}$.

		We can illustrate this simply by applying the Watson-Sommerfeld transformation to (\ref{eq:hypSum}) to derive the Mellin representation (\ref{eq:mellin2f1}) for ${}_2 F_1(a, b\es ; \es c\es ; \es z)$. The most obvious interpolation is
		\begin{equation}
			\frac{\Gamma(a+n) \Gamma(b+n)}{\Gamma(c+n)} \frac{z^n}{n!} \to \frac{\Gamma(a+s) \Gamma(b+s)}{\Gamma(c+s) \Gamma(s+1)} (\minus z)^s \e^{\sminus i \pi s}\,,\label{eq:hypInterp}
		\end{equation} 
		and so, with the kernel $k(s) = \minus \e^{i \pi s} \Gamma(\minus s) \Gamma(s+1)$, we can rewrite the sum as
		\begin{equation}
			{}_2 F_1(a, b \es ;\es c\es ;\es z) = -\frac{\Gamma(c)}{\Gamma(a) \Gamma(b)} \int_{\mathcal{C}}\frac{\ud s}{2 \pi i} \frac{\Gamma(a + s) \Gamma(b+s) \Gamma(\minus s)}{\Gamma(c+s)} (\minus z)^s \label{eq:hypMelIntermediate}
		\end{equation}
		where the contour $\mathcal{C}$ sandwiches the positive real axis, coming from $s = \infty + i \epsilon$ to $s = i \epsilon$, enclosing the pole at $s = 0$, and then going off to $s = \infty - i \epsilon$. 

		Unfortunately, the integral in (\ref{eq:hypMelIntermediate}) still is only absolutely convergent for $|z| < 1$, and so nothing has yet been accomplished. To analytically continue to $|z| \geq 1$, we must deform $\mathcal{C}$ into a contour along which the integral decays more rapidly as $|s| \to \infty$. The main factor to pay attention to is
		\begin{equation}
			|(\minus z)^s| = \rho^k \e^{\sminus \vartheta t}\,,
		\end{equation}  
		where we have written $\minus z = \rho \es\es \e^{i \vartheta}$ and $s = k + i t$. This tells us that the integrand is additionally exponentially suppressed if we approach the point $s = \infty$ in the upper (lower) half-plane when $\vartheta > 0$ ($\vartheta < 0$). Ideally, we could take advantage of this exponential suppression and arrive at an absolutely convergent integral representation for ${}_2 F_1(a, b; c;z)$ by deforming $\mathcal{C}$ into a contour $\gamma$ that lies parallel to the imaginary axis. However, this would also require that the interpolation decays quickly enough in the opposite direction---the lower or upper half-plane when $\vartheta > 0$ or $\vartheta < 0$, respectively---that we can ignore the resulting arcs at infinity. 

		Using (\ref{eq:gammaAsymp}), the integrand behaves as
		\begin{equation}
			\left|\frac{\Gamma(a+s) \Gamma(b+s)\Gamma(\minus s)}{\Gamma(c+s)} (\minus z)^s \right| \sim 2 \pi \rho^k \, t^{\lab{Re}(a + b -c) - 1} \e^{-t \vartheta - \pi |t|}\,,\mathrlap{\quad |t| \to \infty\,,}
		\end{equation}
		we see that the integrand is always exponentially suppressed as $t \to \pm \infty$ as long as $\vartheta \neq \minus \pi$. This is where Carlson's theorem comes in: any entire function that vanishes on the positive integers and grows slower than $\e^{\pi |s|}$ as $|s| \to \infty$ must be identically zero. So, if we instead modify the interpolation by adding to it an arbitrary regular function $f(s)$ multiplied by $\sin \pi s$, designed to vanish at all positive integers, this ruins the interpolation's asymptotic behavior as $|s| \to \infty$ and it instead \emph{grows} exponentially at least as fast as $\e^{\pi |s|}$ as $|s| \to \infty$. This interpolation is thus useless for defining an analytic continuation for $|z| \geq 1$ whereas (\ref{eq:hypInterp}), unique in its asymptotic behavior, is the one that must be used to go beyond the original domain of convergence~$|z| < 1$. An analogous discussion applies to the Gegenbauer polynomial expansions like~(\ref{eq:propInt}), since asymptotically (\ref{eq:gegAsymp}) they have the same polynomial dependence on their argument.

		\newpage

	\section{UV Divergences of de Sitter Self-Energies} \label{app:uvDiv}

		Both the bubble and sunset diagrams are UV divergent and, in this appendix, we regularize them via dimensional regularization. Specifically, we first evaluate these diagrams at an $\alpha = (3 - \epsilon)/2$ for which they converge. For the bubble and sunset diagrams, this naively requires that $\lab{Re}\, \alpha < 1$ and $\lab{Re}\, \alpha < \frac{5}{6}$, respectively. We then analytically continue these results to $\alpha = \frac{3}{2}$ and $\epsilon \to 0$, keeping the masses $m_\varphi$ and $m_\sigma$ fixed. That they diverge when $\alpha = \frac{3}{2}$ is represented by a set of poles in $\epsilon$, which we then subtract with our mass and kinetic counterterms. 

		Unfortunately, while we can efficiently extract the self-energies' singularities in $J$  from the representations derived in the main text (\ref{eq:cyclopsIR}) and (\ref{eq:sunsetIR}), there is a sort of conservation of trouble and it is quite difficult to extract their $\epsilon \to 0$ behavior for $\alpha = \frac{3}{2}$, or in any even spacetime dimension with $D > 2$. To see the essence of this problem, we can consider the free field propagator in the form
		\begin{equation}
			\begin{aligned}
				G(\imd) &= \frac{\Gamma\big(\alpha - \tfrac{1}{2} \big)}{(4 \pi)^{\alpha + \frac{1}{2}}} \left(-\frac{1}{\imd}\right)^{\frac{1}{2} - \alpha} \!\!\tFo{\frac{1}{2} + \alpha - \Delta}{\frac{1}{2} - \alpha + \Delta}{\frac{3}{2}- \alpha}{\minus \frac{1}{\imd}} \\
				&+ \frac{\Gamma(\Delta)}{(4 \pi)^{\alpha + \frac{1}{2}}}\,  \G{\frac{1}{2} - \alpha\,,\,2 \alpha - \Delta}{\frac{1}{2} + \alpha - \Delta\,,\, \frac{1}{2} - \alpha + \Delta}\es\es \tFo{\Delta}{2 \alpha - \Delta}{\alpha + \frac{1}{2}}{\minus \frac{1}{\imd}}\,
			\end{aligned}\,, \label{eq:propBad}
		\end{equation}
		which is useful for constructing a series expansion in the deep UV, $\imd \to \infty$. Unfortunately, both terms in (\ref{eq:propBad}) are individually singular as $\alpha \to \frac{3}{2}$, even though their sum is well-defined. These cancellations are necessary in order for the series expansion of (\ref{eq:propBad}) to the logarithms of $\imd$ that appear in (\ref{eq:propBad}) when $\alpha = \frac{3}{2}$\,. To make matters worse, these cancellations occur between terms at different orders in the series expansion around $\imd = \infty$, which makes it exceedingly difficult to track which terms in (\ref{eq:propBad}) dominate the bubble and sunset diagrams as $\epsilon \to 0$, and to efficiently extract their divergences.

		Instead, we will rely on the Mellin representation of these self-energies to extract their UV behavior. Such representations have found wide use in the study of flat space Feynman diagrams---for instance, see~\cite{Smirnov:2004ym} for a pedagogical review---but also the study of de Sitter correlation functions in the flat slicing~\cite{Sleight:2019mgd,Sleight:2019hfp,Sleight:2021plv,Premkumar:2021mlz}, though it will be helpful to also review them here in the context of Euclidean de Sitter correlation functions. We will first illustrate these techniques using the propagator (\ref{eq:propSplit}) and the bubble diagram, where we have exact results to which we can compare. Finally, we apply them to the sunset diagram. 

		\newpage
		\subsection{The Propagator}

			Let us consider one-half of the propagator (\ref{eq:propSplit})
			\begin{equation}
				\begin{aligned}
					&\quad [\mathcal{G}_{\Delta}]_J = \mathcal{N}_{J, \Delta} \int_{0}^{\infty}\!\ud \imd\, \imd^{J + \Delta - 1}\,  \tFo{J+\alpha + \frac{1}{2}}{J+1}{2J + 2 \alpha +1 }{\minus \imd} \tFo{\Delta}{\Delta - \alpha + \frac{1}{2}}{2 \Delta - 2 \alpha + 1}{\minus \imd} 
				\end{aligned} \label{eq:propUVSplit}
			\end{equation}
			with the coefficient $\mathcal{N}_{J, \Delta}$ given by (\ref{eq:propLInvCoeff}) and  the full propagator $[G]_J = [\mathcal{G}_{\Delta}]_J + [\mathcal{G}_{\bar{\Delta}}]_J$ given by this contribution plus its shadow $\bar{\Delta} = 2 \alpha - \Delta$.  This integral can be evaluated exactly and is singular when $\alpha = (3-\epsilon)/2  \to \frac{3}{2}$, with pole
			\begin{equation}
				[\mathcal{G}_{\Delta}]_J \sim -\frac{\tan \pi \Delta}{\pi (J+1)(J+2)} \frac{1}{\epsilon}\,,\mathrlap{\qquad \epsilon \to 0\,.} \label{eq:propUVDiv}
			\end{equation}
			Of course, $[G]_{J}$ is regular as $\epsilon \to 0$, and (\ref{eq:propUVDiv}) cancels against an equal and opposite contribution from its shadow $[\mathcal{G}_{\bar{\Delta}}]_J$.

			\begin{figure}
				\centering
				\begin{subfigure}[t]{0.495\textwidth}
					\centering
					\includegraphics{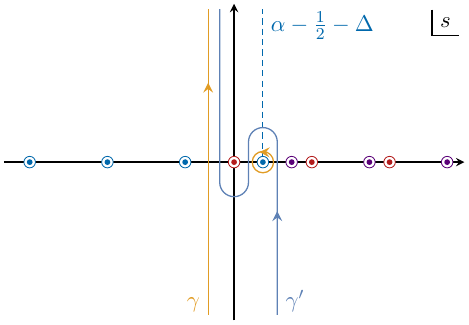}
				\end{subfigure} 
				\hfill
				\begin{subfigure}[t]{0.495\textwidth}
					\centering
					\includegraphics{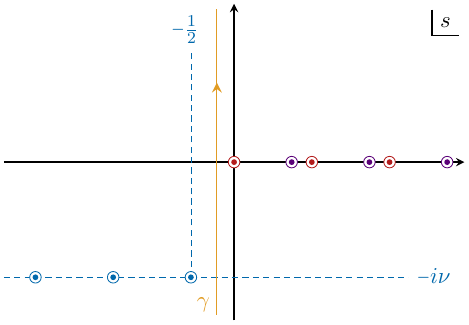}
				\end{subfigure}
				\caption{Mellin-Barnes integration contours for the $\mathcal{G}_{\Delta}(\imd)$ for a light [\emph{left}] and heavy [\emph{right}] fields. For light fields, the original integration contour $\gamma'$ must snake around the ``left'' pole at $s = \alpha - \frac{1}{2} - \Delta$ in order to completely separate the left and right poles. We will deform this contour to one that runs parallel to the imaginary $s$-axis, denoted $\gamma$, by including the contribution from this pole. This is unnecessary for both heavy fields and the Mellin-Barnes representation of $\mathcal{G}_{\bar{\Delta}}$. \label{fig:mellinProp}}
			\end{figure}

			Our basic strategy is to replace the hypergeometric function from the propagator in (\ref{eq:propUVSplit}) with its Mellin representation (\ref{eq:mellin2f1}), which we write as
			\begin{equation}
				\begin{aligned}
					&\tFo{\Delta}{\Delta - \alpha + \frac{1}{2}}{2 \Delta - 2 \alpha + 1}{\minus \imd} = \\
					& \qquad\qquad\qquad \G{2 \Delta - 2 \alpha + 1}{\Delta\,,\, \Delta - \alpha + \frac{1}{2}}\int_{\gamma} \frac{\ud s}{2 \pi i}\,  \G{\Delta - \alpha + \frac{1}{2} + s\,,\, \minus  s\,, \, \alpha - \frac{1}{2} - s}{\Delta - \alpha + \frac{1}{2} - s} \imd^{\es \alpha - \Delta - \frac{1}{2} - s}\,.
				\end{aligned} \label{eq:mellinProp}
			\end{equation}
			We can then integrate over $\imd$.
			As pictured in Figure~\ref{fig:mellinProp}, the integrand of (\ref{eq:mellinProp}) has a set of ``left'' poles at $s = \alpha - \frac{1}{2} - \Delta - k$ and two sets of ``right'' poles at $s = k$ and $s = \alpha -\frac{1}{2} + k$, with $k \in \mathbb{N}$ a non-negative integer. The contour $\gamma$ is chosen such that it separates these left and right poles. For a heavy field, this means that the contour $\gamma$ can be chosen to run parallel to the imaginary axis in the strip $\minus \frac{1}{2} < \lab{Re}\, s < 0$, from $\lab{Im}\, s = \minus \infty$ to $\lab{Im}\, s = + \infty$. The same is true for light fields with dimensions $\alpha - \frac{1}{2}< \Delta < \alpha$, though here the strip is reduced to $\alpha-\frac{1}{2} -\Delta < \lab{Re}\, s < 0$. For light fields with dimensions below $\Delta < \alpha - \frac{1}{2}$, the right-most left pole is actually to the right of the left-most right pole, and so the contour can no longer be purely parallel to the imaginary axis and separate the left and right poles. However, we can deform this contour to be parallel to the imaginary axis and lie in the strip $-\frac{3}{2} + \sqrt{\alpha^2 - m^2} < \lab{Re} \, s < 0$, at the cost of including the residue of the pole at $s = \alpha - \frac{1}{2} - \Delta$,  which is thankfully just $1$.\footnote{We will not consider ``conformally-coupled'' scalar fields with $m^2 =  \alpha^2 - \frac{1}{4}$ or equivalently dimension $\Delta = \alpha - \frac{1}{2}$. In this case, the left-most right pole and right-most left pole overlap, and the Mellin representation is not so useful. However, in this case, the propagator takes such a simple form $G(\imd) = (\minus 1/\imd)^{\frac{1}{2} - \alpha} \Gamma\big(\alpha - \frac{1}{2}\big)/(4 \pi)^{\alpha + \frac{1}{2}}$ that the inversion formula (\ref{eq:lInvForm}) can be done exactly, and so this more complicated analysis is unnecessary.} The upside is that we will always choose the contour $\gamma$ to be parallel to the $\lab{Im}\, s$ axis, with a small negative real part.

		With (\ref{eq:mellinProp}) inserted (\ref{eq:propUVSplit}), we can perform the integral over $\imd$ to find
		\begin{equation}
			[\mathcal{G}_{\Delta}]_{J} =  \int_{\gamma}\frac{\ud s}{2 \pi i}\, \Gamma\big(\tfrac{3}{2} - \alpha + s\big)\Gamma(\minus s) \mathcal{F}(s) \label{eq:propMellin}
		\end{equation}
		with
		\begin{equation}
			\mathcal{F}(s) = \frac{\sin \pi \Delta}{2 \pi \sin \pi(\alpha - \Delta)}\, \G{ 1 + s\,,\, \alpha - \frac{1}{2} - s\,,\, \Delta - \alpha + \frac{1}{2} + s\,,\, J + \alpha - \frac{1}{2} - s}{J + \alpha + \frac{3}{2} + s\,,\, \Delta - \alpha + \frac{1}{2} - s}\,.
		\end{equation}
		The integral over $\imd$ converges when $\lab{Re}\, s >  \alpha - \frac{3}{2}$ and $\lab{Re}\,(J - s + \alpha) > 1$, which is satisfied by the contour $\gamma$ as long as $\alpha < \frac{3}{2}$, or for $\epsilon > 0$.\footnote{We will assume throughout this analysis that $\lab{Re}\, J$ is always large enough to avoid any potential $J$ singularities. Relaxing this assumption would not change our conclusions but only make the analysis more tedious, and it is easier to extract the singularity structure in $J$ using the techniques presented in the main text.} 

		\begin{figure}
				\centering
				\begin{subfigure}[t]{0.495\textwidth}
					\centering
					\includegraphics{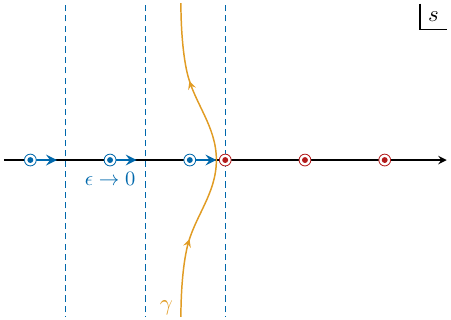}
				\end{subfigure} 
				\hfill
				\begin{subfigure}[t]{0.495\textwidth}
					\centering
					\includegraphics{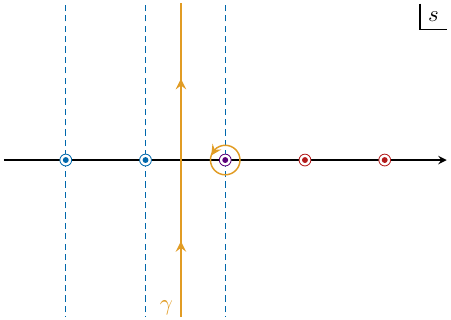}
				\end{subfigure}
				\caption{A singularity may develop in (\ref{eq:propMellin}) if the contour $\gamma$ is ``pinched'' between left and right poles as $\epsilon \to 0$ [\emph{left}]. We can compute the resulting singularity by moving the contour $\gamma$ past the pinching pole and picking up its residue [\emph{right}]. The resulting integral is regular as $\epsilon \to 0$, so the $\epsilon$ singularity is fully contained within this residue and may be easily calculated. \label{fig:mellinPinched}}
		\end{figure}

		The integral (\ref{eq:propMellin}) is well-defined and thus non-singular as long as the contour $\gamma$ separates the left poles from the right poles. The only way a singularity can develop is if the contour $\gamma$ gets ``pinched'' by the left-most right pole at $s = 0$ and the right-most left pole at $s = \alpha - \frac{3}{2} = \minus \epsilon/2$, as illustrated in Figure~\ref{fig:mellinPinched}. As $\epsilon \to 0$, we can extract this singularity by deforming the contour past it, at the price of picking up its residue,
		\begin{equation}
			[\mathcal{G}_{\Delta}]_J = \underset{{s= -\epsilon/2}}{\lab{res}}\left[ \Gamma\big(\tfrac{\epsilon}{2} - s\big)\Gamma(\minus s)  \mathcal{F}(s)\right] + \int_{\gamma}\frac{\ud s}{2 \pi i}\,  \Gamma^{/\{0\}}\big(\tfrac{\epsilon}{2} - s\big) \Gamma(\minus s) \mathcal{F}(s)\,, \label{eq:propContourCommuted}
		\end{equation}
		where we use $\Gamma^{/\{0\}}(z)$ to denote that the contour $\gamma$ is deformed to avoid the $z = 0$ singularity of $\Gamma(z)$. Similarly, we will use $\Gamma^{/\{0, \minus 1, \ldots\}}(z)$ to denote that the contour $\gamma$ avoids the $z = 0, \minus 1, \ldots$ singularities of $\Gamma(z)$. The remaining integral in (\ref{eq:propContourCommuted}) is completely regular as $\epsilon \to 0$ and so the only divergence comes from the residue in (\ref{eq:propContourCommuted}),
		\begin{equation}
			[\mathcal{G}_{\Delta}]_J \sim -\frac{\sec \pi \alpha \sin \pi \Delta}{2 \sin \pi(\alpha - \Delta)}  \, \G{J+1\,,\, \Delta - 1}{J+2 \alpha\,,\, \Delta - 2 \alpha +2 } \sim -\frac{\tan \pi \Delta}{\pi (J+1)(J+2)} \frac{1}{\epsilon}\,,
		\end{equation}
		as $\epsilon \to 0$, which recovers the UV divergence (\ref{eq:propUVDiv}) from the exact result.

	\subsection{The Bubble}

		A similar strategy works for the bubble diagram, though it will be more convenient to sum over the dimensions and their shadows at the outset. For instance, using (\ref{eq:mellinProp}) in (\ref{eq:cyclopsLinv}), we have 
		\begin{equation}
			\begin{aligned}
				[G^{\sigma} G^{\varphi}]_{J} = \int\!\es \frac{\ud s_1}{2 \pi i} \frac{\ud s_2}{2 \pi i}& \,\,  \Gamma\big(2 - 2 \alpha + s_1 + s_2\big)\Gamma\big(\tfrac{3}{2} - \alpha + s_1 + s_2\big) \\
				\times &\,\, \Gamma\big[ \minus s_1\,,\,\minus s_2\,,\,\alpha - \tfrac{1}{2} - s_1\,,\,\alpha - \tfrac{1}{2} - s_2\big] \mathcal{F}(s_1, s_2)
			\end{aligned}\,,
		\end{equation}
		where the $s_1$ and $s_2$ contours (which we denote $\gamma_1$ and $\gamma_2$, respectively, but will suppress in our expressions going forward) are chosen to run parallel to the imaginary $s_1$ and $s_2$ axes, with small negative real parts, as described in the previous section. Here, we have also defined the function 
		\begin{equation}
			\begin{aligned}
				\mathcal{F}(s_1, s_2) &= \frac{(4 \pi)^{-\alpha - \frac{3}{2}} \sin \pi(\Delta_\sigma + \Delta_\varphi)}{\sin \pi(\alpha - \Delta_\sigma) \sin \pi(\alpha - \Delta_\varphi)}\frac{\Gamma(J+2 \alpha - 1 - s_1 - s_2)}{\Gamma(J+2 + s_1 + s_2)} \\
				&\qquad\qquad \times \GB{\Delta_\varphi- \alpha + \frac{1}{2} + s_1\,,\, \Delta_\sigma - \alpha + \frac{1}{2} + s_2}{\Delta_\varphi-\alpha + \frac{1}{2} - s_1\,,\, \Delta_\sigma -\alpha + \frac{1}{2} -s_2} + \cdots
			\end{aligned}\,, \label{eq:fBubble}
		\end{equation}
		where the $\cdots$ denote the three other permutations under $\Delta_\sigma \to \bar{\Delta}_\sigma = 2 \alpha - \Delta_\sigma$ and $\Delta_{\varphi} \to \bar{\Delta}_{\varphi} = 2 \alpha - \Delta_\varphi$.
		This Mellin representation is valid (for large enough $\lab{Re}\, J$) as long as $\alpha$ is such that the contours can sit to the left of the right poles generated by $\Gamma(\minus s_1)$ and $\Gamma(\minus s_2)$, and to the right of the left poles generated by $\Gamma(2 -2 \alpha + s_1 +s_2)$ and $\Gamma\big(\tfrac{3}{2} - \alpha + s_1 + s_2\big)$. This is possible as long as $\alpha < 1$, and we will need to analytically continue this result from $\alpha < 1$ to $\alpha \to \frac{3}{2}$ to extract the bubble's UV divergences.

		\begin{figure}
				\centering
				\begin{subfigure}[t]{0.495\textwidth}
					\centering
					\includegraphics{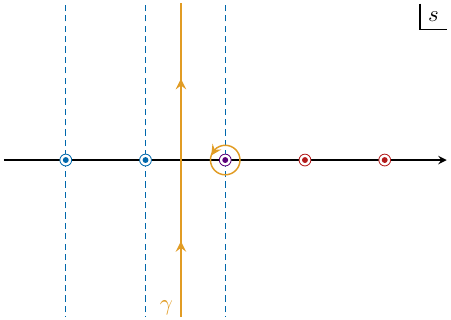}
				\end{subfigure} 
				\hfill
				\begin{subfigure}[t]{0.495\textwidth}
					\centering
					\includegraphics{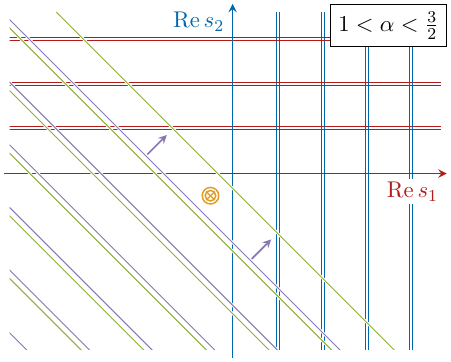}
				\end{subfigure}
			\caption{As $\alpha \to 1$ [\emph{left}], the right-most left pole [\emph{\color{cornellGreen}green\es}] traps the integration contour, here shown as a crossed circle in {[\emph{\color{Mathematica2}yellow\es}]}, between the right poles [\emph{\color{cornellRed}red}/\emph{\color{cornellBlue}blue}]. These poles form lines in the $\lab{Re}\, s_1$--$\lab{Re}\, s_2$ plane, with $\lab{Im}\, s_1 = \lab{Im}\, s_2 = 0$. To analytically continue beyond $\alpha = 1$, we again move the contour around the pinching poles, picking up the residue at that pole. As described in the text, this happens again as $\alpha \to \frac{3}{2}$ [\emph{right}] for two sets of poles in [\emph{\color{cornellGreen}green}/\emph{\color{Mathematica5}purple\es}]. \label{fig:mellinPinch2d}}
		\end{figure}

		As illustrated in Figure~\ref{fig:mellinPinch2d}, the ``right'' singularities generated by the factor $\Gamma\big[\minus s_1\,,\, \minus s_2\,,\, \alpha - \frac{1}{2} - s_1\,,\, \alpha - \frac{1}{2} - s_2\big]$ form horizontal and vertical lines in the $\lab{Re}\, s_1$ -- $\lab{Re}\, s_2$ plane (in red and blue), while the ``left'' singularities generated by $\Gamma(2 -2 \alpha + s_1 + s_2) \Gamma(\tfrac{3}{2} - \alpha + s_1 + s_2)$ form diagonal lines (green and purple) that move towards the upper right as $\alpha \to \tfrac{3}{2}$. As they move, they trap the integration contour (in yellow) between the lines of the right singularities and pinch it. As $\alpha \to \tfrac{3}{2}$ from $\alpha < 1$, the contour is first pinched by the $2 - 2 \alpha + s_1 + s_2 = 0$ singularity of $\Gamma(2 - 2 \alpha + s_1 + s_2)$. As before, we need to move the contour past this singularity. We do this first by replacing the $s_1$ integral with its residue and singularity-subtracted integral, and then similarly the $s_2$ integral. Fortunately, the residue at $s_1 = -(2 - 2 \alpha + s_2)$ vanishes identically when we include the sum over shadows in (\ref{eq:fBubble}), so we may write
		\begin{equation}
			\begin{aligned}
				[G^{\sigma} G^{\varphi}]_{J} = \int\!\es \frac{\ud s_1}{2 \pi i} \frac{\ud s_2}{2 \pi i}& \,\,  \Gamma^{/\{0\}}\big(2 - 2 \alpha + s_1 + s_2\big)\Gamma\big(\tfrac{3}{2} - \alpha + s_1 + s_2\big) \\
				\times &\,\, \Gamma\big[ \minus s_1\,,\,\minus s_2\,,\,\alpha - \tfrac{1}{2} - s_1\,,\,\alpha - \tfrac{1}{2} - s_2\big] \mathcal{F}(s_1, s_2)
			\end{aligned}\,. \label{eq:bubbleMellinStep2}
		\end{equation}
		We can now continue this expression to $\alpha \to \tfrac{3}{2}$. The integration contour is first pinched by the $\tfrac{3}{2} - \alpha + s_1 + s_2 = 0$ singularity and then the $2 - 2 \alpha + s_1 + s_2 = -1$ singularity. Treating each pinch in turn, we find that the residue again vanishes for the $2 - 2 \alpha + s_1 + s_2 = -1$ singularity and so we are left with a single contribution from the $\tfrac{3}{2} - \alpha + s_1 + s_2 = 0$ singularity,
		\begin{equation}
			\begin{aligned}
				[G^{\sigma} G^{\varphi}]_{J} &= -\pi \sec \pi \alpha \, \Gamma\big(\tfrac{1}{2}-\alpha\big)\mathcal{F}\big(0, \alpha - \tfrac{3}{2}\big) + \cdots \\
				&= -\frac{\Gamma \big(\tfrac{1}{2}-\alpha \big) \csc \pi  \Delta_\sigma \cos \pi  (\alpha -\Delta_\sigma) \csc \pi  (2 \alpha -\Delta_\sigma )}{4 (4 \pi)^{\alpha -\frac{1}{2}}\Gamma (2-\Delta_\sigma ) \Gamma (2-2 \alpha +\Delta_\sigma)} + \cdots
			\end{aligned} \label{eq:bubbleUVAnswer}
		\end{equation}
		where the $\cdots$ denote the remaining integral terms that are regular as $\alpha \to \frac{3}{2}$. Taking $\alpha = (3 - \epsilon)/2$ and $\epsilon \to 0$, we are left with a suprisingly simple result for the full bubble diagram,
		\begin{equation}
			[G^{\sigma} G^{\varphi}]_{J} \sim \frac{1}{8 \pi^2 \epsilon}\,, \label{eq:bubbleUVDiv}
		\end{equation}
		which agrees with the divergence extracted from the exact result (\ref{eq:mmResult}). 

		Before we move on, we should note that this derivation treated $\varphi$ as if it had mass $m_\varphi^2 > \alpha^2 - \frac{1}{4}$ so that we did not need to ``straighten out'' the $s_1$ contour to lie parallel to the imaginary axis. If the field is light enough, $m_\varphi^2 < \alpha^2 - \frac{1}{4}$, we need to include a $1$ in our resolution (\ref{eq:mellinProp}) of the $\Delta_\varphi$-dependent hypergeometric function that appears in $[\mathcal{G}_{\Delta_\varphi} G^{\sigma}]_{J}$. We are then left with an additional Mellin-type integral with an integrand that is proportional to $\Gamma\big(s_2 + \tfrac{3}{2} - \alpha -\Delta_\varphi\big) \Gamma(\minus s_2)$. As $\alpha \to \frac{3}{2} - \Delta_\varphi = \frac{3}{2} - \frac{1}{3} m_\varphi^2 +  \cdots$, the left-most pole of $\Gamma\big(s_1 + \tfrac{3}{2} - \alpha - \Delta_\varphi\big)$ and the right-most pole of $\Gamma(\minus s_1)$ pinch the integration contour. However, this does not contribute to the $\epsilon$-divergence of $[G^{\sigma} G^{\varphi}]_J$ for two reasons. First, while the residue of $[\mathcal{G}_{\Delta_\sigma} \mathcal{G}_{\Delta_{\varphi}}]_J$ at this pole is non-zero, it is exactly canceled by its shadow contribution from $[\mathcal{G}_{\smash{\bar{\Delta}_\sigma}} \mathcal{G}_{\Delta_\varphi}]_J$. Second, even if this residue did not vanish, this contribution is necessarily regular as $\epsilon \to 0$, even if it diverges as $\epsilon \to 2 m_\varphi^2/3$. Thus, (\ref{eq:bubbleUVDiv}) applies when $\varphi$ is either light or heavy. Similar logic holds for other diagrams with light fields. Intuitively, this is obvious---straightening out this contour represents a modification in the deep IR, which should not affect the UV structure of these diagrams. We now turn our attention to the sunset diagram.

	\subsection{The Sunset}

		Finally, let us turn our attention to the sunset diagram. It will be helpful to consider an arbitrary sunset diagram with fields $\varphi_1$, $\varphi_2$, and $\varphi_3$ with dimensions $\Delta_1$, $\Delta_2$, and $\Delta_3$ and propagators $G^{\varphi_1}$, $G^{\varphi_2}$ and $G^{\varphi_3}$, respectively, 
		\begin{equation}
			\begin{aligned}
				\big[G^{\varphi_1} G^{\varphi_2} G^{\varphi_3}\big]_J &= \int \!\es \frac{\ud s_1}{2 \pi i} \, \frac{\ud s_2}{2 \pi i} \,  \frac{\ud s_3}{2 \pi i} \, \, \Gamma\big(\tfrac{5}{2} - 3 \alpha + s_1 + s_2 + s_3\big)\Gamma\big(2 - 2 \alpha + s_1 + s_2 + s_3\big) \\
				&\times \,\, \Gamma\big[ \minus s_1\,,\,\minus s_2\,,\, \minus s_3\,,\,\alpha - \tfrac{1}{2} - s_1\,,\,\alpha - \tfrac{1}{2} - s_2\,,\, \alpha - \tfrac{1}{2} - s_3\big]\mathcal{F}(s_1, s_2, s_3)\,,
			\end{aligned} \label{eq:sunsetMellin}
		\end{equation}
		where we have defined the function 
		\begin{equation}
			\begin{aligned}
				\mathcal{F}(s_1, s_2, s_3) &= \frac{(4 \pi)^{-2\alpha - 2} \sin \pi(\Delta_1 + \Delta_2 + \Delta_3) }{2 \sin \pi(\alpha - \Delta_1) \sin \pi(\alpha - \Delta_2) \sin \pi(\alpha - \Delta_3)}\,\, \GB{J + 3 \alpha - \frac{3}{2} - s_1 - s_2 -s_3}{J - \alpha + \frac{5}{2} + s_1 + s_2 + s_3}   \\
				&\times \GB{\Delta_1 - \alpha + \frac{1}{2} + s_1}{\Delta_1 - \alpha + \frac{1}{2} - s_1}\GB{\Delta_2 - \alpha + \frac{1}{2} + s_2}{\Delta_2 - \alpha + \frac{1}{2} - s_2}\GB{\Delta_3 - \alpha + \frac{1}{2} + s_3}{\Delta_3 - \alpha + \frac{1}{2} - s_3} + \cdots
			\end{aligned}\,,
		\end{equation}
		where the $\cdots$ denote the other 7 permutations generated by taking $\Delta_i  \to \bar{\Delta}_i =  2 \alpha - \Delta_i$ for $i = 1, 2, 3$.
		The integral over $\imd$ from which this is derived naively converges only when $\alpha < \frac{5}{6}$, and so we must analytically continue (\ref{eq:sunsetMellin}) from $\alpha < \frac{5}{6}$ towards $\alpha \to \frac{3}{2}$ in order to extract its singularity structure as~$\epsilon \to 0$.

		The first singularity we encounter is at at $\frac{5}{2} - 3 \alpha + s_1 + s_2 + s_3 = 0$ as $\alpha \to \frac{5}{6}$. Fortunately, like with the bubble diagram, the residue of the integrand at $s_1 = 3 \alpha - \frac{5}{2} - s_2 -s_3$ vanishes identically once we include the shadow contributions and so $[G^{\varphi_1} G^{\varphi_2} G^{\varphi_3}]_J$ is actually regular as $\alpha \to \frac{5}{6}$. The first true singularity we encounter is from $2 - 2 \alpha + s_1 + s_2 + s_3 = 0$ as $\alpha \to 1$, in which we can write
		\begin{equation}
			\begin{aligned}
				&\big[G^{\varphi_1} G^{\varphi_2} G^{\varphi_3}\big]_J = \Gamma(2 - 2 \alpha) \Gamma\big(\alpha - \tfrac{1}{2}\big)^2 \Gamma\big(\tfrac{3}{2} - \alpha\big) \Gamma\big(\tfrac{1}{2} - \alpha\big) \mathcal{F}(0, 0, 2 \alpha - 2)\\
				&+ \int\!\es\frac{\ud s_3}{2 \pi i}\,\, \Gamma^{/\{0\}}\big(2 - 2\alpha + s_3\big)\Gamma\big( \tfrac{3}{2} -\alpha + s_3\big) \\
				&\qquad\quad\,\times \, \Gamma\big[\minus s_3\,,\, \alpha - \tfrac{1}{2} - s_3\,,\, \alpha -\tfrac{1}{2} \,,\, \tfrac{1}{2} - \alpha\big] \mathcal{F}(0, 2 \alpha -2  - s_3, s_3)  \\
				&+ \int\!\es\frac{\ud s_2}{2 \pi i} \, \frac{\ud s_3}{2 \pi i}\,\,  \Gamma^{/\{0\}}(2 - 2 \alpha + s_2 + s_3) \Gamma\big(\tfrac{3}{2} - \alpha + s_2 + s_3\big) \\
				&\qquad\qquad\,\,\,\,\, \times\,  \Gamma\big[\minus s_2\,,\, \minus s_3\,,\, \alpha - \tfrac{1}{2} - s_2\,,\, \alpha - \tfrac{1}{2}\big] \mathcal{F}\big(2\alpha - 2 - s_2 -s_3, s_2, s_3\big) \\
				&+\int\!\es \frac{\ud s_1}{2 \pi i} \, \frac{\ud s_2}{2 \pi i} \,  \frac{\ud s_3}{2 \pi i} \, \, \Gamma^{/\{0\}}\big(\tfrac{5}{2} - 3 \alpha + s_1 + s_2 + s_3\big)\Gamma^{/\{0\}}\big(2 - 2 \alpha + s_1 + s_2 + s_3\big) \\
				&\qquad\qquad\qquad \,\,\,\, \times \, \Gamma\big[ \minus s_1\,,\,\minus s_2\,,\, \minus s_3\,,\,\alpha - \tfrac{1}{2} - s_1\,,\,\alpha - \tfrac{1}{2} - s_2\,,\, \alpha - \tfrac{1}{2} - s_3\big]\mathcal{F}(s_1, s_2, s_3)\,
			\end{aligned}\,. \label{eq:sunsetMellinPart2}
		\end{equation}
		The next relevant singularity is from $\frac{5}{2} - 3 \alpha + s_1 + s_2 + s_3 = \minus 1$ as $\alpha \to \frac{7}{6}$, but fortunately there the residue at $s_1 = 3\alpha - \frac{7}{2} - s_2 - s_3$ also vanishes identically. Finally, there are four singularities that become relevant as $\alpha \to \frac{3}{2}$, those from $\frac{5}{2} - 3 \alpha + s_1 + s_2 + s_3 = \minus 2$, $2 - 2 \alpha + s_1 + s_2 + s_3 = \minus 1$, and $\tfrac{3}{2} - \alpha + s_2 + s_3= 0$ or $\tfrac{3}{2} - \alpha + s_3 = 0$. Luckily, the residue of the fourth line in (\ref{eq:sunsetMellinPart2}) vanishes identically at $s_1 = 3 \alpha - \frac{9}{2} - s_2 - s_3$ after including the shadows, and so we end up with six terms
		\begin{align}
				[G^{\varphi_1} &G^{\varphi_2} G^{\varphi_3}]_J = \Gamma\big(\tfrac{3}{2}-\alpha\big) \Gamma\big(\alpha -\tfrac{1}{2}\big) \Gamma \big(\tfrac{1}{2}-\alpha\big)^2\left[ \mathcal{F}\big(0,\alpha -\tfrac{1}{2},\alpha -\tfrac{3}{2}\big) + \mathcal{F}\big(\alpha -\tfrac{1}{2},0,\alpha -\tfrac{3}{2}\big) \right] \nonumber \\
				&-\Gamma (3-2 \alpha ) \Gamma \big(\tfrac{5}{2}-\alpha \big) \Gamma \big(\alpha -\tfrac{3}{2}\big) \Gamma \big(\alpha -\tfrac{1}{2}\big) \Gamma \big(\tfrac{1}{2}-\alpha \big) \left[\mathcal{F}\big(0,1,2 \alpha -3\big)+ \mathcal{F}\big(1,0,2 \alpha -3\big)\right] \nonumber \\
				&+\Gamma \big(\tfrac{3}{2}-\alpha \big) \Gamma \big(\alpha -\tfrac{1}{2}\big)^2 \Gamma \big(2-2 \alpha \big) \Gamma \big(\tfrac{1}{2}-\alpha \big) \mathcal{F}\big(0,0,2 \alpha -2\big)\\
				&-\Gamma (3-2 \alpha ) \Gamma \big(\minus \alpha -\tfrac{1}{2}\big) \Gamma \big(\tfrac{5}{2}-\alpha \big) \Gamma \big(\alpha -\tfrac{1}{2}\big)^2 \mathcal{F}(0,0,2 \alpha -3) + \cdots \nonumber
		\end{align}
		where the $\cdots$ again denote the remainder integral terms that are regular as $\alpha \to \frac{3}{2}$. In the limit $\epsilon \to 0$, we have
		\begin{equation}
			\begin{aligned}
				[G^{\varphi_1} G^{\varphi_2} G^{\varphi_3}]_J \sim -\frac{J(J+3)}{2 (4 \pi)^4 \epsilon} - \frac{1}{64 \pi^4 \epsilon} + \mathcal{U}(\Delta_1) + \mathcal{U}(\Delta_2) + \mathcal{U}(\Delta_3) \label{eq:sunsetUVDiv}
			\end{aligned}
		\end{equation}
		where we have defined
		\begin{equation}
			\mathcal{U}(\Delta) = -\frac{2(\Delta - 1)(\bar{\Delta} - 1)}{(4 \pi)^4 \epsilon} \left[\frac{1}{\epsilon} - \psi\big(\Delta-1\big) - \psi\big(\bar{\Delta}-1\big) + \log 4 \pi \e^{\sminus \gamma_\slab{e}} + \frac{3}{2} \right]\,. \label{eq:uDef}
		\end{equation}
		Clearly the first divergence in (\ref{eq:sunsetUVDiv}) will be absorbed by the kinetic counterterm, while the remaining divergences will be absorbed by the mass counterterm.

		As a check, it will be helpful to compare (\ref{eq:sunsetUVDiv}) to its flat space limit. For simplicity, we consider the sunset diagram of the $\frac{1}{4!}g_\varphi \varphi^4$ theory, which behaves as~\cite{Kleinert:2001ax}
		\begin{equation}
			\begin{tikzpicture}[baseline=-3pt, thick]
					\def\lSize{0.9}
					\def\lSizeA{1}
					\def\circSize{0.3}
					\def\circSizeCT{0.08}
					\draw[cornellBlue] (-\lSize, 0) -- (\lSize, 0);
					\draw[cornellBlue] (0, 0)+(0.5, 0) arc (0:180:0.5);
					\draw[cornellBlue] (0, 0)+(0.5, 0) arc (0:-180:0.5);
					\fill[cornellBlue, draw=white] (-0.5,0) circle (0.06);
					\fill[cornellBlue, draw=white] (0.5,0) circle (0.06);
				\end{tikzpicture} = \frac{1}{3!}(\minus g_\varphi)^2 \left[-\frac{k^2}{2(4 \pi)^4 \epsilon} - \frac{6 m_\varphi^2}{(4 \pi)^2 \epsilon} \left[\frac{1}{\epsilon} + \log 4 \pi \e^{\sminus \gamma_\slab{e}} + \log \mu^2/m_\varphi^2 + \frac{3}{2}\right]\right] \label{eq:sunsetFlatComp}
		\end{equation}
		with $D = (4 - \epsilon)$, with $k^2$ the square of the external four-momentum. The same diagram in de Sitter yields
		\begin{equation}
			\begin{tikzpicture}[baseline=-3pt, thick]
					\def\lSize{0.9}
					\def\lSizeA{1}
					\def\circSize{0.3}
					\def\circSizeCT{0.08}
					\draw[cornellBlue] (-\lSize, 0) -- (\lSize, 0);
					\draw[cornellBlue] (0, 0)+(0.5, 0) arc (0:180:0.5);
					\draw[cornellBlue] (0, 0)+(0.5, 0) arc (0:-180:0.5);
					\fill[cornellBlue, draw=white] (-0.5,0) circle (0.06);
					\fill[cornellBlue, draw=white] (0.5,0) circle (0.06);
				\end{tikzpicture} = \frac{1}{3!}(\minus g_\varphi)^2 \left[-\frac{J(J+3)}{2(4\pi)^4 \epsilon} - \frac{1}{64 \pi^4 \epsilon} + 3 \,\mathcal{U}(\Delta_\varphi)\right]\,. \label{eq:sunsetdeSitterComp}
		\end{equation}
		Throughout this paper, we have worked in Hubble units $H = 1$. We can restore dimensions by multiplying the sunset by a factor of $H^2$. The flat space limit then corresponds to taking $H \to 0$. All fields are ``heavy'' in this limit, and so $\Delta_\varphi = \alpha + i \sqrt{(m_\varphi/H)^2 - \alpha^2}$ and $\Delta_\varphi \bar{\Delta}_\varphi = (m_\varphi/H)^2$. With $H^2 J(J+3) \to k^2$, since both are eigenvalues of the Laplacian, we find that
		\begin{equation}
			-\frac{J(J+3)}{2(4\pi)^4 \epsilon} - \frac{1}{64 \pi^4 \epsilon} + 3 \,\mathcal{U}(\Delta_\varphi) \to  -\frac{k^2}{2(4\pi)^4 \epsilon} - \frac{6 m_\varphi^2}{(4 \pi)^2 \epsilon}\left[\frac{1}{\epsilon} +\log 4 \pi \e^{\sminus \gamma_\slab{e}}+ \log H^2/m_\varphi^2 + \frac{3}{2}\right]
		\end{equation}
		and so (\ref{eq:sunsetdeSitterComp}) exactly reproduces the flat space divergence (\ref{eq:sunsetFlatComp}) when the renormalization scale is taken to be Hubble $\mu = H$. This, however, is an automatic consequence of working in Hubble units, and so (\ref{eq:sunsetdeSitterComp}) exactly matches (\ref{eq:sunsetFlatComp}) in the flat-space limit $H \to 0$.

	\section{UV Contributions in the Light Limit} \label{app:UVLight}

		In the main text, we argued that the UV contributions to the bubble and sunset diagrams were subleading as $m_\varphi$ for all $J$, and we justify that statement in this appendix. As an example, we will study the UV contribution to the bubble diagram, given by
		\begin{equation}
			[G^{\sigma} G^{\varphi}]_J^\slab{uv} = [\mathcal{G}_{\Delta_\sigma} \mathcal{G}_{\Delta_\varphi}]_J^{\slab{uv}} + [\mathcal{G}_{\smash{\bar{\Delta}_\sigma}} \mathcal{G}_{\Delta_\varphi}]_J^{\slab{uv}} + [\mathcal{G}_{\Delta_\sigma} \mathcal{G}_{\smash{\bar{\Delta}_\varphi}}]_J^{\slab{uv}} + [\mathcal{G}_{\smash{\bar{\Delta}_\sigma}} \mathcal{G}_{\smash{\bar{\Delta}_\varphi}}]_J^{\slab{uv}} \label{eq:bubbleUVFull}
		\end{equation}
		in the limit $\Delta_\varphi \to 0$, where
		\begin{equation}
			\begin{aligned}
				[\mathcal{G}_{\Delta_\sigma} \mathcal{G}_{\Delta_\varphi}]^{\slab{uv}}_{J} = \mathcal{N}_{J, \Delta_{\sigma} \Delta_{\varphi}} &\int_{1}^\infty \!\ud \imd \, \imd^{J + \Delta_\varphi + \Delta_\sigma - 1} \, \tFo{J+\alpha + \frac{1}{2}}{J+1}{2 J + 2 \alpha + 1}{\minus \imd} \\  & \qquad \times \tFo{\Delta_{\sigma}}{\Delta_\sigma - \alpha + \frac{1}{2}}{2 \Delta_\sigma - 2 \alpha + 1}{\minus \imd} \tFo{\Delta_{\varphi}}{\Delta_\varphi - \alpha + \frac{1}{2}}{2 \Delta_\varphi - 2 \alpha + 1}{\minus \imd} \label{eq:bubbleUVContrib}
			\end{aligned}
		\end{equation}
		and the other terms in (\ref{eq:bubbleUVFull}) are just (\ref{eq:bubbleUVContrib}) with the dimensions appropriately replaced with their shadows. A similar analysis holds for the sunset diagrams, though since it is virtually identical to the bubble we will not repeat it.

		The main trouble with naively expanding (\ref{eq:bubbleUVContrib}) in $\Delta_\varphi$ is that, as $\imd \to \infty$, all three hypergeometric functions conspire so that the integrand behaves as $\imd^{2 \alpha -3}$ as $\imd \to \infty$. Of course, this integral diverges as $\alpha \to \frac{3}{2}$ but, as discussed in the previous section, we will assume that we have already regularized these integrals using dimensional regularization. But, it is important to expand this integral in a way that does not disturb its structure as $\imd \to \infty$. This can be accomplished by writing the integral as
		\begin{equation}
			\begin{aligned}
				&\mathcal{N}_{J, \Delta_{\sigma}\Delta_{\varphi}}^{\sminus 1} [\mathcal{G}_{\Delta_\sigma} \mathcal{G}_{\Delta_\varphi}]^{\slab{uv}}_{J} = \\
				&\qquad \,\G{\frac{1}{2} - \alpha\,,\, 2 \Delta_\varphi - 2 \alpha + 1}{\Delta_\varphi - \alpha + \frac{1}{2}\,,\, \Delta_\varphi - 2 \alpha + 1} \int_{1}^\infty \!\ud \imd \, \imd^{J + \Delta_\sigma - 1} \, \tFo{J+\alpha + \frac{1}{2}}{J+1}{2 J + 2 \alpha + 1}{\minus \imd} \\  & \,\,\,\qquad\qquad\qquad\qquad \times \tFo{\Delta_{\sigma}}{\Delta_\sigma - \alpha + \frac{1}{2}}{2 \Delta_\sigma - 2 \alpha + 1}{\minus \imd} \tFo{\Delta_{\varphi}}{2\alpha - \Delta_\varphi}{\alpha + \frac{1}{2}}{\minus \frac{1}{\imd}} \, + \\
				& \qquad \G{\alpha - \frac{1}{2}\,,\, 2 \Delta_\varphi - 2 \alpha + 1}{\Delta_\varphi\,,\, \Delta_\varphi- \alpha + \frac{1}{2}} \int_{1}^\infty \!\ud \imd \, \imd^{J + \Delta_\sigma + \alpha - \frac{3}{2}} \, \tFo{J+\alpha + \frac{1}{2}}{J+1}{2 J + 2 \alpha + 1}{\minus \imd} \\  & \,\,\,\qquad\qquad\qquad\qquad\times \tFo{\Delta_{\sigma}}{\Delta_\sigma - \alpha + \frac{1}{2}}{2 \Delta_\sigma - 2 \alpha + 1}{\minus \imd} \tFo{\frac{1}{2} - \Delta_\varphi + \alpha }{\frac{1}{2} + \Delta_\varphi- \alpha}{\frac{3}{2} - \alpha}{\minus \frac{1}{\imd}}\,.
			\end{aligned} \label{eq:complicatedUV}
		\end{equation}
		Both integrals are now well-behaved if we expand in $\Delta_\varphi$ and the second line is clearly subleading to the first as $\Delta_\varphi \to 0$, since in this limit $1/\Gamma(\Delta_\varphi) \sim \Delta_\varphi$. 

		To leading order in $\Delta_\varphi$, we can thus approximate this UV contribution as
		\begin{equation}
			\begin{aligned}
				[\mathcal{G}_{\Delta_\sigma} \mathcal{G}_{\Delta_\varphi}]^{\slab{uv}}_{J} \approx \mathcal{N}_{J, \Delta_{\sigma} \Delta_{\varphi}} \int_{1}^\infty \!\ud \imd &\, \imd^{J + \Delta_\sigma - 1} \, \tFo{J+\alpha + \frac{1}{2}}{J+1}{2 J + 2 \alpha + 1}{\minus \imd} \tFo{\Delta_{\varphi}}{\Delta_\varphi - \alpha + \frac{1}{2}}{2 \Delta_\varphi - 2 \alpha + 1}{\minus \imd} \,,
			\end{aligned}
		\end{equation}
		which is indeed the original expression (\ref{eq:bubbleUVContrib}) with $\Delta_\varphi = 0$. The benefit of (\ref{eq:complicatedUV}) is that it makes this approximation well-controlled, as the subleading terms are clearly of order $\mathcal{O}(1)$ and we can use it to construct a controlled series expansion in $\Delta_\varphi$.

		Combining this with the IR contribution, $[\mathcal{G}_{\Delta_\sigma} \mathcal{G}_{\Delta_\varphi}]_J = [\mathcal{G}_{\Delta_\sigma} \mathcal{G}_{\Delta_\varphi}]^{\slab{ir}}_J + [\mathcal{G}_{\Delta_\sigma} \mathcal{G}_{\Delta_\varphi}]^{\slab{uv}}_J$ yields
		\begin{equation}
			\begin{aligned}
				&[\mathcal{G}_{\Delta_\sigma} \mathcal{G}_{\Delta_\varphi}]_{J} \approx \frac{\Gamma(\alpha)\, \mathcal{N}_{J, \Delta_\sigma}}{4 \pi^{\alpha+1} \Delta_\varphi}  \int_{0}^{1}\!\ud\imd\, \imd^{J + \Delta_\sigma - 1} \, \!\big(\imd^{\Delta_\varphi} - 1\big)\, \tFo{J+\alpha + \frac{1}{2}}{J+1}{2 J + 2 \alpha + 1}{\minus \imd} \\
				& \qquad\qquad\qquad\qquad\qquad\qquad\qquad \times \tFo{\Delta_{\sigma}}{\Delta_\sigma - \alpha + \frac{1}{2}}{2 \Delta_\sigma - 2 \alpha + 1}{\minus \imd} \\
				& + \frac{\Gamma(\alpha)\mathcal{N}_{J, \Delta_\sigma}}{4 \pi^{\alpha+1} \Delta_\varphi}  \int_{0}^{\infty}\!\ud \imd \, \imd^{J+\Delta_\sigma -1} \tFo{J+\alpha + \frac{1}{2}}{J+1}{2 J + 2 \alpha + 1}{\minus \imd} \tFo{\Delta_{\sigma}}{\Delta_\sigma - \alpha + \frac{1}{2}}{2 \Delta_\sigma - 2 \alpha + 1}{\minus \imd}\,.
			\end{aligned}
		\end{equation}
		To this we add $[\mathcal{G}_{\smash{\bar{\Delta}_{\sigma}}} \mathcal{G}_{\Delta_\varphi}]_J$, which is of the same order as $\Delta_\varphi \to 0$, and find with (\ref{eq:bubbleUVDiv}) that
		\begin{equation}
			\begin{aligned}
				[G^{\sigma} G^{\varphi}]_{J} &\approx \frac{\Gamma(\alpha)}{4 \pi^{\alpha+1}}\frac{1}{\Delta_\varphi} \sum_{k = 0}^{\infty} \left[\frac{ \mathcal{N}_{J, \Delta_\sigma} c_k(J, \Delta_\sigma)}{J + \Delta_\sigma + \Delta_\varphi + k}+ \frac{ \mathcal{N}_{J, \smash{\bar{\Delta}}_{\sigma}} c_k(J, \bar{\Delta}_\sigma)}{J + \bar{\Delta}_\sigma + \Delta_\varphi + k}\right]   + \frac{1}{8 \pi^2 \epsilon} \\
				& + \frac{\Gamma(\alpha)}{4 \pi^{\alpha+1}} \frac{1}{\Delta_\varphi} \left[\frac{1}{(J+\Delta_\sigma)(J + \bar{\Delta}_{\sigma})} - \sum_{k = 0}^{\infty} \left[\frac{ \mathcal{N}_{J, \Delta_\sigma} c_k(J, \Delta_\sigma)}{J + \Delta_\sigma + k}  + \frac{ \mathcal{N}_{J, \smash{\bar{\Delta}}_{\sigma}} c_k(J, \bar{\Delta}_\sigma)}{J + \bar{\Delta}_\sigma + k}\right]\right] 
			\end{aligned}
		\end{equation}
		where $c_k(J, \Delta_\sigma)$ are the series coefficients (\ref{eq:propagatorExpCoeff}). 
		Since $c_{k}\big(\minus[ \Delta_\sigma + k], \Delta_\sigma\big) = \delta_{k, 0}$, the second term  is purely analytic in $J$, and thus subleading near $J = \Delta_\sigma$. Furthermore, this also implies that $c_{k}\big(\minus[ \Delta_\sigma + \Delta_\varphi + k], \Delta_\sigma\big) = \delta_{k, 0} + \mathcal{O}(\Delta_\varphi)$ and so, to leading order in $\Delta_\varphi \propto m_\varphi^2$, the bubble drastically simplifies to
		\begin{equation}
			[G^\sigma G^\varphi]_{J} \approx \left[\frac{\Gamma(\alpha+1)}{2 \pi^{\alpha + 1}} \frac{1}{m_\varphi^2}\right] \frac{1}{(J + \Delta_\sigma + \Delta_\varphi)(J + \bar{\Delta}_{\sigma} + \Delta_\varphi)} + \frac{1}{ 8 \pi^2 \epsilon}\,.
		\end{equation}
		A similar story applies to the sunset diagram discussed in \S\ref{sec:sunset}.

\phantomsection
\addcontentsline{toc}{section}{References}
\bibliographystyle{utphys}
\bibliography{light.bib}

\end{document}